\documentclass{elsarticle}
\usepackage{hyperref}
\usepackage{graphicx}  
\usepackage{dcolumn}   
\usepackage{bm}        
\usepackage{amssymb}   
\usepackage{amsfonts}
\usepackage{graphics}
\usepackage{grffile}   
\usepackage{epsfig}
\usepackage{siunitx}
\usepackage[normalem]{ulem} 
\usepackage[utf8]{inputenc}
\usepackage[T1]{fontenc}
\usepackage{amsmath}
\usepackage{subfigure}
\usepackage{capt-of}
\usepackage{float}
\biboptions{numbers,sort&compress}

\usepackage{siunitx}
\usepackage{multirow}
\usepackage{multicol}
\usepackage{arydshln} 
\usepackage{threeparttable}
\usepackage{cleveref}
\usepackage{upgreek}

\DeclareSIUnit\ele{e^{\text{-}}}

\usepackage{lineno}
\usepackage{comment}
\usepackage{booktabs}
\usepackage{xcolor}

\journal{Nucl. Instrum. Methods Phys. Res. A}
\begin{document}

\begin{frontmatter}
\title{Characterisation of analogue Monolithic Active Pixel Sensor test structures implemented in a 65 nm CMOS imaging process}

\author[1]{Gianluca Aglieri Rinella}
\author[9]{Giacomo Alocco}
\author[31]{Matias Antonelli}
\author[31]{Roberto Baccomi}
\author[10]{Stefania Maria Beole}
\author[2]{Mihail Bogdan Blidaru}
\author[2]{Bent Benedikt Buttwill}
\author[1]{Eric Buschmann}
\author[5,31]{Paolo Camerini}
\author[1]{Francesca Carnesecchi\corref{cor1}}
\author[15]{Marielle Chartier}
\author[19]{Yongjun Choi}
\author[13]{Manuel Colocci}
\author[5,31]{Giacomo Contin}
\author[1]{Dominik Dannheim}
\author[20]{Daniele De Gruttola}
\author[22,28]{Manuel Del Rio Viera}
\author[24]{Andrea Dubla}
\author[1]{Antonello di Mauro}
\author[2]{Maurice Calvin Donner}
\author[4]{Gregor Hieronymus Eberwein}
\author[3]{Jan Egger}
\author[3]{Laura Fabbietti}
\author[22]{Finn Feindt}
\author[14,23]{Kunal Gautam}
\author[3]{Roman Gernhaeuser}
\author[8]{James Julian Glover}
\author[8]{Laura Gonella}
\author[1]{Karl Gran Grodaas}
\author[22,28]{Ingrid-Maria Gregor}
\author[1]{Hartmut Hillemanns}
\author[22]{Lennart Huth}
\author[14]{Armin Ilg}
\author[25]{Artem Isakov}
\author[15]{Daniel Matthew Jones}
\author[1]{Antoine Junique}
\author[30]{Jetnipit Kaewjai}
\author[1]{Markus Keil}
\author[17]{Jiyoung Kim}
\author[1]{Alex Kluge}
\author[30]{Chinorat Kobdaj}
\author[7,26]{Artem Kotliarov}
\author[16]{Kritsada Kittimanapun}
\author[7]{Filip K\v{r}\'i\v{z}ek}
\author[1]{Gabriela Kucharska}
\author[7]{Svetlana Kushpil}
\author[11,27]{Paola La Rocca}
\author[30]{Natthawut Laojamnongwong}
\author[1,3]{Lukas Lautner}
\author[32]{Roy Crawford Lemmon}
\author[1,12]{Corentin Lemoine}
\author[8]{Long Li}
\author[27]{Francesco Librizzi}
\author[15]{Jian Liu}
\author[14]{Anna Macchiolo}
\author[1]{Magnus Mager}
\author[9]{Davide Marras}
\author[1]{Paolo Martinengo}
\author[2,24]{Silvia Masciocchi}
\author[18]{Serena Mattiazzo}
\author[1]{Marius Wilm Menzel}
\author[9]{Alice Mulliri}
\author[15]{Mia Rose Mylne}
\author[1]{Francesco Piro}
\author[31]{Alexandre Rachevski}
\author[11,27]{Marika Ras\`a}
\author[1]{Karoliina Rebane}
\author[1]{Felix Reidt}
\author[20]{Riccardo Ricci}
\author[22,28]{Sara Ruiz Daza}
\author[27]{Gaspare Sacc\`a}
\author[1,3]{Isabella Sanna}
\author[9]{Valerio Sarritzu}
\author[22,29]{Judith Schlaadt}
\author[2]{David Schledewitz}
\author[13]{Gilda Scioli}
\author[12]{Serhiy Senyukov}
\author[22,28]{Adriana Simancas}
\author[1]{Walter Snoeys}
\author[22]{Simon Spannagel}
\author[1]{Miljenko \v{S}ulji\'c}
\author[21,27]{Alessandro Sturniolo}
\author[1]{Nicolas Tiltmann}
\author[21,27]{Antonio Trifir\`o}
\author[9]{Gianluca Usai}
\author[22]{Tomas Vanat}
\author[1]{Jacob Bastiaan Van Beelen}
\author[3]{Laszlo Varga}
\author[5,31]{Michele Verdoglia}
\author[22,28]{Gianpiero Vignola}
\author[5,31]{Anna Villani}
\author[22]{Haakan Wennloef}
\author[15,6]{Jonathan Witte}
\author[14]{Rebekka Bettina Wittwer}

\affiliation[1]{organization={European Organisation for Nuclear Research (CERN)}, city={Geneva}, country={Switzerland}}
\affiliation[9]{organization={University and INFN of Cagliari}, city={Cagliari}, country={Italy}}
\affiliation[10]{organization={University and INFN of Torino}, city={Torino}, country={Italy}}
\affiliation[2]{organization={Ruprecht Karls Universit\"at Heidelberg}, city={Heidelberg}, country={Germany}}
\affiliation[19]{organization={University of Pusan}, city={Pusan}, country={South Korea}}
\affiliation[13]{organization={University and INFN of Bologna}, city={Bologna}, country={Italy}}
\affiliation[5]{organization={University of Trieste}, city={Trieste}, country={Italy}}
\affiliation[31]{organization={INFN of Trieste}, city={Trieste}, country={Italy}}
\affiliation[20]{organization={University and INFN of Salerno}, city={Salerno}, country={Italy}}
\affiliation[24]{organization={GSI Helmholtzzentrum f{\"u}r Schwerionenforschung}, city={Darmstadt}, country={Germany}}
\affiliation[4]{organization={University of Oxford}, city={Oxford}, country={United Kingdom}}
\affiliation[3]{organization={Technical University Munich}, city={Munich}, country={Germany}}
\affiliation[22]{organization={Deutsches Elektronen-Synchrotron (DESY)}, city={Hamburg}, country={Germany}}
\affiliation[14]{organization={University of Zurich}, city={Zurich}, country={Switzerland}}
\affiliation[23]{organization={Vrije Universiteit Brussel}, city={Brussels}, country={Belgium}}
\affiliation[8]{organization={University of Birmingham},city = {Birmingham}, country ={United Kingdom}}
\affiliation[7]{organization={Nuclear Physics Institute of the Czech Academy of Sciences}, city={Rez}, country={Czechia}}
\affiliation[16]{organization={Synchrotron Light Research Institute}, city={Nakhon Ratchasima}, country={Thailand}}
\affiliation[17]{organization={University of Inha}, city={Inha}, country={South Korea}}
\affiliation[11]{organization={University of Catania}, city={Catania}, country={Italy}}
\affiliation[15]{organization={ University of Liverpool}, city={Liverpool}, country={United Kingdom}}
\affiliation[18]{organization={University and INFN of Padova}, city={Padova}, country={Italy}}
\affiliation[12]{organization={Université de Strasbourg, CNRS, IPHC UMR 7178}, city={Strasbourg}, country={France}}
\affiliation[21]{organization={University of Messina}, city={Messina}, country={Italy}}
\affiliation[25]{organization={Nikhef National institute for subatomic physics}, city={Amsterdam}, country={Netherlands}}
\affiliation[26]{organization={Faculty of Nuclear Sciences and Physical Engineering, Czech Technical University},city={Prague}, country={Czechia}}
\affiliation[27]{organization={INFN of Catania}, city={Catania}, country={Italy}}
\affiliation[28]{organization={Rheinische Friedrich-Wilhelms-Universität Bonn}, city={Bonn}, country={Germany}}
\affiliation[29]{organization={Johannes Gutenberg-Universität Mainz}, city={Mainz},  country={Germany}}
\affiliation[30]{organization={Suranaree University of Technology}, city={Nakhon Ratchasima},  country={Thailand}}
\affiliation[6]{organization={University of Tübingen}, city={Tübingen}, country={Germany}}
\affiliation[32]{organization={STFC Daresbury Laboratory}, city={Daresbury}, country={United Kingdom}}
\cortext[cor1]{Corresponding author}

\begin{abstract}
Analogue test structures were fabricated using the Tower Partners Semiconductor Co. CMOS 65 nm ISC process. The purpose was to characterise and qualify this process and to optimise the sensor for the next generation of Monolithic Active Pixels Sensors for high-energy physics. The technology was explored in several variants which differed by: doping levels, pixel geometries and pixel pitches (10--25~\textmu m). These variants have been tested following exposure to varying levels of irradiation up to 3~MGy and $10^{16}$~1~MeV~n$_\text{eq}$~cm$^{-2}$. 
Here the results from prototypes that feature direct analogue output of a 4$\times$4 pixel matrix are reported, allowing the systematic and detailed study of charge collection properties. Measurements were taken both using $^{55}$Fe X-ray sources and in beam tests using minimum ionizing particles.
The results not only demonstrate the feasibility of using this technology for particle detection but also serve as a reference for future applications and optimisations.

\end{abstract}


\end{frontmatter}

%
%
\section{Introduction}
\label{sec:intro}

The use of commercial CMOS imaging sensor technologies for particle sensors in High-Energy Physics (HEP) was successfully demonstrated over the last decade with the notable examples of STAR PIXEL~\cite{star_2012, MAPS_CONTIN2018} (based on ULTIMATE also known as MIMOSA28 sensors),
and the ALICE ITS2~\cite{ALPIDE-proceedings-1,ALPIDE-proceedings-2,MAPS_REIDT2022} (based on ALPIDE sensors).

This motivated the ALICE experiment and CERN EP R$\&$D~\cite{CERN_EP} to set out a program to investigate the use of the 65 nm CMOS Image Sensor (65~nm~ISC)   technology of Tower Partners Semiconductor Co. (TPSCo)~\cite{TPSCo} for future vertex and tracking detectors. Following a sensor optimization employing the same principle as for the 180 nm technology~\cite{SNOEYS201790}, several test chips were designed and fabricated in a first run to study charge collection properties and qualify the 65 nm technology for HEP.

The foreseen upgrade of the ALICE Inner Tracking System, ITS3~\cite{ITS3_loi}, proposed for installation in 2026--2028 during the Long Shutdown 3 (LS3) of the Large Hadron Collider (LHC) at CERN, will be the first application of this technology in HEP. Important requirements for this application are a spatial resolution better than 5 \textmu m, particle detection efficiency larger than~99\% after an exposure to $\sim$10~kGy of Total Ionizing Dose (TID) and 10$^{13}$~1~MeV~n$_\text{eq}$~cm$^{-2}$ of Non Ionizing Energy Loss (NIEL). 
The ALICE 3 Vertex Tracker proposed to be installed during the LHC LS4 in 2033--2034, will impose even more agressive requirements for Monolithic Active Pixels Sensors (MAPS), including a radiation tolerance of $\sim$300 kGy and 1.5 $\times$ 10$^{15}$~1~MeV~n$_\text{eq}$~cm$^{-2}$ per year of operation~\cite{ALICE3_loi}.
\section{The Analog Pixel Test Structure, APTS}
\label{sec:apts_chip}
The APTS is a 1.5~mm $\times$ 1.5~mm prototype chip, produced in the TPSCo CMOS 65nm ISC technology, containing a matrix of 4$\times$4 pixels. Each pixel is equipped with an analogue output individually buffered and connected to an output pad, to provide full access to the time evolution of the signal. The matrix of these 16 pixels is surrounded by a ring of dummy pixels to minimise distortion of the electric field. The chip was produced with four different pixel pitches: 10, 15, 20 and 25~\textmu m.

\begin{figure}[htpb!!]
    \begin{minipage}{0.38\hsize}
\includegraphics[width=1\hsize]{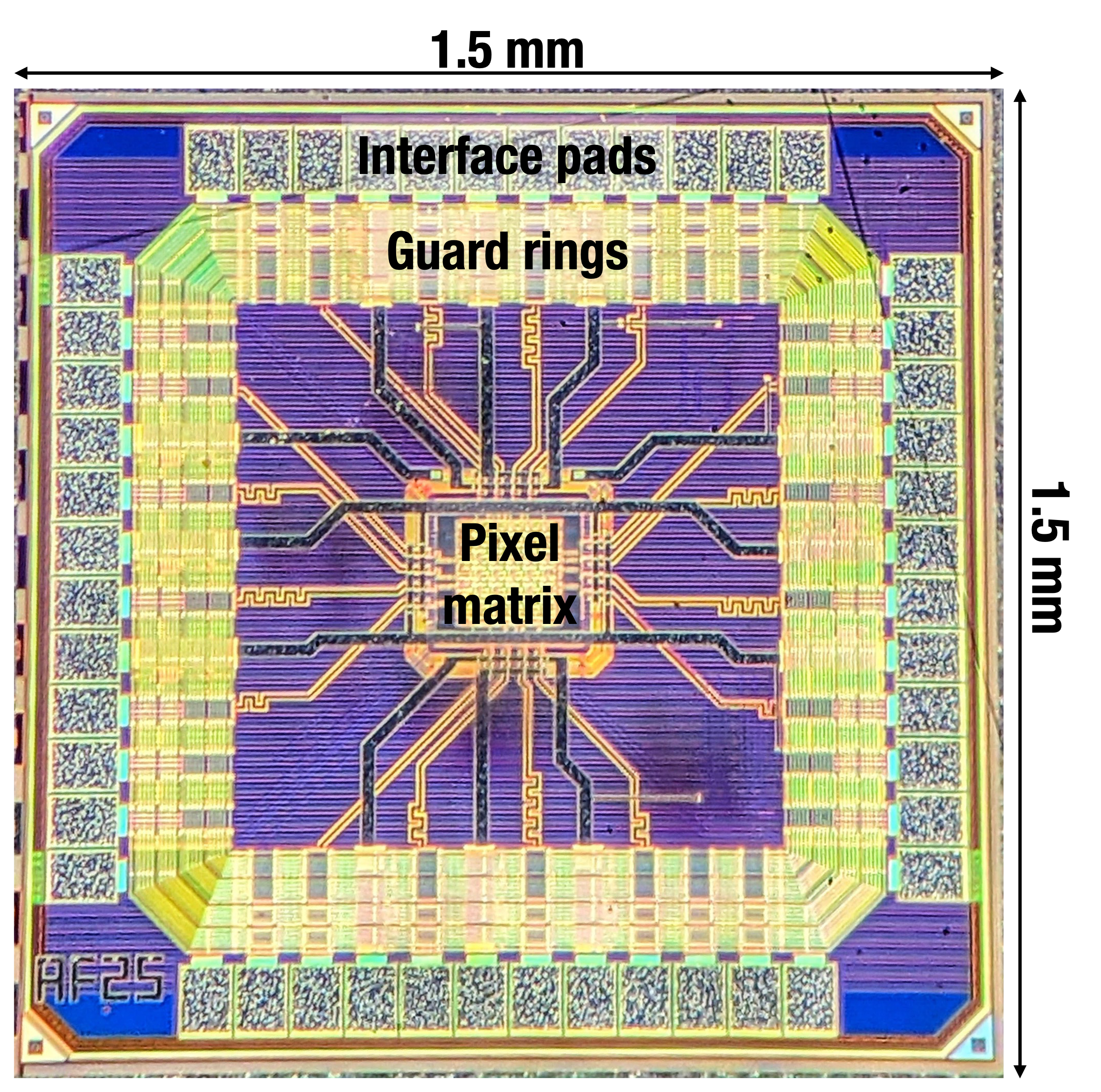}
  \captionof{figure}{Photo of the APTS chip under a microscope.}
  \label{fig:apts}
\end{minipage}
    \hfill
\begin{minipage}{0.61\hsize}
\small
\setlength{\tabcolsep}{3pt}
\begin{tabular}{lc}
 \toprule
 & APTS \\
\midrule
\textbf{die size} & 1.5 mm $\times$ 1.5 mm  \\
\textbf{matrix} & 6$\times$6 pixels  \\
\textbf{readout} & direct analogue of central 4$\times$4 pixels   \\
\textbf{pitch} & 10, 15, 20 or 25 \textmu m \\
\textbf{design} & standard, mod., mod. with gap \\
\textbf{split} & 1, 2, 3, 4 \\
\textbf{variant} & ref., larger n-well, smaller p-well, finger p-well \\
\bottomrule
\end{tabular}
\captionof{table}{Main characteristics of the APTS silicon sensors.}
    \label{tab:table_char}
\end{minipage}
    \end{figure}
The APTS allows reverse substrate voltages in the range of 0 to -5~V. Figure~\ref{fig:apts} shows a photo of the chip under the microscope, and Table~\ref{tab:table_char} reports its main characteristics, together with its several versions detailed below.

The pixel and process optimisation to enhance the sensor performance in the 180~nm TowerJazz imaging technology~\cite{SNOEYS201790,Munker_2019,Dyndal_2020} formed the basis for the process optimisation in the TPSCo CMOS 65~nm~ISC technology. Figure~\ref{fig:variants} shows schematic cross-sections of three different  sensor \textit{designs} implemented in different versions of the APTS:
 \begin{figure}[h]
        \centering%
        \subfigure[\label{fig:standard}]%
          {\includegraphics[height=4.33cm]{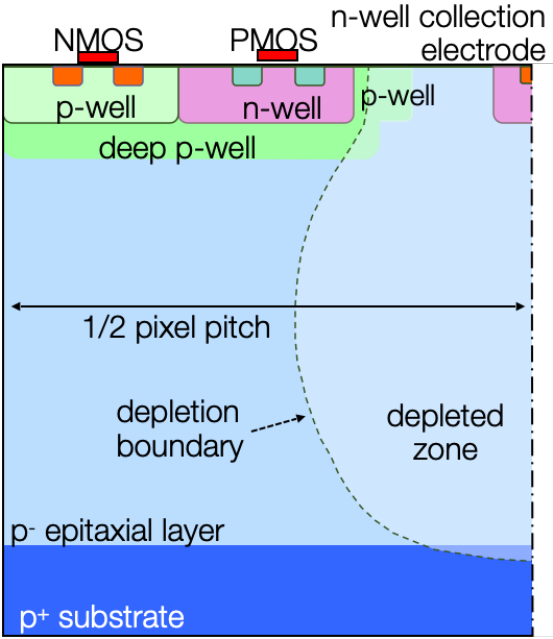}}
        \centering
        \subfigure[\label{fig:blanket}]%
          {	\includegraphics[height=4.33cm]{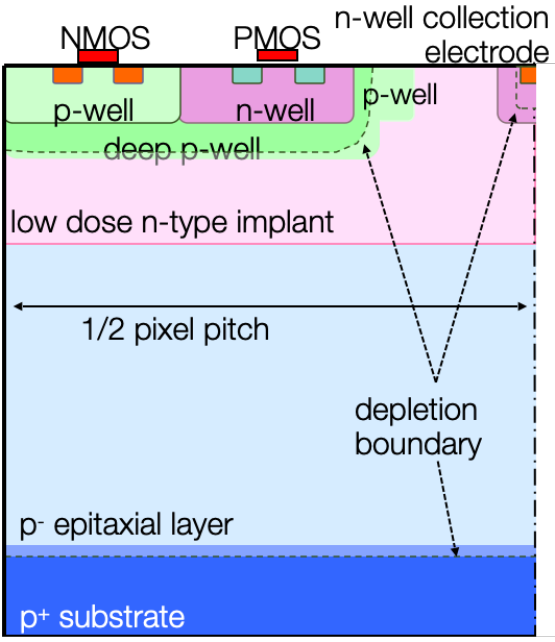}}
        \centering
        \subfigure[\label{fig:p_mod}]%
          {	\includegraphics[height=4.37cm]{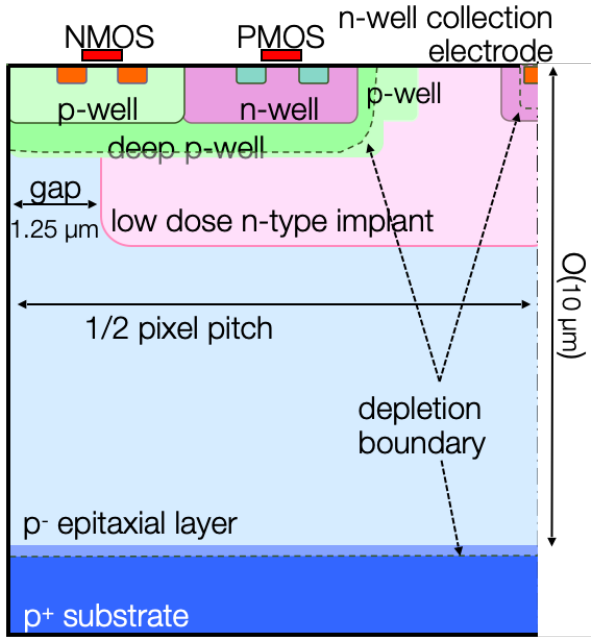}}
\caption{\label{fig:variants}Cross-sections of three different pixel designs implemented: (a) standard: no additional low-dose implant, (b) modified: with deep blanket low dose n-type implant, and (c) modified with gap: with a gap in the deep low dose n-type implant at the pixel borders.}
\end{figure}
\begin{itemize}
    \item \textbf{Standard}, Figure~\ref{fig:standard}: this pixel design features a collection electrode, formed by an n-well diffusion on a high-resistivity p-type epitaxial layer, grown on top of a low-resistivity p-type substrate (ALPIDE-like design). The in-pixel circuits are placed outside the collection electrode and inside a deep p-well, which shields n-wells within the circuitry from the epitaxial layer, preventing them from collecting charge and thus enabling full CMOS circuitry. The depletion zone is balloon-shaped extending from the junction at the collection electrode without reaching the pixel edges.
    The charge generated outside the depletion region is collected primarily by diffusion,  but this collection is relatively slow and subject to charge trapping in defects generated by exposure to non-ionizing radiation.
    \item \textbf{Modified}, Figure~\ref{fig:blanket}: a deep low-dose n-type implant is added under the full pixel area and creates a planar junction deep in the sensor, separated from the collection electrode extending over the full pixel area. With some reverse substrate bias, this allows full depletion of the epitaxial layer and signal charge collection by drift.
    \item \textbf{Modified with gap}, Figure~\ref{fig:p_mod}: a gap of 2.5~\textmu m is introduced in the low-dose n-type implant at the pixel boundaries. This creates a vertical junction to increase the lateral electric field pushing the charge from the pixel boundary towards the collection electrode, thus reducing the charge sharing among neighbouring pixels. 
\end{itemize} 

Moving from the pixel design~\ref{fig:standard} to~\ref{fig:p_mod}, charge collection by drift becomes largely dominant, reducing charge sharing among pixels. This results in a larger fraction of the charge collected by a single pixel and, consequently, in an increased signal to noise ratio (S/N) providing greater operational margins. 

To optimise the sensor for ionizing-particle detection, where signal charge is generated over the full depth, the modified with gap design has been produced in four so-called \textit{splits} (named 1, 2, 3, 4), gradually modifying the doping levels of various implants~\cite{TPSCo65_2023}.

\begin{figure}[H]
        \centering%
        \subfigure[\label{fig:mux_ref_sh}]%
          {\includegraphics[height=4.4cm]{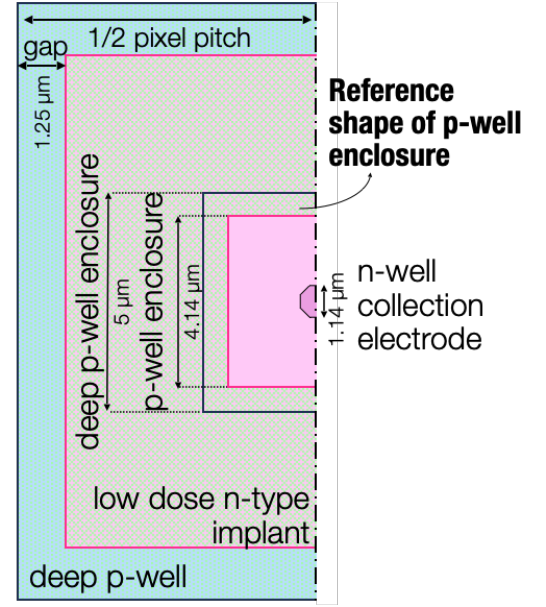}}
        \centering
        \subfigure[\label{fig:mux_fing_sh}]%
          {	\includegraphics[height=4.4cm]{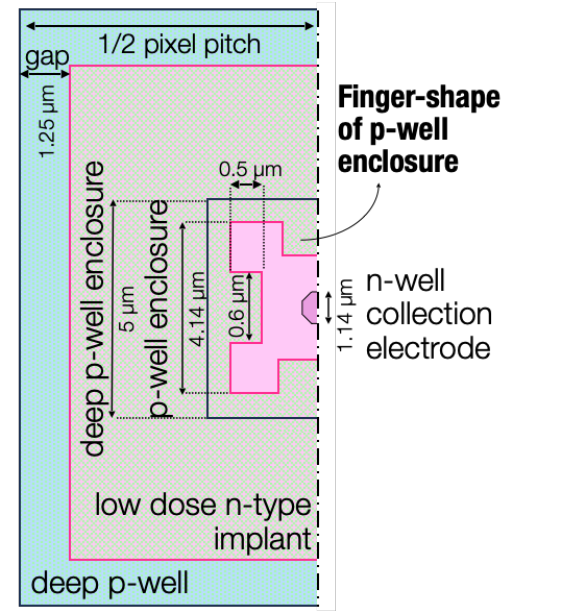}}
        \caption{\label{fig:mux} Top view of half a pitch of a modified with gap pixel design for different shapes of the deep p-well: (a) reference shape and (b) finger-shaped p-well enclosure.}
\end{figure}

In addition, to explore the influence on capacitance and charge collection, the geometry and size of both n-well collection electrode and the p-well were implemented in four different \textit{variants}: 
\begin{itemize}
    \item \textbf{Reference}: n-well collection electrode size of 1.14~\textmu m and a square p-well enclosure of 4.14~\textmu m, see Figure~\ref{fig:mux_ref_sh} for the top view.
    \item \textbf{Smaller p-well enclosure}: p-well enclosure of 3.14~\textmu m.
    \item \textbf{Larger n-well collection electrode}: n-well collection electrode of 2.28~\textmu m.
    \item \textbf{Finger-shaped p-well enclosure}: the shape of the p-well enclosure is changed as sketched in Figure~\ref{fig:mux_fing_sh} resulting in p-well fingers further approaching the collection electrode.
\end{itemize} 

Table ~\ref{tab:sensor_tests} lists a summary of all the sensors studied in this paper, including the various reverse substrate voltages applied, and NIEL and TID irradiation levels\footnote{The NIEL irradiation was performed with neutrons at JSI Ljubljana. The TID irradiation was carried out with 10 keV X-rays from a tungsten target at CERN.} tested.

\begin{table}[htpb!!]
  \centering
  \begin{threeparttable}
  \caption{Summary table of all the APTS types and irradiation levels reported in this paper. NIEL unit is 1~MeV~n$_\text{eq}$~cm$^{-2}$, TID unit is Gy.}
  \label{tab:sensor_tests}
  \footnotesize
  \setlength{\tabcolsep}{2pt}
  \begin{tabular}{ccccccc}
    \toprule
    \textbf{Pitch(\textmu m)} & \textbf{Design} & \textbf{Split} & \textbf{Variant} & \textbf{$V_\text{sub}$(V)} & \textbf{NIEL}  & \textbf{TID}\\
    \midrule
    \multirow{3}{*}{10} & \multirow{1}{*}{standard} & 4 & ref. & -1.2 & None\tnote{*} & None\tnote{*}\\
    & \multirow{1}{*}{modified} & 4 & ref.& -1.2  & None & None\\
    & \multirow{1}{*}{mod. with gap} & 4 & ref. & 0 -- -4.8 & None,  10$^{15}$, 2$ \times$10$^{15}$\tnote{**} & None\\
    \hline
    \multirow{6}{*}{15} & \multirow{1}{*}{standard} & 4 & ref. & 0 -- -4.8 & None & None\\
    & \multirow{1}{*}{modified} & 4 & ref.& 0 -- -4.8  & None & None\\
    \cdashline{2-5}
    & \multirow{4}{*}{mod. with gap} & 1 & ref. & -4.8 & None\tnote{*}  & None\tnote{*}\\
    & & 2 & ref. & -4.8 & None\tnote{*}  & None\tnote{*}\\
    & & 3 & ref. & -4.8 & None\tnote{*}  & None\tnote{*}\\
    & & \multirow{2}{*}{4} & \multirow{2}{*}{ref.} & \multirow{2}{*}{0 -- -4.8}& None, 10$^{13}$, 10$^{14}$, 10$^{15}$, & \multirow{2}{*}{ 3$ \times$10$^6$}\\
    & & & & & 2$ \times$10$^{15}$\tnote{*}, 5$ \times$10$^{15}$\tnote{*}, 10$^{16}$\tnote{*}  & \\
    \hline
    \multirow{5}{*}{\begin{tabular}{@{}c@{}}20\end{tabular}} & \multirow{1}{*}{standard} & 4 & ref.& -1.2  & None\tnote{*} & None\tnote{*}\\
    \cdashline{2-5}
    & \multirow{4}{*}{\begin{tabular}{@{}c@{}}mod. with gap\end{tabular}} & \multirow{4}{*}{\begin{tabular}{@{}c@{}}4\end{tabular}} & ref. & -1.2 & None, 10$^{13}$\tnote{**} , 10$^{14}$\tnote{**} , 10$^{15}$& None\\
    & & & larger n-well & -1.2 & None\tnote{*} & None\tnote{*}\\
    & & & finger p-well & -1.2 & None\tnote{*} & None\tnote{*}\\
    & & & smaller p-well  & -1.2 & None\tnote{*} & None\tnote{*}\\
    \hline
    \multirow{3}{*}{25} & \multirow{1}{*}{standard} & 4 & ref. & -1.2 & None\tnote{*} & None\tnote{*}\\
    & \multirow{1}{*}{modified} & 4 & ref.& -1.2  & None & None\\
    & \multirow{1}{*}{mod. with gap} & 4 & ref. & -1.2  & None, 10$^{13}$\tnote{**} , 10$^{14}$\tnote{**} , 10$^{15}$   & None\\
    \bottomrule
  \end{tabular}
   \begin{tablenotes}
             \item[*] Results reported only for laboratory setup.
            \item[**] Results reported only for beamtest setup.
        \end{tablenotes}
  \end{threeparttable}
\end{table}

\subsection{Readout circuitry}
\label{subsec:pixel_FE}
\begin{figure}[hbt!!]
\centering
\includegraphics[width=0.85\textwidth]{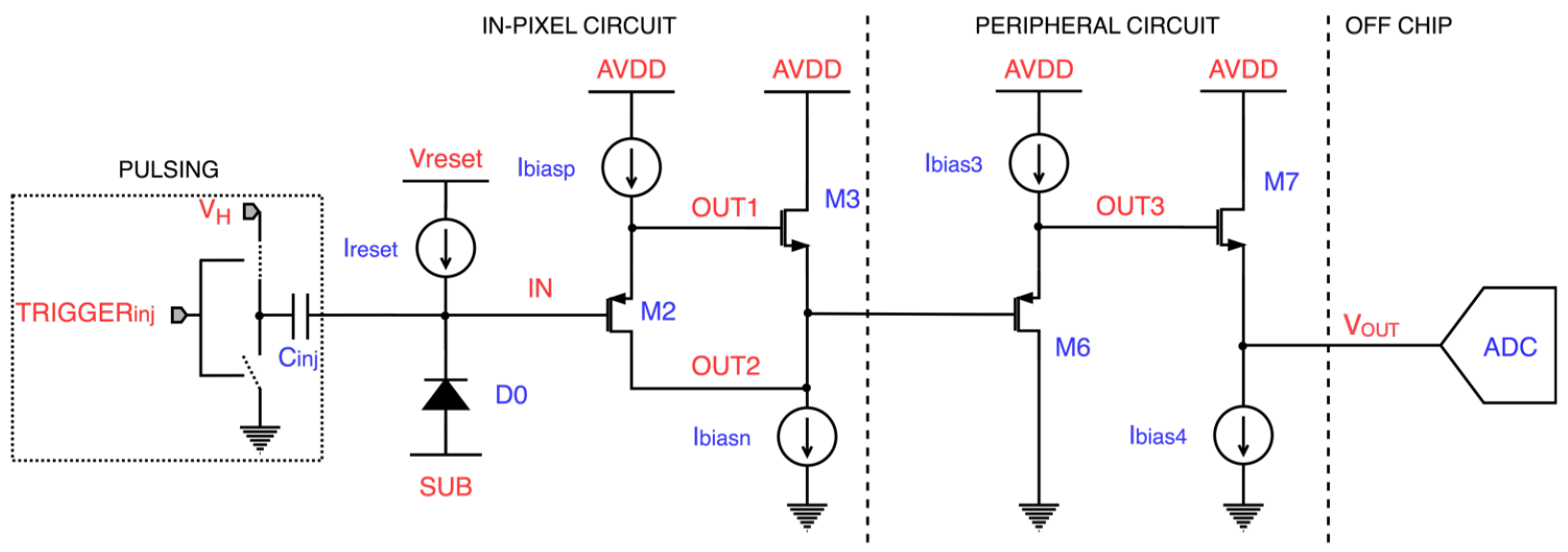}
\caption{A schematic of the front-end chain of APTS.}
\label{fig:schematics}
\end{figure}

The schematic of the readout chain of each pixel in the APTS is shown in Figure~\ref{fig:schematics}. 
The sensor is represented by the diode D0. The buffer chain is partially implemented in-pixel and partially in the periphery. 
The collection electrode is biased and reset using a constant current mechanism. At rest the current source $I_\text{reset}$ only compensates the sensor leakage current, and the DC voltage on the collection electrode is close to $V_\text{reset}$. Only after a hit the current source delivers the constant current $I_\text{reset}$, which has to be larger than the sensor leakage: the difference between the two is used to reset the pixel. 

The collection electrode is DC-connected to the in-pixel circuit. The latter is composed of two source-follower stages, a PMOS (the current source $I_\text{biasp}$ and M2) and an NMOS follower (M3 and the current source $I_\text{biasn}$). The output of the second follower is connected to the drain of the input transistor in the first stage. Because of this connection, both the source and the drain of the input transistor of the readout chain follow the voltage on the collection electrode which reduces the capacitive load on the collection electrode. The two source-follower stages in the peripheral circuit buffer the output signal of each individual pixel in the matrix and send it via an analogue output pad to an off-chip ADC. 

A pulsing circuit inside the pixel allows the injection of charge through an injection capacitor $C_\text{inj}$ = 242~aF (nominal value) into the collection electrode, when the TRIGGER$_\text{inj}$ is given.
 The injected charge can be tuned through the voltage setting $V_\text{H}$.
\section{Laboratory results}
\label{sec:lab_results}

The chip is operated and read out by a custom test system consisting of an FPGA-based data acquisition board and a proximity card. The latter provides power and biases to the chip and hosts ADCs that can sample all 16 pixels at 4~MSPS, corresponding to a sampling period of 250~ns. More details on the test system can be found in~\cite{Sarritzu_2023}.

\subsection{Signal shape and extraction}
\label{subsec:waveforms_signal}
\begin{figure}[h]
\centering
\includegraphics[width=0.6\textwidth]{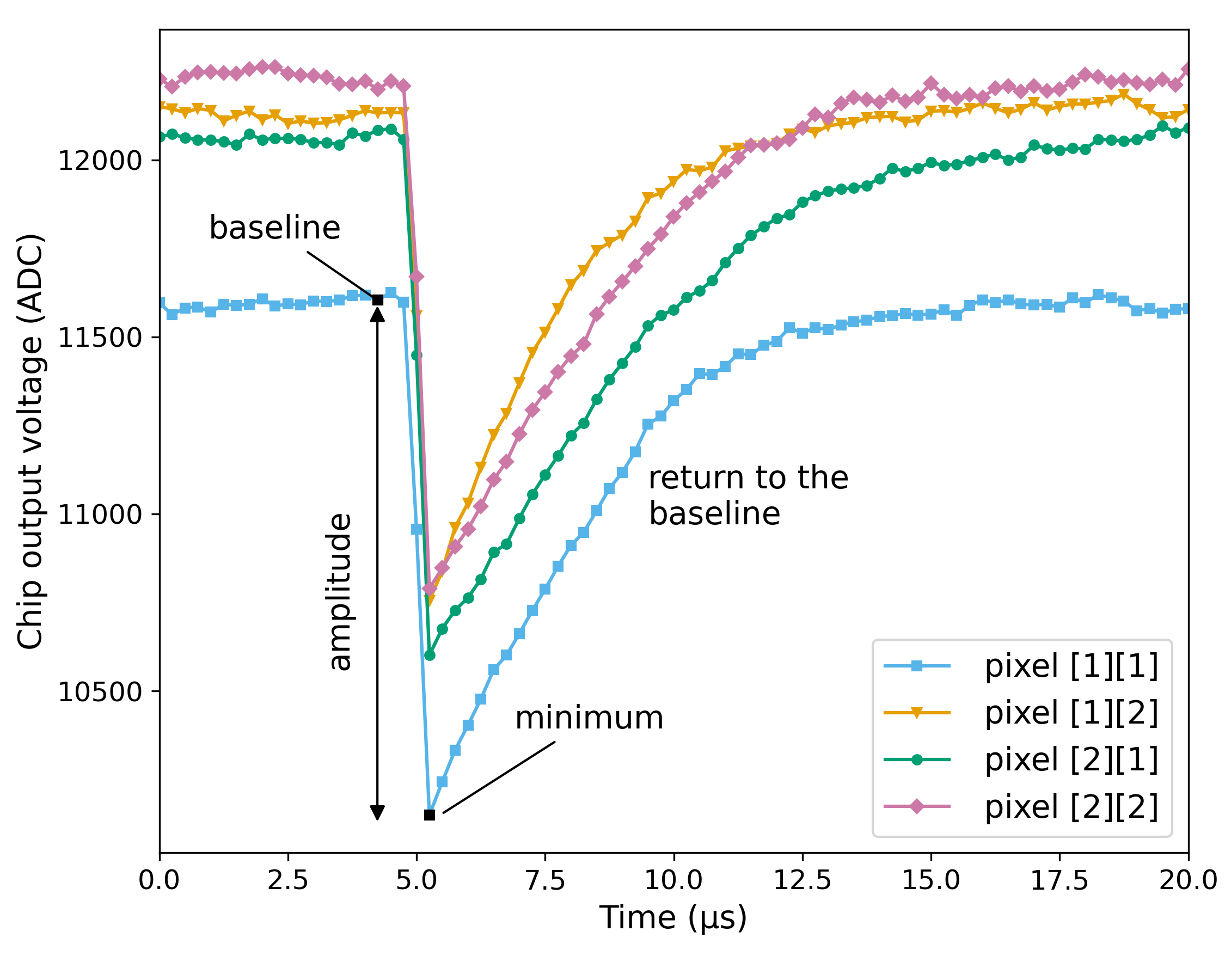}
\caption{Typical uncalibrated chip output signal from the 4 innermost pixels of an APTS chip pulsed with $V_\text{H}$ = 1.2~V. APTS with 15 \textmu m pitch, modified with gap, split 4, reference variant, $V_\text{sub}$~= -1.2~V.}
\label{fig:waveforms}
\end{figure}

Figure~\ref{fig:waveforms} shows typical uncalibrated (conversion factor, see ~\cite{Sarritzu_2023}, and front-end gain factor, see in the following, not applied) output signals $V_\text{OUT}$ obtained by pulsing the chip with a charge of around 1800 electrons at a reverse substrate voltage of 1.2~V.
The signal amplitude is defined as the difference between the baseline and the minimum of the signal. The baseline value is evaluated as the output in the 4$^\text{th}$ sampling before the minimum. 
This is because a significant short-range auto-correlation was observed among close samples, progressively decreasing for more distant ones (up to 24\% degradation on the energy resolution depending on the frames chosen for the baseline definition). 
The 4$^\text{th}$ sample has been defined to be as close as possible to the minimum, without being influenced by the signal edge itself.\\
 The pixels have different baselines, but similar signal amplitudes and shapes.
The return to the baseline of the output signal depends on $I_\text{reset}$, and it takes around 10 \textmu s in these conditions at the nominal operating point.

In order to evaluate the noise of the chip, the baseline fluctuations have been studied. They were compatible among all the pixels.
The average noise ranges from 23 to 36 electrons (for the conversion to electrons, see Sect.~\ref{subsec:iron55}), depending on the design and the reverse substrate voltage applied. A detailed study is reported in Figure~\ref{fig:eff_summary}.


The front-end settings were tuned for signal-to-noise (S/N), gain and linearity. The standard settings used in this paper are $I_\text{biasn}$=5~\textmu A, $I_\text{biasp}$=0.5~\textmu A, $I_\text{bias3}$=200~\textmu A, $I_\text{bias4}$=150~\textmu A,  $I_\text{reset}$=100~pA, $V_\text{reset}$=500~mV, unless stated differently. The resulting gain between the IN and $V_\text{OUT}$ nodes shown in Figure \ref{fig:schematics}, is around 0.6 for an input signal range from 200~mV to 600~mV.

\subsection{Leakage current}
\label{subsec:leak}
The leakage current of the chip is expected to increase with reverse substrate voltage, irradiation level and temperature. The increased leakage current effectively reduces the fraction of $I_\text{reset}$ available as reset current, and influences the baseline. 
 \begin{figure}[!hbt]
        \centering
	   \includegraphics[width=1\textwidth]{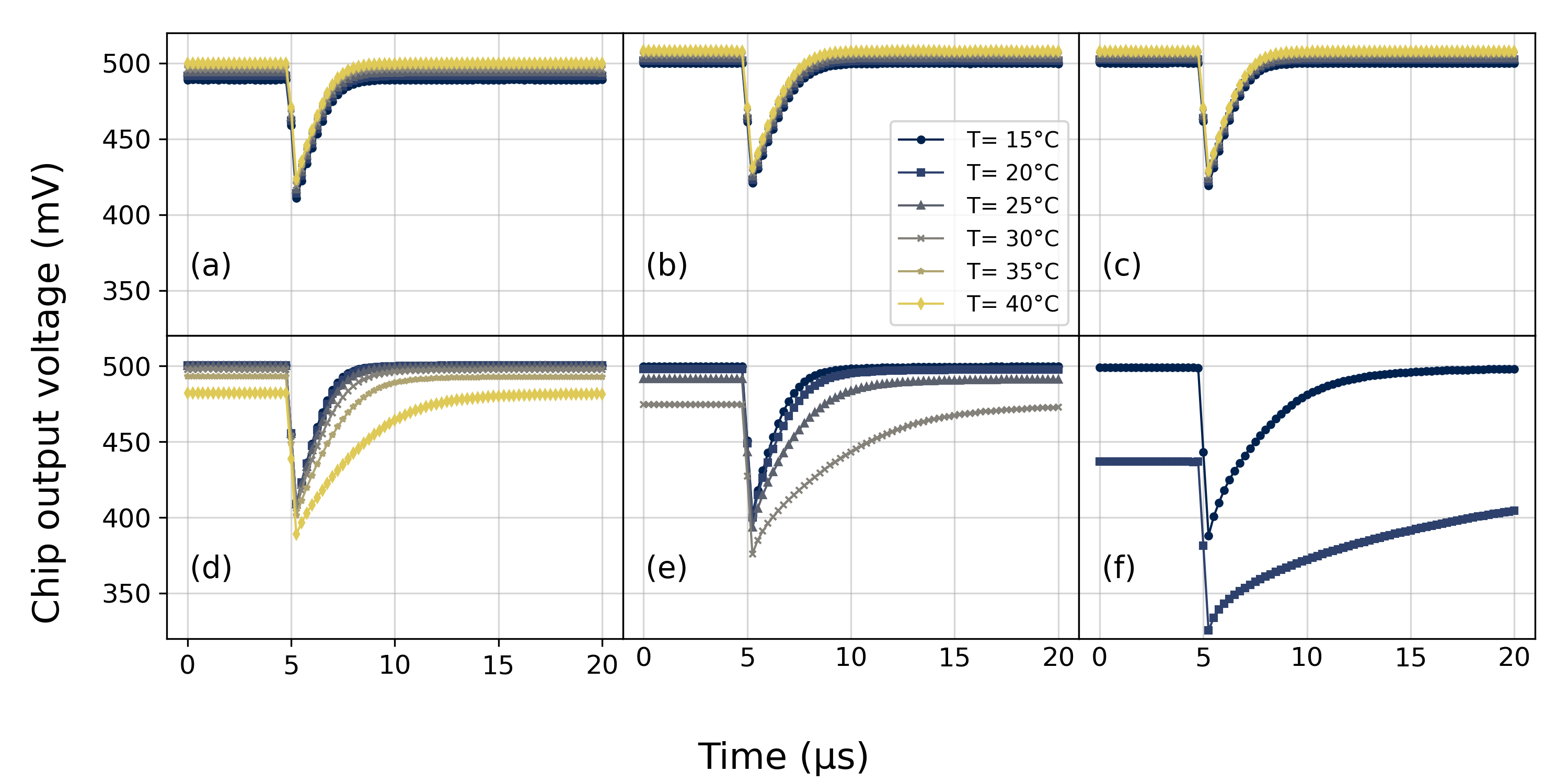}
        \caption{\label{fig:tempirrir25} Comparison at different temperatures of the signal waveform pulsed with $V_\text{H}$ = 1.2~V, averaged over 1000 events, for chips at different irradiation levels, using an optimized working point with increased $I_\text{reset}$ of 250~pA in order to compensate for leakage current: (a) Not irradiated, (b) $10^{13}$~1~MeV~n$_\text{eq}$~cm$^{-2}$, (c) $10^{14}$~1~MeV~n$_\text{eq}$~cm$^{-2}$, (d) $10^{15}$~1~MeV~n$_\text{eq}$~cm$^{-2}$, (e) $2\times10^{15}$~1~MeV~n$_\text{eq}$~cm$^{-2}$ and (f) $10^{16}$~1~MeV~n$_\text{eq}$~cm$^{-2}$. APTS with 15~\textmu m pitch, modified with gap, split 4, reference variant, $V_\text{sub}$~= -1.2~V.}
\end{figure}
Therefore a higher leakage current has to be compensated by increasing $I_\text{reset}$. Larger $I_\text{reset}$  values lead to increased noise (approximately 15\% for $I_\text{reset} = 250$~pA w.r.t. the standard value).
In Figure~\ref{fig:tempirrir25}, the waveforms for several NIEL irradiation doses are reported at different temperatures, with $I_\text{reset}$ = 250~pA\footnote{Higher values of $I_\text{reset}$ are not supported by the present test system so this is the best optimization achievable with the present setup.}. \\
All the temperature values reported in this paper refer to the cooling water temperature set to the chiller; the actual temperature on the chip can be higher by about 1--3~$^{\circ}$C depending on the environment.

\begin{figure}[hbt!!]
\centering
\includegraphics[width=0.6\textwidth]{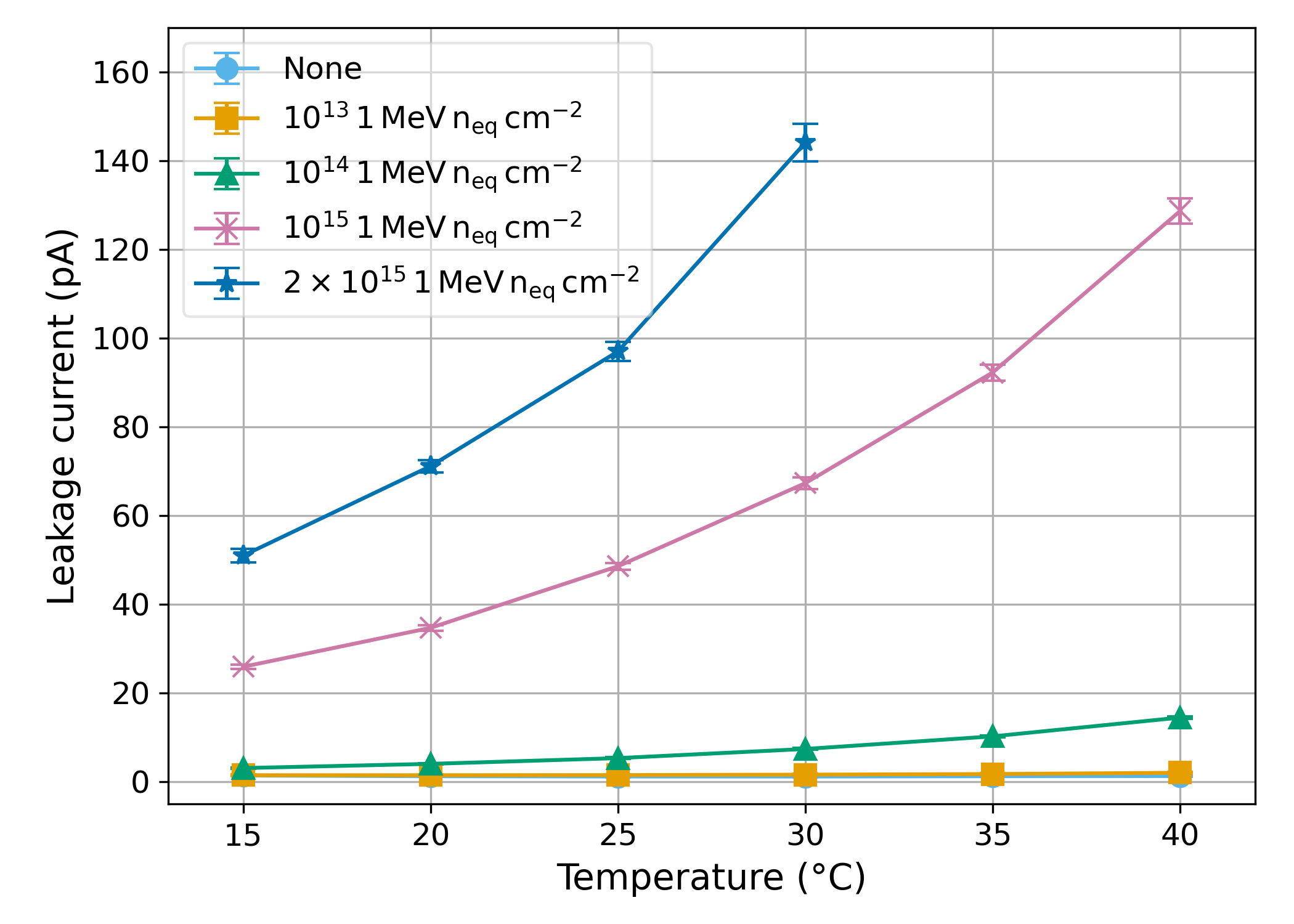}
\caption{Leakage current versus temperature for different irradiation levels at a reverse substrate bias of 1.2 V. APTS with 15 \textmu m pitch, modified with gap, split 4, reference variant.}
\label{fig:leak_temp}
\end{figure}

A direct evaluation of the leakage current can be done by fitting the return to the baseline of the waveform~\cite{2023_david}. The leakage current values obtained are plotted in Figure~\ref{fig:leak_temp} at different temperatures and irradiation levels.
For an unirradiated APTS pixel, a leakage current of around 2~pA has been obtained. 
The leakage current increases with temperature and irradiation becoming appreciable beyond $10^{14}$~1~MeV~n$_\text{eq}$~cm$^{-2}$. 
Therefore all the measurements on the irradiated chips reported in Section~\ref{subsec:iron55} were taken at a temperature of 14~$^{\circ}$C and, for a level of irradiation higher than $10^{15}$~1~MeV~n$_\text{eq}$~cm$^{-2}$, with $I_\text{reset}$ = 250~pA.

\subsection{$^{55}$Fe measurements}
\label{subsec:iron55}

An extensive measurement campaign was performed using a $^{55}$Fe source, to evaluate the performance of the chip in terms of charge collection and energy resolution, and to compare the performance of all the available versions. 
\begin{figure}[hbpt!]
    \subfigure[\label{fig:spectrum}]{{\includegraphics[width=0.45\textwidth]{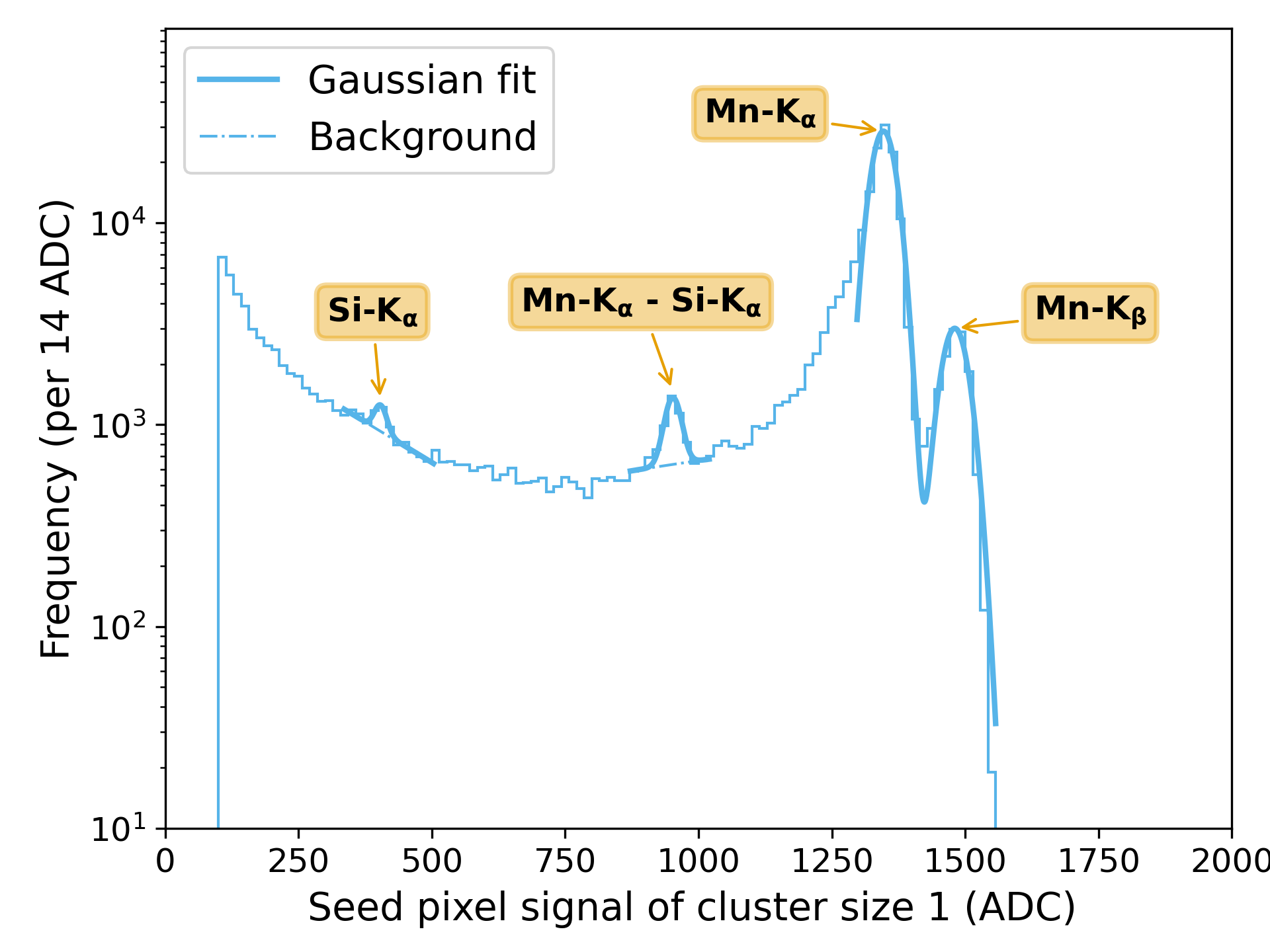} }}%
    \hfill
    \subfigure[\label{fig:en_calib}]{{\includegraphics[width=0.5\textwidth]{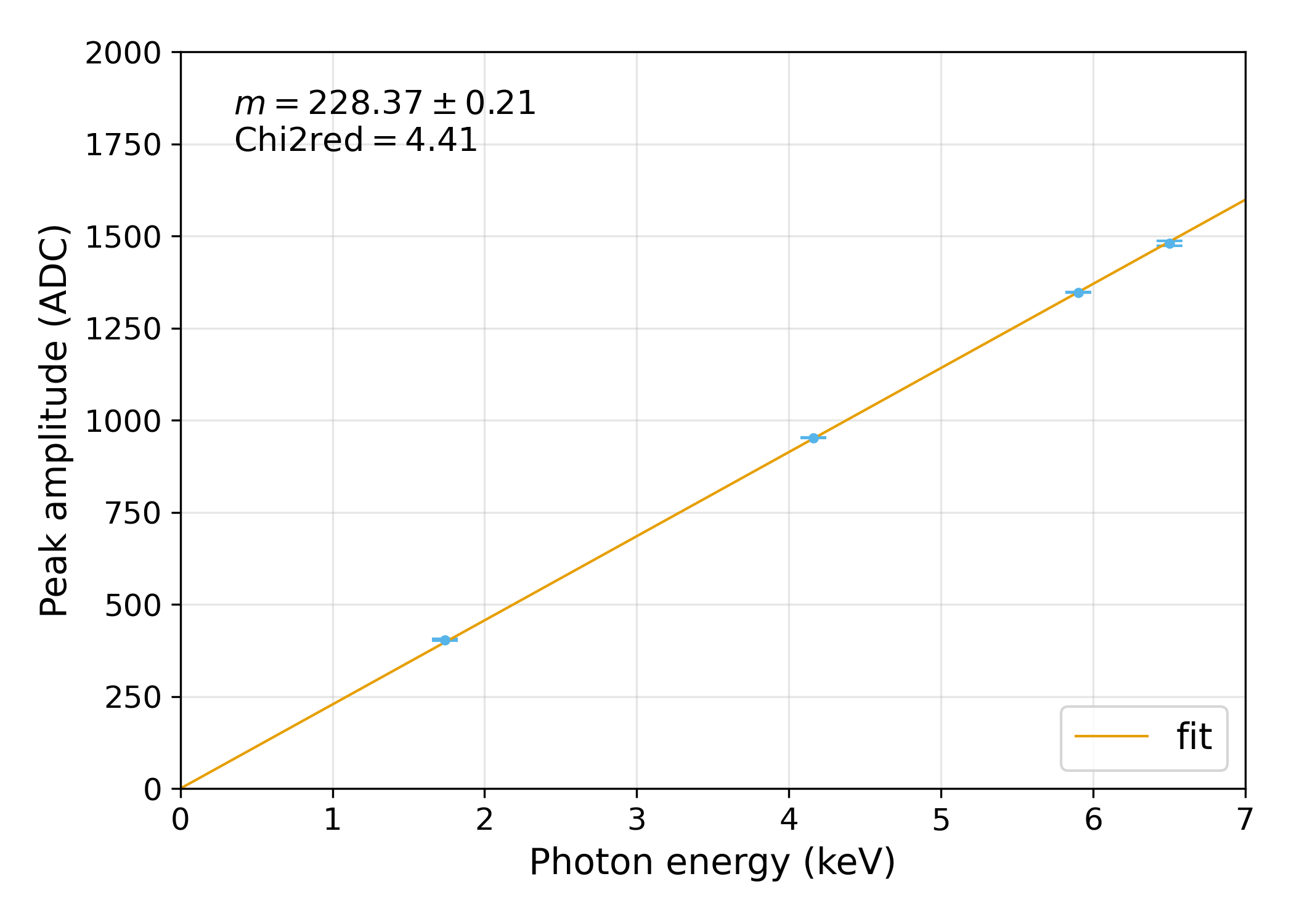} }}%
    \hfill
    \caption{(a) $^{55}$Fe spectrum for events with cluster size 1. The seed distribution is filled with the seed signals of the 4 central pixels of the matrix. (b) Energy calibration was obtained using the peak means from the Gaussian fits on (a). Data are fitted with a linear function, where $m$ is the slope parameter. The errors refer to the ones obtained from the fits. All these measurements were taken at $I_\text{reset}$ = 10~pA for a better energy resolution, as explained in \ref{app:reset_readout}. APTS with 15~\textmu m pitch, modified with gap, split 4, reference variant, $V_\text{sub}$~= -1.2~V.}%
    \label{fig:spectrum_calib}
\end{figure}

An example of an $^{55}$Fe spectrum for the seed signal with a cluster size of 1 is shown in Figure~\ref{fig:spectrum}. The cluster is defined as a set of adjacent pixels inside a 3$\times$3 matrix centered around the pixel with the largest amplitude (called seed) which collected an amplitude higher than a given threshold.
In order to avoid edge effects and have access to all pixels in the cluster, only clusters having as seed one of the central 4 pixels are considered.

The peaks from the spectrum in Figure~\ref{fig:spectrum} were fitted with Gaussian functions and the value of the mean is used for the energy calibration in Figure~\ref{fig:en_calib}. The correlation between the amplitude of the seed pixel signal and the corresponding photon energy is fitted with a linear function, setting the intercept parameter at 0. Asserting the linearity of the energy calibration was a fundamental step to allow the conversion factor method explained in the following.

For every DUT (Device Under Test), the seed signal distribution for every cluster size is fitted using a double Gaussian function, one for each peak of the spectrum, Mn-K$_\alpha$ and Mn-K$_\beta$.
The mean of the most prominent peak, $V_{\text{Mn-K}_\alpha}$, is used to convert mV or ADC units into electrons ($e^-$).
\begin{figure}[h]
    \centering
    \subfigure[\label{fig:seedvbb_mv_fit}]{{\includegraphics[width=0.7\textwidth]{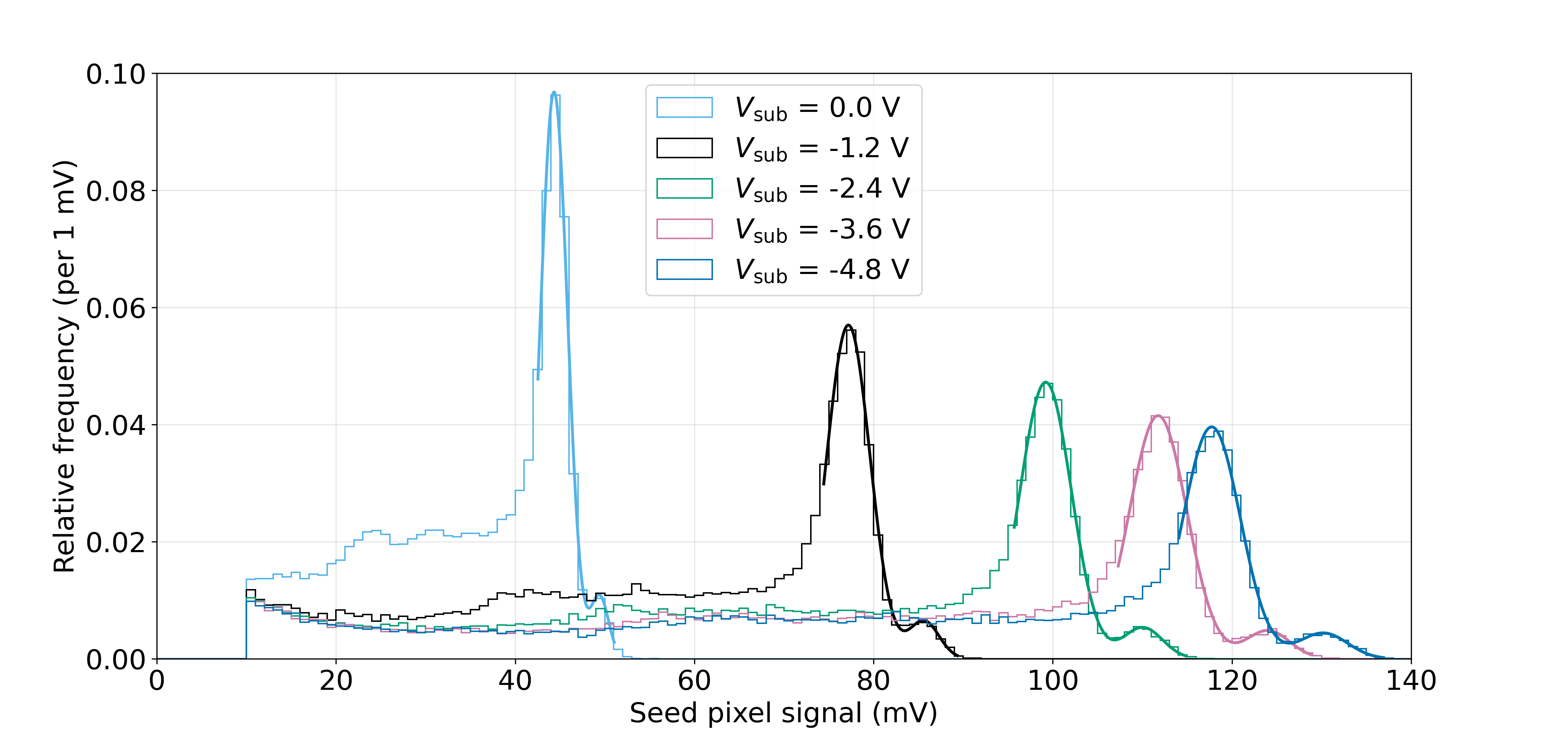} }}
    \\
    \subfigure[\label{fig:seedvbb_el}]{{\includegraphics[width=0.7\textwidth]{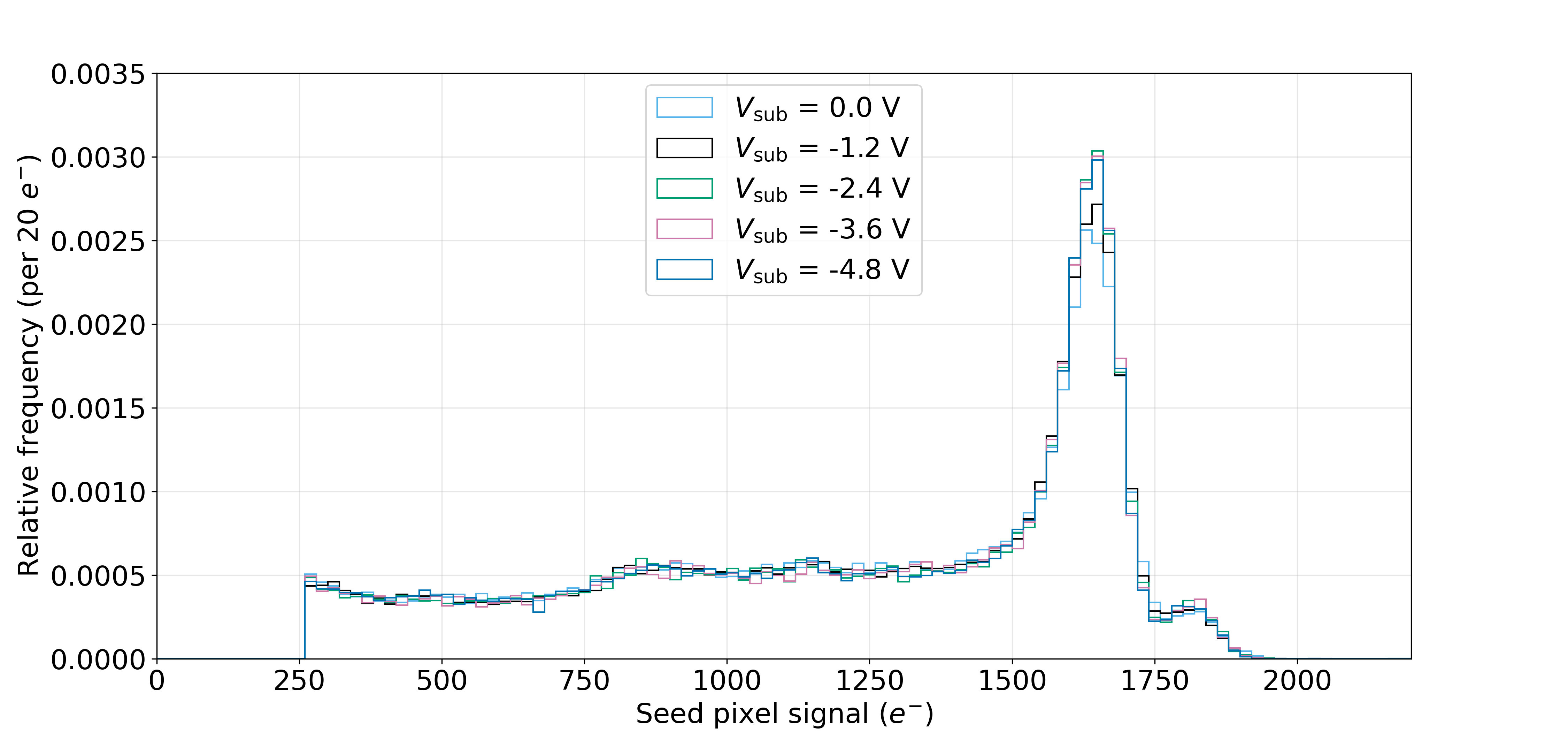} }}
    \caption{Seed signal distribution comparison among different reverse substrate voltages, both in mV (a) and in electrons (b). APTS with 15~\textmu m pitch, modified with gap, split 4, reference variant.}
    \label{fig:seedvbb_fit}
\end{figure}
Figure~\ref{fig:seedvbb_fit} shows an example of an $^{55}$Fe spectrum for the seed pixel signal at different reverse substrate voltages, and illustrates how the amplitude increases with the reverse substrate voltage, due to the increase of depletion region and consequently lower capacitance.
Plotting the same data in electrons eliminates the effect of the capacitance, and illustrates, as in Figure~\ref{fig:seedvbb_el}, that no visible change in the charge distribution  with the reverse substrate voltage can be observed.
For both these plots, the distributions were normalized by the total number of events.

\begin{figure}[!hbt!]
    \centering
	   \includegraphics[width=1.\textwidth]{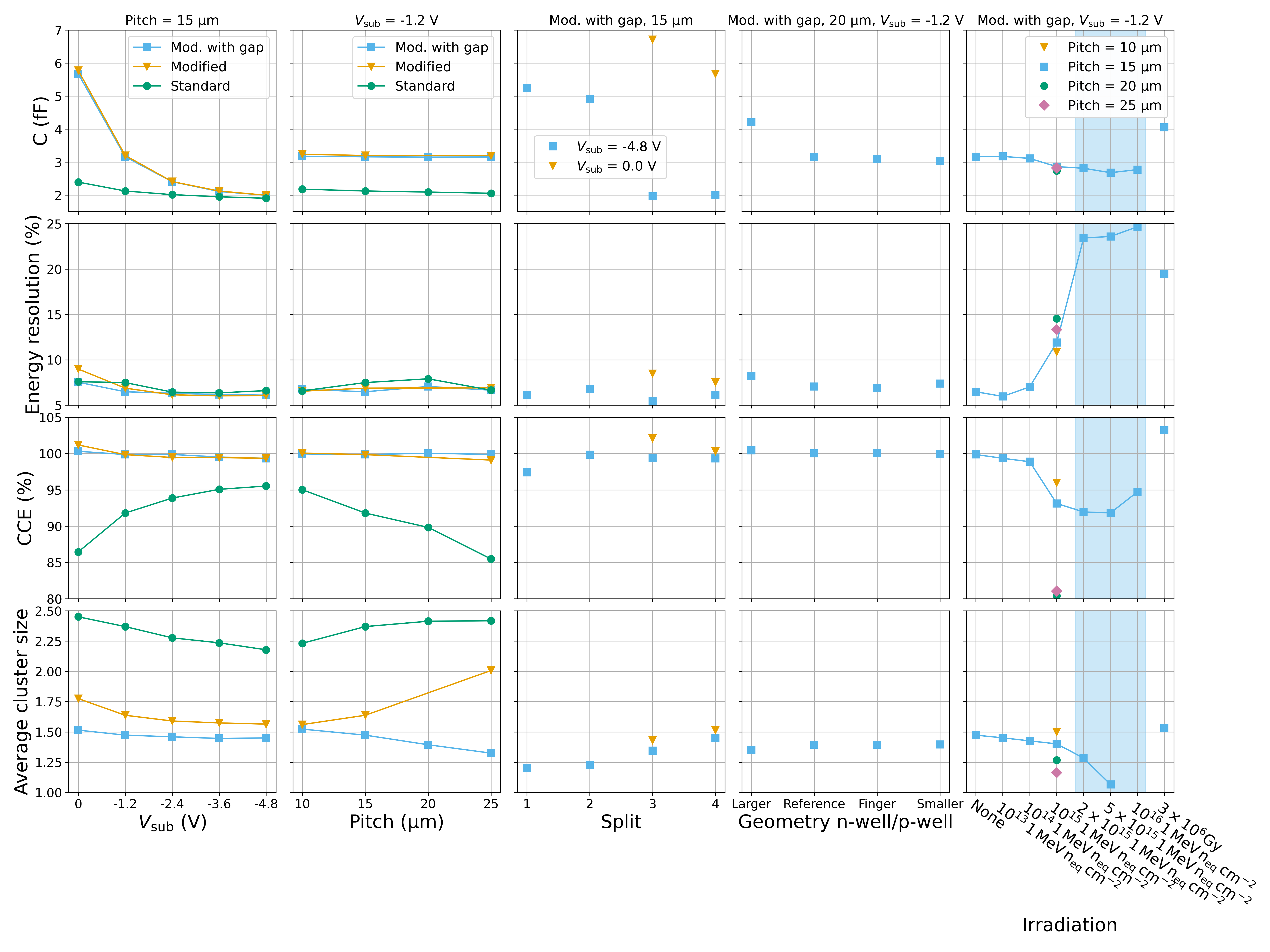}	
    \caption{Comparison of the main characteristics extracted from the $^{55}$Fe spectra. The light blue window in the last column refers to the data points acquired with $I_\text{reset}=~$250~pA.}
    \label{fig:fe55_summary}
\end{figure}

The APTS versions reported in Table~\ref{tab:sensor_tests} have been tested with $^{55}$Fe \footnote{All the spectra are reported in the supplementary material.}. The following quantities have been extracted from the seed distributions and reported in Figure~\ref{fig:fe55_summary}:
\begin{itemize}
    \item \textbf{Input capacitance} (first row): it is related to the measured capacitance:
    \begin{equation}
    C_\text{input} = C_\text{measured} - C_\text{injection}= \frac{n_\text{el}\cdot q_\text{el}}{V_{\text{Mn-K}_\alpha}} - C_\text{injection}
    \end{equation}
    with $V_{\text{Mn-K}_\alpha}$ (mV) the mean of the Gaussian fit around the Mn-K$_\alpha$ peak, $n_\text{el}$ = 1640~electrons, $q_\text{el}$ the elementary charge and $C_\text{injection}\sim$242~aF (ranging from 214 to 269~aF depending on the DUT).
    A smaller input capacitance is very important as it increases the voltage excursion at the input of the front-end circuit for a given charge. This improves the signal with respect to the noise of the front-end circuit, and also increases its reaction speed. Therefore this improves the signal-to-noise ratio and analog performance at the same power consumption, or allows to reduce the front-end circuit power for the same performance.
    \item \textbf{Energy resolution} (second row): obtained as the FWHM (Full Width at Half Maximum) of the Mn-K$_\alpha$ peak divided by its mean, $V_{\text{Mn-K}_\alpha}$.
    \item \textbf{Charge Collection Efficiency ratio, CCE} (third row): the ratio of the most probable value of the 3$\times$3-pixel matrix signal distribution to the most probable value of the signal distribution for cluster size of one.
    \item \textbf{Average cluster size} (fourth row): at a threshold of 150 electrons.
\end{itemize}

Several trends can be observed from the plots in Figure~\ref{fig:fe55_summary}:
\begin{itemize}
    \item 
     In the first column, first row, the sensor capacitance dependence on the reverse substrate bias is reported. The depletion region starts extending from the junction. Therefore, in the standard design, (see Fig.~\ref{fig:variants}), it expands from the collection electrode outward, and reaches the p-well at low reverse bias. Due to the high p-well and n-well doping the depletion cannot extend much further. In addition, the depletion boundary extending into the epitaxial layer is at that point already significantly larger than the undepleted part of the n-well, making the dimension of the latter dominant in the capacitance calculation.
    This explains the small influence of the reverse bias on the sensor capacitance in the standard design. 
    
    For the modified and modified with gap designs, the depletion starts at the junction formed by the low-dose deep n-type implant, and does not yet reach the n-well electrode at zero reverse bias. With increasing reverse bias, it extends towards the n-well electrode and significantly reduces the size of the undepleted region around the collection electrode, dominant for the capacitance value.  Ultimately, the depletion also converges to near the n-well boundary as in the standard design, resulting in the same capacitance value of about 2fF. 

    \item In the first column, last two rows, a lower CCE is observed for the standard type together with a higher average cluster size; this was expected considering that the charge is collected more by drift  for the modified designs with consequent lower charge losses.
    \item In the second column, fixing the voltage at $V_\text{sub}$~= -1.2~V and varying the pitch, lower CCE values can be observed for the standard design for larger pitches; this is attributed to the larger charge transport path together with the  expected smaller (lateral) extension of the depletion region relative to the pixel pitch. Moreover, an increase in average cluster size is observed, both for the standard and modified design, as expected from the increased charge diffusion for larger pitches. The pattern is opposite for the modified with gap design, due to the smaller relative size of the gap region for larger pitches, and the fact that the field induced by the gap counteracts or prevents charge sharing for hits closer to the pixel center.
    \item In the third column, different splits are also compared. To compare all the splits, the reverse substrate voltage has been fixed at $V_\text{sub}$~= -4.8~V. For the splits 1 and 2 the signal only becomes visible at reverse substrate voltages larger than 3.6~V, due to the larger capacitance causing a reduction of the signal amplitude. The splits 3 and 4 are not subject to this larger capacitance and are considered better choices in terms of chip performance. Since split 4 was the most optimized~\cite{TPSCo65_2023}, it was used for all studies reported from this point onwards, with the benefit of a slightly smaller capacitance at $V_\text{sub}$~= 0~V.
    \item In the fourth column the performance of the different geometric variants are compared. All of them show similar behaviour, with the exception of energy resolution and capacitance slightly worse for the larger n-well collection electrode driven by the larger geometry itself.
    \item In the last column, different levels of NIEL irradiation are compared with a non-irradiated chip and with a TID irradiated chip. For NIEL irradiation levels of 2~$\times 10^{15}$~1~MeV~n$_\text{eq}$~cm$^{-2}$ and higher, the measurements have been taken with $I_\text{reset}$ = 250~pA to accommodate the larger sensor leakage (see Section \ref{subsec:leak}). The different $I_\text{reset}$ setting, together with the limited sampling frequency of the readout, might have an influence on the extracted values, further discussed  in~\ref{app:reset_readout}.
    
    Remarkably, for all the irradiation levels, the  DUT continues to work. No major difference is observed between the performance of an unirradiated and a $10^{13}$~1~MeV~n$_\text{eq}$~cm$^{-2}$ chip, satisfying the ALICE ITS3 radiation hardness requirement.
    
    For higher levels of NIEL irradiation, the capacitance reduces, mainly due to a reduction of the radiation-induced effective active doping. 
    For TID on the contrary a larger capacitance is observed. As expected, for higher levels of NIEL irradiation, the energy resolution degrades, due to increase of noise from leakage current and to charge losses from an increase of the radiation-induced trapping centers. These losses also cause a decrease in CCE and average cluster size. 
    It has to be noted that for high levels of irradiation the uncertainty on the CCE calculation method is larger due to the signal distribution deterioration. 
    \item In the last column the comparison of different pixel pitches for a fixed NIEL irradiation of $10^{15}$~1~MeV~n$_\text{eq}$~cm$^{-2}$ is also shown. The energy resolution is more affected by irradiation for larger pixel pitches, again due to a larger shot noise contribution as leakage is collected from a larger volume, and larger charge losses due to the longer collection path, also reflected in the CCE and cluster size.
\end{itemize}

\section{Testbeam setup}
\label{sec:testbeam_setup}
\begin{figure}[!hbt]
\centering
	\includegraphics[width=0.8\textwidth]{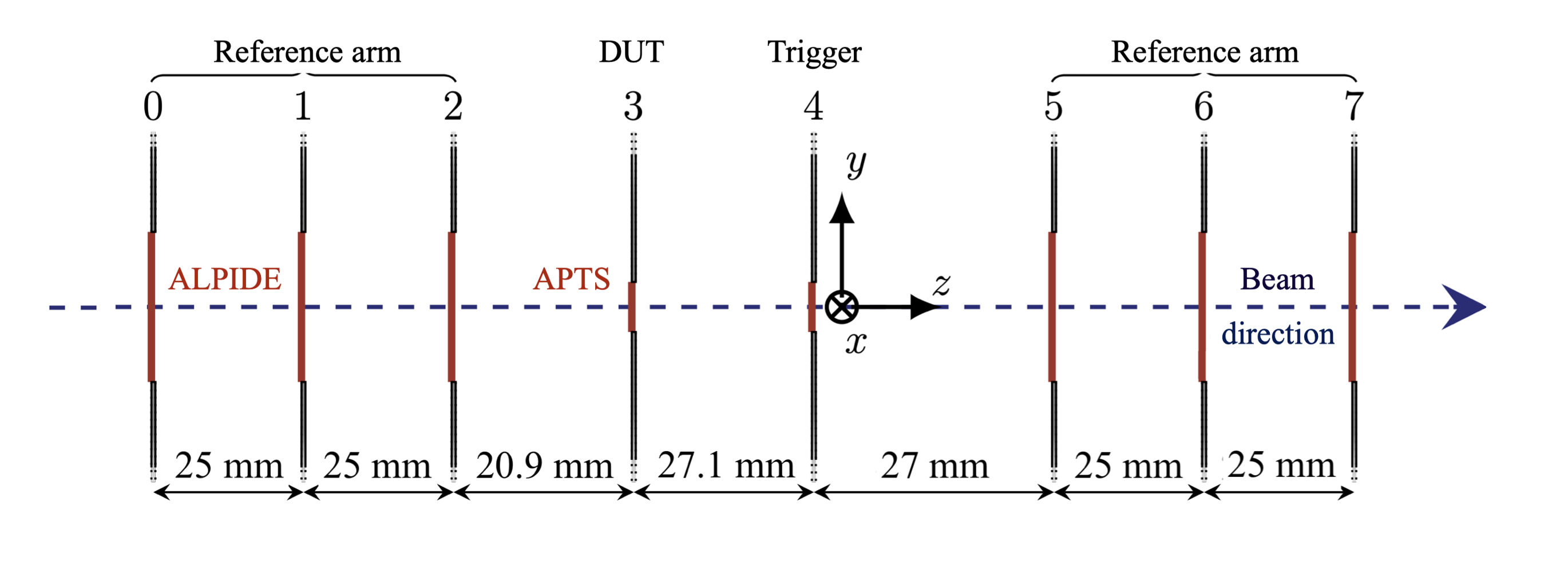}	
\caption{Scheme of the telescope setup used at the SPS test beams. The planes 3 and 4 were made with APTS chips, all others with ALPIDE chips. Not to scale.}
\label{fig:SPS_telescope}
\end{figure}

The APTS chips were also tested with a particle beam of 120~GeV/$c$ positive hadrons, at the CERN--SPS H6~\cite{SPS}, to study the detector charge distribution, detection efficiency and spatial resolution. This was made possible by using a telescope made of ALPIDE chips as reference planes, providing a tracking resolution of $\sigma_\text{track}$ = 2.1~\textmu m~\cite{optimizer}.
The setup was placed inside a box in a light tight environment. Two APTS sensors were mounted between the reference planes, as shown in Figure~\ref{fig:SPS_telescope}. The leftmost APTS was the DUT, while the other was mounted on a movable stage and used as trigger device.
The DUT temperature was kept constant using a cooling jig, which was cooled with constantly circulating water at a temperature of $T$~=~15~°C.\\

\subsection{Analysis tools and methods}
\label{sec:testbeam_results_analysis_methods}
The data were analyzed using the Corryvreckan~\cite{corryvreckan} track reconstruction framework, using the GBL (General Broken Lines) model~\cite{blobel2006new}. Event and track quality selection criteria were applied to ensure a clean data sample: one track per event, track $\chi^2 /n_\text{dof}$~<~5, and track points on each reference plane.
The detection efficiency and the spatial resolution have been computed by associating the DUT clusters to tracks passing inside a search radius of 75~\textmu m around the track intercept point on the DUT and only the tracks passing through the 4 central pixels have been considered, in order to avoid border effects.
The signal and noise extraction has been done as described in Section~\ref{subsec:waveforms_signal}.

The spatial resolution was measured both by treating the information as digital, i.e. hit/no hit, and by leveraging the whole analogue information. The former mimics the usual digital readout of pixel sensors, with the cluster position given by the center of mass of the pixels in the cluster, whereas the latter shows the full potential of the technology.
For the analogue measurement, the so-called $\eta$-algorithm~\cite{turchetta1993spatial,bugiel2021high} has been used to take into account non-linear charge sharing, which is expected to be more and more enhanced going from the standard to modified with gap design.

For both the hit/no-hit and analogue measurements, the residuals have been obtained as the distance between the intercept of the track on the DUT plane and the associated cluster position, both in $x$ and $y$ directions. From the two distributions, the standard deviation was computed and $\sigma_{\text{track}}$ was quadratically subtracted to obtain $\sigma_{\text{x(y)}}$.
The final spatial resolution, $\sigma_{\rm res}$, has been defined as the arithmetic average of $\sigma_{\rm x}$ and $\sigma_{\rm y}$.


All the results are plotted for thresholds above 3 times the RMS of the noise distribution, in order to report only efficiency values not biased by the noise.
\section{Testbeam results}
\label{sec:testbeam_results_new}
\subsection{Charge distribution}
\label{sec:testbeam_results_ch}
\begin{figure}[hbt!!]
\centering
\includegraphics[width=1\textwidth]{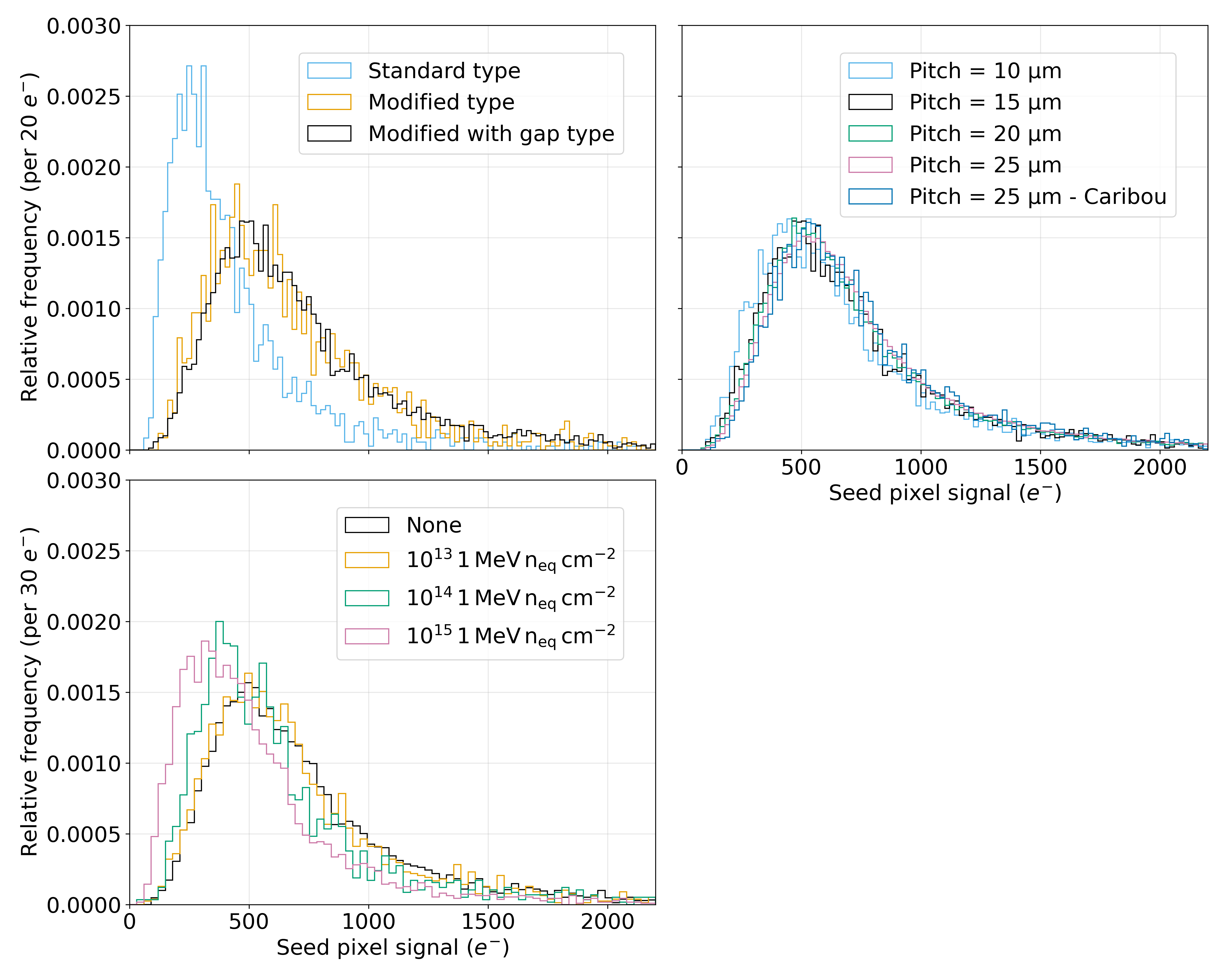}	
\caption{(Top left) Comparison between different designs of the seed pixel signal charge distribution for a pitch of 15~\textmu m. (Top right) Comparison between different pixel pitch sizes of the seed pixel signal charge distribution for modified with gap sensors. For the Caribou measurement, an $I_\text{bias4}$ = 546~\textmu A has been set. (Bottom) Comparison between different NIEL irradiation doses of the seed pixel signal charge distribution for a modified with gap sensor with a pitch 15~\textmu m. APTS with split 4, reference variant, $V_\text{sub}$ = -1.2~V.}
\label{fig:landau_comparison}
\end{figure}

In Figure~\ref{fig:landau_comparison}, the seed pixel signal distributions for the three APTS designs at $V_\text{sub}$ = -1.2~V are shown.
It can be observed in the top left figure that by moving from the standard design to the modified and modified with gap, the Most Probable Value (MPV) for the seed signal increases to higher values, improving the signal-to-noise ratio by a factor~2. 
This behaviour was expected from the lower charge sharing observed in the $^{55}$Fe measurements (see Figure~\ref{fig:fe55_summary}).
Considering an MPV of 500 electrons for the modified with gap, the epitaxial thickness can be estimated to $\sim$ 10 \textmu m \cite{Riegler_2017}.

The seed pixel signal distribution has also been studied for different pitches, as shown in the top right figure. For comparison, a measurement done with a different setup and using the Caribou readout system~\cite{caribou2020}, in a 4~GeV/c electron beam is included. The measurements obtained with both setups are in good agreement; for details on the setup, system and analysis method see~\cite{DESY_beam,eudet_2016,Simancas_2023,aida2019}.
It can be seen that the seed signal becomes slightly higher for increasing pitch, with the MPV moving slightly to the right, in accordance with the decreases in average cluster size with increasing pitch, observed in Figure~\ref{fig:fe55_summary}.

The seed pixel signal distributions for different NIEL irradiation levels are reported in the bottom left figure. The MPV shifts towards lower values with increasing irradiation levels, due to the worse charge collection which becomes evident from $10^{14}$~1~MeV~n$_\text{eq}$~cm$^{-2}$ onwards. As expected from $^{55}$Fe measurements (see Figure~\ref{fig:fe55_summary}), at $10^{13}$~1~MeV~n$_\text{eq}$~cm$^{-2}$ no degradation is observed. 

\subsection{Detection efficiency}
\label{sec:testbeam_results_eff}
\begin{figure}[hbt!]
\centering
    \includegraphics[width=0.8\textwidth]{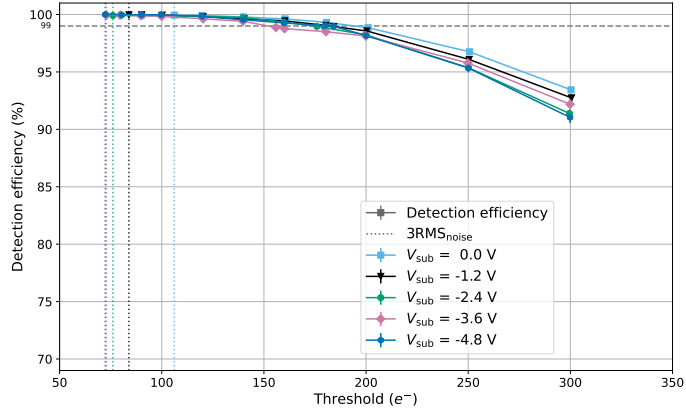}	
\caption{Efficiency comparison between different reverse substrate voltages as a function of the applied seed threshold. The dotted lines indicate the threshold corresponding to 3 times the RMS noise value. APTS with 15~\textmu m pitch, modified with gap, split 4, reference variant.}
\label{fig:eff_AF15P_s}
\end{figure}

Figure~\ref{fig:eff_AF15P_s} shows the detection efficiency of the 15~\textmu m modified with gap sensor for different values of $V_\text{sub}$  as a function of the threshold in electrons ($e^{-}$), for threshold values larger than 3 times the noise RMS.
The detection efficiency is found to nearly entirely depend on the charge threshold, with only a weak dependence on the reverse substrate voltage.
However, a more negative $V_\text{sub}$ yields lower input capacitance and hence a lower noise level, allowing lower thresholds.
The dotted, vertical lines in Figure~\ref{fig:eff_AF15P_s} (the starting point for each curve) illustrate how for more negative voltage the operational region can be extended to lower signal thresholds, increasing the operating margin.

\begin{figure}[!hbt]
    \centering
    \includegraphics[width=1\textwidth]{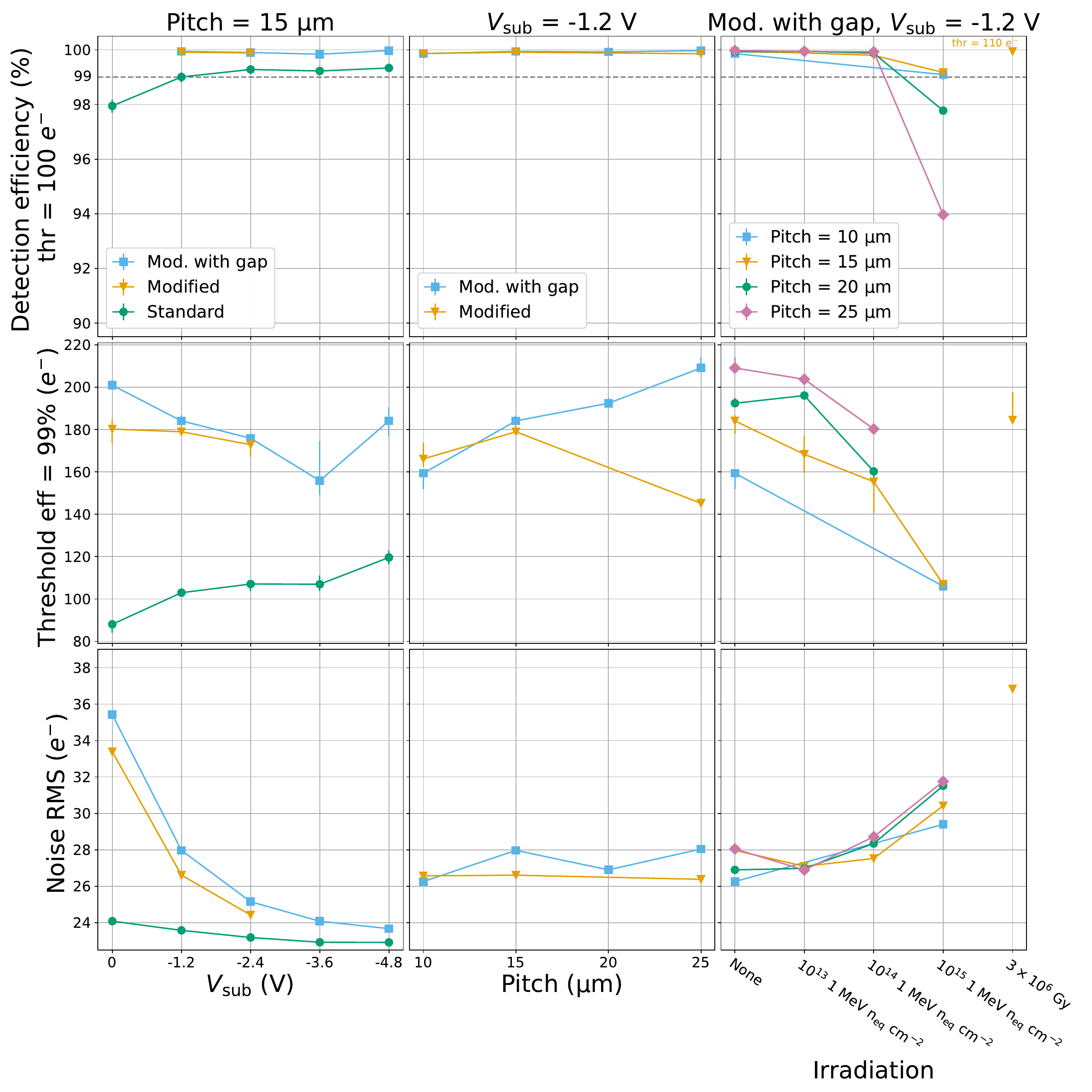}	
    \caption{Comparison of the main characteristics extracted from the efficiency plots versus threshold. APTS with split 4, reference variant.}
    \label{fig:eff_summary}
\end{figure}

All APTS versions reported in Table~\ref{tab:sensor_tests} have been tested\footnote{All the efficiency plots are reported in the supplementary material.}. The following quantities have been extracted from the efficiency plots and reported in Figure~\ref{fig:eff_summary}:
\begin{itemize}
    \item \textbf{Efficiency at a threshold of 100 electrons} (first row): if no value is reported in the plot, it means that 3 RMS$_\text{noise}$ > 100 electrons.
    \item \textbf{Thresholds} (second row): threshold at which the chip achieves 99$\%$ detection efficiency.
    \item \textbf{Noise RMS} (third row): is the noise of the DUT. The distance between the noise and the threshold value at 99$\%$ gives an indication of the trend of the range of operability of the DUT.
\end{itemize}

Several trends can be observed:
\begin{itemize}
    \item In the first column, in the comparison between designs at different V\( _{\text{sub}} \), the modified designs show a larger operational margin w.r.t. the standard type. For higher substrate reverse bias this is more and more evident, due to the significantly reduced noise for modified and modified with gap sensors. This was expected from the larger CCE and smaller average cluster size observed for both the modified and modified with gap designs in Figure~\ref{fig:fe55_summary}.
    \item In the second column, different pixel pitches for the modified and modified with gap designs are compared at $V_\text{sub}$~=~-1.2~V. 
    All the pitches and designs reach an efficiency of more than 99\%. 
    In the modified design the largest pitch shows the smallest operational margin, while the modified with gap shows the opposite. This was expected from the average cluster size trend reported in Figure~\ref{fig:fe55_summary}. Since larger operating margins are observed for the modified with gap sensors, this design has been chosen as a reference for the rest of the study.
    As a further validation, independent measurements of the 25~\textmu m  pitch, using the Caribou system, lead to compatible results. 
    \item In the last column, as for the other chips, all irradiated sensors were operated at a fixed chiller temperature of 15~$^\circ$C. All sensors, regardless of the pitch and for all the reverse substrate voltages studied, can withstand a NIEL irradiation level up to $10^{14}$~1~MeV~n$_\text{eq}$~cm$^{-2}$, higher than the ALICE ITS3 requirement ($10^{13}$~1~MeV~n$_\text{eq}$~cm$^{-2}$). For all pitches, a higher NIEL irradation reduces the operation margin. Up to $10^{14}$~1~MeV~n$_\text{eq}$~cm$^{-2}$, larger pitches show a larger operational margin, but at $10^{15}$~1~MeV~n$_\text{eq}$~cm$^{-2}$ they 
    show the lowest detection efficiency, while the pitches of 10 and 15~\textmu m still reach 
  an efficiency higher than 99\%. For TID irradiations of up to $3 \times 10^6$~Gy (much higher than the ALICE ITS3 requirement and reaching the ALICE 3 requirement), the DUT has reached an efficiency of 99\%, though over a smaller range of operation due to the higher noise. All these results are in agreement with the observations reported in Figure~\ref{fig:fe55_summary}.
\end{itemize}

Figure~\ref{fig:res_summary_10p} compares the influence of reverse substrate bias on the detection efficiency and the noise of sensors which {received high NIEL irradiation dose. For the detection efficiency, the threshold is set to three times the RMS noise. 
\begin{figure}[h]
    \centering
	   \includegraphics[width=0.8\textwidth]{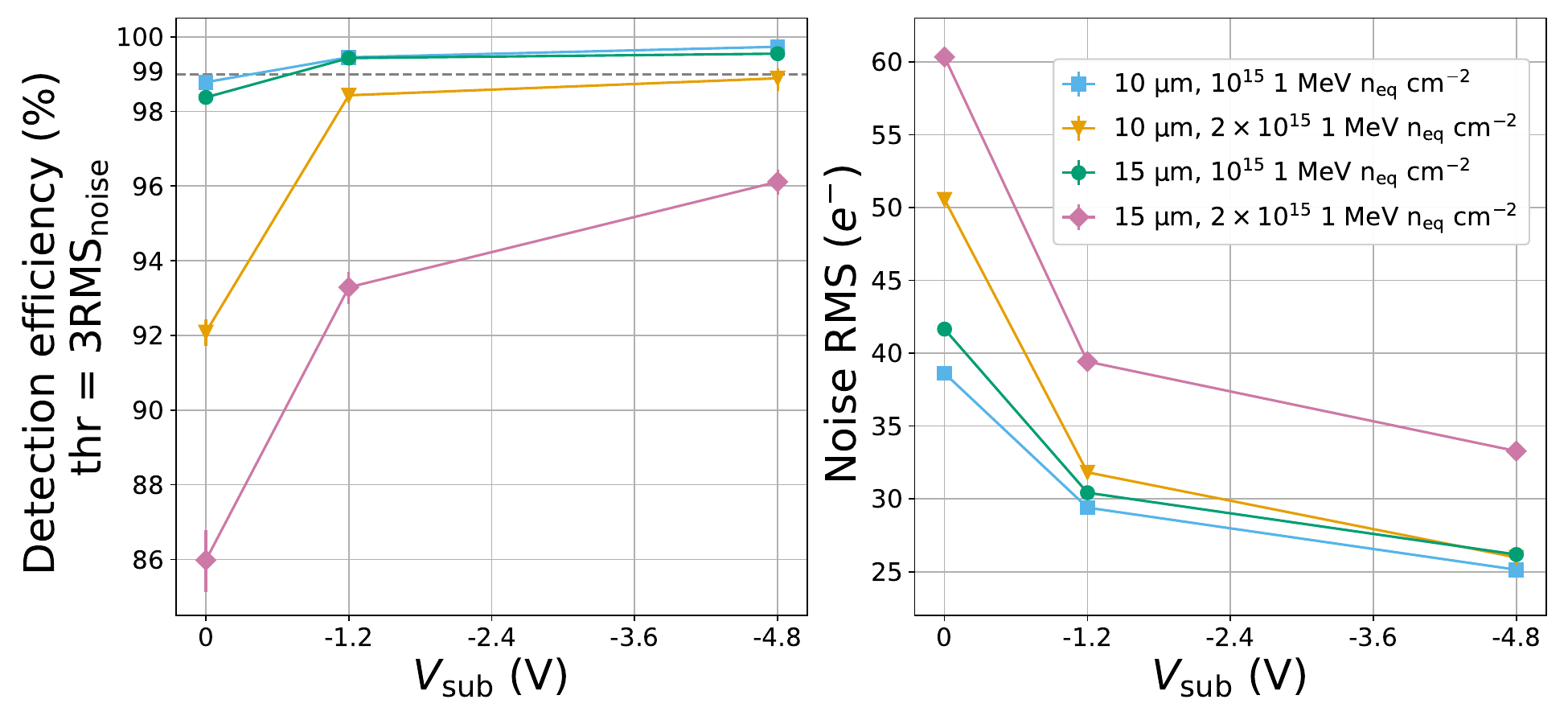}	
    \caption{Comparison of efficiency and noise versus substrate reverse bias for different pitches and radiation levels. APTS with split 4, modified with gap, reference variant.}
    \label{fig:res_summary_10p}
\end{figure}
For the smallest pitch, even at 0~V an efficiency of almost 99\% can be reached for a NIEL of $10^{15}$~1~MeV~n$_\text{eq}$~cm$^{-2}$. This observed radiation hardness, near room temperature, can be attributed to a smaller collection volume resulting in a lower leakage current and hence lower noise, and a lower charge losses for smaller pitches under high radiation dose. For a voltage of -4.8~V and a 10~\textmu m pitch, an efficiency compatible with 99\% can be reached even up to a radiation of 2 $ \times$ $10^{15}$~1~MeV~n$_\text{eq}$~cm$^{-2}$.

\subsection{Spatial resolution and cluster size}
\label{sec:testbeam_results_pos}
In Figure~\ref{fig:res_p_vbb} the spatial resolution and cluster size of a 15~\textmu m modified with gap sensor for different values of $V_\text{sub}$ are reported as a function of the threshold in $e^{-}$. The threshold was required to be greater than 3 times the noise RMS.
\begin{figure}[!hbt!!]
\centering
\includegraphics[width=0.9\textwidth]{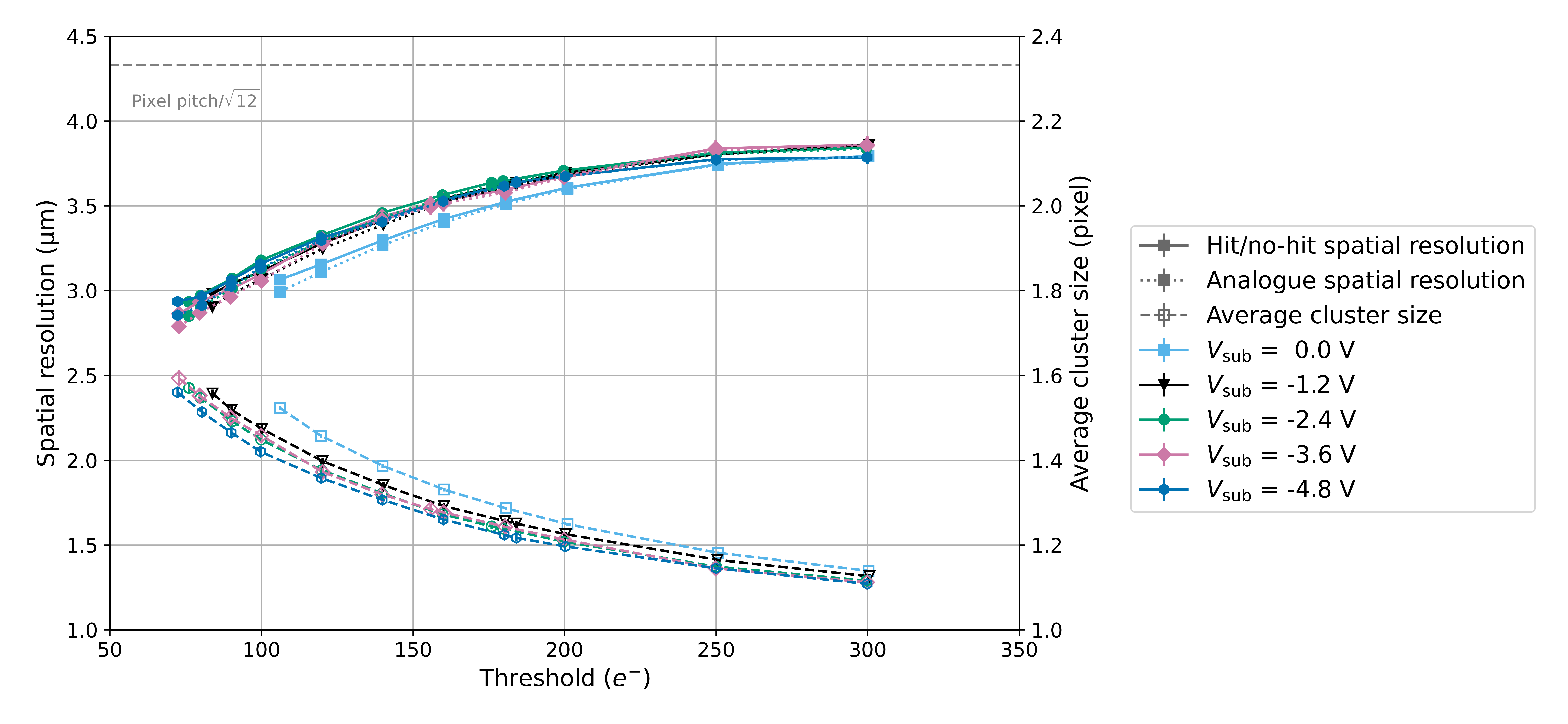}
\caption{Hit/no-hit resolution (solid lines), analogue resolution (dotted lines) and average cluster size (dashed lines) vs threshold at different $V_\text{sub}$ values. The chiller temperature was set to $T =$ 15~°C. APTS with 15~\textmu m pitch, modified with gap, split 4, reference variant.}
\label{fig:res_p_vbb}
\end{figure}
As expected, the cluster size decreases for higher thresholds and leads to a slight deterioration of the spatial resolution.
The average cluster size shows a small decrease going to more negative $V_\text{sub}$. The largest difference is observed when going from $V_\text{sub}$ = 0~V to -1.2~V, which is when most of the depletion process happens. 
As expected from the small cluster size, the difference between the analogue and hit/no-hit resolution is small and noticeable only at low thresholds.

The APTS versions listed in Table~\ref{tab:sensor_tests} have all been tested\footnote{All the resolution plots are reported in the supplementary material.}. The following quantities have been extracted from the resolution plots and reported in Figure~\ref{fig:res_summary}:
\begin{itemize}
    \item \textbf{Spatial resolution and average cluster size at a threshold of 100 electrons} (second and third rows): if a value is not reported, it means that 3 $\times$ RMS$_\text{noise}$ > 100~electrons or that efficiency is smaller than 99$\%$. 
    \item \textbf{Spatial resolution when the efficiency is equal to 99$\%$} (first row): by definition, the corresponding threshold value is larger than 100 electrons (if reported). It indicates, in the range of operation (up to 99$\%$), how much the spatial resolution degrades.
\end{itemize}

Several trends can be observed:
\begin{itemize}
    \item For all the DUTs, the spatial resolution degrades for higher thresholds, as expected due to the decrease of the average cluster size. Operating the chips at lower thresholds improves spatial resolution and efficiency.
    \item For all the DUTs, the resolution is better than the pixel pitch divided by $\sqrt{12}$.
\begin{figure}[H]
    \centering
\includegraphics[width=1\textwidth]{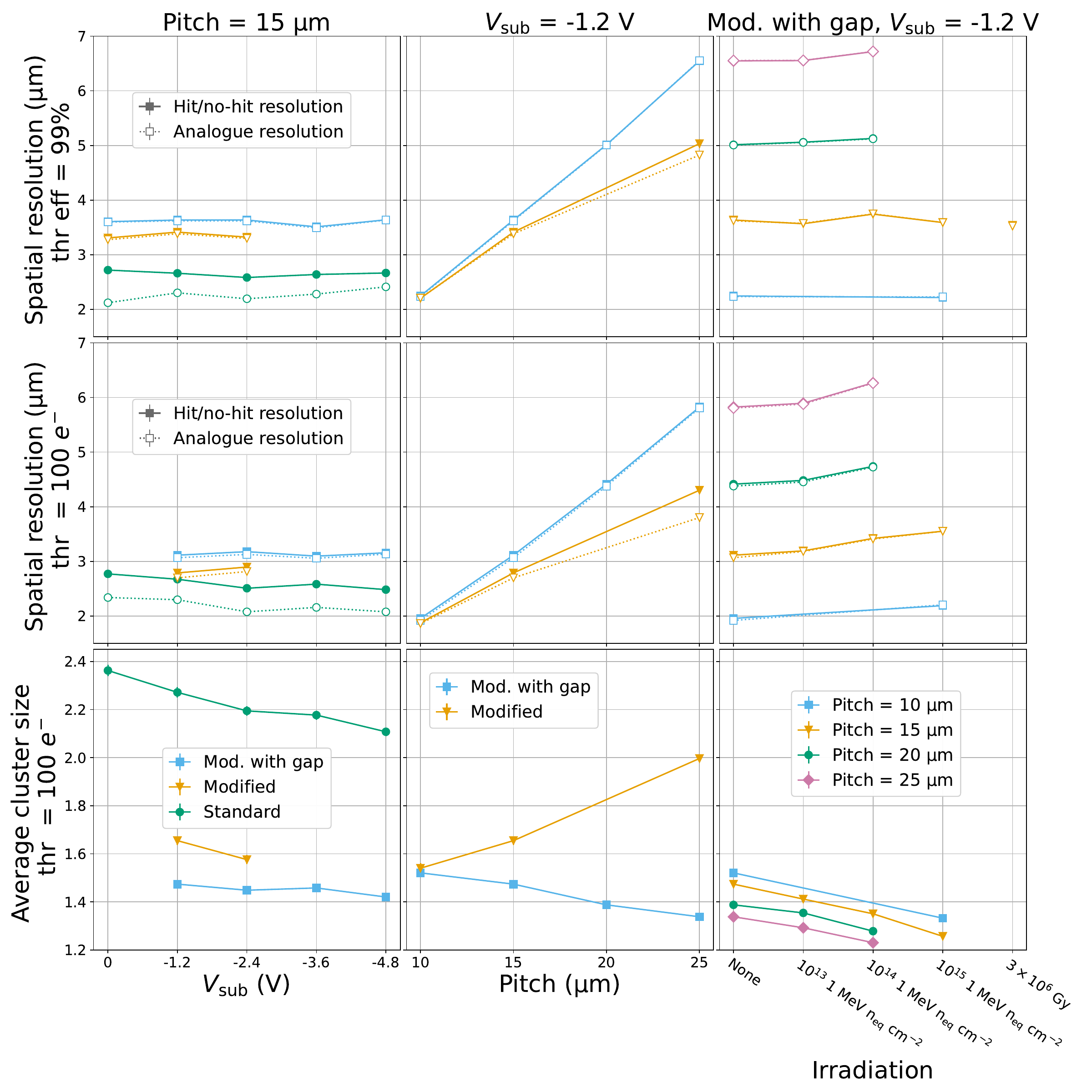}	
    \caption{Comparison of the main characteristics extracted from the resolution plots versus threshold. APTS with split 4, reference variant.}
    \label{fig:res_summary}
\end{figure}
    \item In the first column, when comparing between designs at different \( V_{\text{sub}} \), no strong dependence of the spatial resolution is observed. Since the fraction of clusters that have a cluster size above one is larger for the standard type, for this variant the largest improvement in using the analogue information w.r.t. the hit/no-hit resolution is observed. As observed from the $^{55}$Fe measurements reported in Figure~\ref{fig:fe55_summary}, the standard design has the largest average cluster size and consequently the best resolution.
    \item In the second column, for the modified with gap design, the average cluster size decreases for larger pitches, in agreement with the $^{55}$Fe (see Figure~\ref{fig:fe55_summary}) and the detection efficiency measurements (see~\Cref{fig:eff_summary}). As expected, for the modified design the trend is the opposite, with a larger average cluster size for the larger pitches and a stronger dependence on the pitch size. Consequently, the improvement in the spatial resolution given by the full analogue information is larger for the 25~\textmu m pixel pitch. The larger cluster sizes allow for better spatial resolutions for pitches above 15~\textmu m w.r.t. the modified with gap type.
    \item In the last column, the average cluster size shows a small decrease with the increasing NIEL dose due to the deterioration of the charge collection efficiency, in agreement with the $^{55}$Fe measurement (see Figure~\ref{fig:fe55_summary}). Due to the lower average cluster size, there is a slight worsening in the resolution for higher irradiation levels, for all pitches. Due to the decrease in the average cluster size, the difference in the hit/no-hit and analogue resolution is even less visible.
\end{itemize}
\section{Conclusions}

The performance of several versions of the first production of analogue MAPS for high energy physics implemented in the TPSCo 65~nm~ISC process has been studied. Various designs (standard, modified and modified with gap), splits, n-well/p-well geometry (variants) and pitches (10 to 25~\textmu m),  have been tested at different reverse substrate bias and NIEL and TID irradiation levels, both with an $^{55}$Fe source and in a beam test setup.

Compared to other designs, the modified with gap design demonstrated a high CCE and little charge sharing, leading to a smaller average cluster size and a higher signal-to-noise ratio, and hence a larger operation margin in terms of efficiency.
As expected the spatial resolution instead is better for the standard than for the modified with gap designs, reaching $\sim$2~\textmu m, due to the increase of charge sharing.
An energy resolution of 4$\%$ for the Mn-K$_\alpha$ peak from $^{55}$Fe has also been obtained for the modified with gap design.
The modified with gap design always reached a detection efficiency higher than 99$\%$, regardless of substrate bias and pitch. Larger pitches provide a slight improvement in operating margin but only up to moderate NIEL levels.

Both $^{55}$Fe and test-beam measurements showed a deterioration of charge collection of the modified with gap design with increasing NIEL or TID irradiation doses. Nevertheless, a signal peak was still visible after $10^{16}$~1~MeV~n$_\text{eq}$~cm$^{-2}$. The detection efficiency still reached 99\% after $3 \times 10^6$~Gy, thus meeting the TID requirements for ALICE ITS3 as well as for the ALICE 3 vertex detector. The efficiency also reached 99\% or higher for all pitches, at a temperature of $\sim$15~$^\circ$C, up to a NIEL of $10^{14}$~1~MeV~n$_\text{eq}$~cm$^{-2}$, above the ALICE ITS3 requirements. This remained the case up to 1 $\times$ $10^{15}$~1~MeV~n$_\text{eq}$~cm$^{-2}$ for a pixel pitch of 15~\textmu m, and up to 2 $\times$ $10^{15}$~1~MeV~n$_\text{eq}$~cm$^{-2}$ for a pixel pitch of 10~\textmu m.

After process modifications and pixel designs based on general principles already explored in the 180 nm technology, this study provides detailed analog information about the sensor signal in this technology and complements the one on the Digital Pixel Test Structure \cite{AGLIERIRINELLA2023168589}. These studies qualify the TPSCo 65 nm ISC technology for HEP after this first run. This smaller feature size technology has significant potential for even further improvement through the exploitation of process features dedicated to imaging, so far unexplored by our community, like pinned diodes, special photodiodes, etc. It offers better circuit density and hence more functionality in the same area. A further improvement in component density at the pixel level may be achieved by stacking a sensor wafer to a 65~nm CMOS readout wafer, or wafers from even finer linewidth technologies, an option recently offered by the foundry. Therefore, this technology provides superior integration possibilities, and hence significant potential for use in future high energy physics experiments.
\newenvironment{acknowledgement}{\relax}{\relax}
\begin{acknowledgement}
\section*{Acknowledgements}






The measurements leading to these results have been performed at the Test Beam Facilities at CERN (Switzerland) and DESY Hamburg (Germany), a member of the Helmholtz Association (HGF). We would like to thank the coordinators at CERN and at DESY for their valuable support of these test beams measurements and for the excellent test beam environment.

B.M. Blidaru acknowledges support by the HighRR research training group [GRK 2058], K. Gautam the support by FWO (Belgium) and SNSF (Switzerland), and F. Krizek the support by the  project LM2023040 of the Ministry of Education, Youth, and Sports of the Czech Republic. This work has also been sponsored by the Wolfgang Gentner Programme of the German Federal Ministry of Education and Research (grant no. 13E18CHA), by Thailand NSRF via PMU-B (grant no. B37G660013) and by the National Science and Technology Development Agency (NSTDA) in Thailand (contract no. JRA-CO-2563-14066-TH).    
\end{acknowledgement}
%

\bibliography{references}

\begin{thebibliography}{10}
\expandafter\ifx\csname url\endcsname\relax
  \def\url#1{\texttt{#1}}\fi
\expandafter\ifx\csname urlprefix\endcsname\relax\def\urlprefix{URL }\fi
\expandafter\ifx\csname href\endcsname\relax
  \def\href#1#2{#2} \def\path#1{#1}\fi

\bibitem{star_2012}
I.~Valin, et~al., {A reticle size CMOS pixel sensor dedicated to the STAR HFT}, JINST 7~(01) (2012) C01102.
\newblock \href {https://doi.org/10.1088/1748-0221/7/01/C01102} {\path{doi:10.1088/1748-0221/7/01/C01102}}.

\bibitem{MAPS_CONTIN2018}
G.~Contin, et~al., {The STAR MAPS-based PiXeL detector}, NIMA 907 (2018) 60--80.
\newblock \href {https://doi.org/https://doi.org/10.1016/j.nima.2018.03.003} {\path{doi:https://doi.org/10.1016/j.nima.2018.03.003}}.

\bibitem{ALPIDE-proceedings-1}
M.~Mager, {ALPIDE, the Monolithic Active Pixel Sensor for the ALICE ITS upgrade}, NIMA 824 (2016) 434 -- 438.
\newblock \href {https://doi.org/https://doi.org/10.1016/j.nima.2015.09.057} {\path{doi:https://doi.org/10.1016/j.nima.2015.09.057}}.

\bibitem{ALPIDE-proceedings-2}
G.~{Aglieri Rinella}, et~al., {The ALPIDE pixel sensor chip for the upgrade of the ALICE Inner Tracking System}, NIMA 845 (2017) 583 -- 587.
\newblock \href {https://doi.org/https://doi.org/10.1016/j.nima.2016.05.016} {\path{doi:https://doi.org/10.1016/j.nima.2016.05.016}}.

\bibitem{MAPS_REIDT2022}
F.~Reidt, et~al., {Upgrade of the ALICE ITS detector}, NIMA 1032 (2022) 166632.
\newblock \href {https://doi.org/https://doi.org/10.1016/j.nima.2022.166632} {\path{doi:https://doi.org/10.1016/j.nima.2022.166632}}.

\bibitem{CERN_EP}
\href{https://ep-rnd.web.cern.ch/topic/monolithic-pixel-detectors}{{CERN EP R$\&$D}} (Last accessed 28/06/2023).
\newline\urlprefix\url{https://ep-rnd.web.cern.ch/topic/monolithic-pixel-detectors}

\bibitem{TPSCo}
\href{http://www.towersemi.com/}{{Tower Partners Semiconductor Co}} (Last accessed 28/06/2023).
\newline\urlprefix\url{http://www.towersemi.com/}

\bibitem{SNOEYS201790}
W.~Snoeys, et~al., {A process modification for CMOS monolithic active pixel sensors for enhanced depletion, timing performance and radiation tolerance}, NIMA 871 (2017) 90--96.
\newblock \href {https://doi.org/https://doi.org/10.1016/j.nima.2017.07.046} {\path{doi:https://doi.org/10.1016/j.nima.2017.07.046}}.

\bibitem{ITS3_loi}
{Letter of Intent for an ALICE ITS Upgrade in LS3}, Tech. rep., CERN, Geneva (2019).
\newblock \href {https://doi.org/10.17181/CERN-LHCC-2019-018} {\path{doi:10.17181/CERN-LHCC-2019-018}}.

\bibitem{ALICE3_loi}
{Letter of intent for ALICE 3: A next generation heavy-ion experiment at the LHC}, Tech. rep., CERN, Geneva (2022).
\newblock \href {https://doi.org/https://doi.org/10.48550/arXiv.2211.02491} {\path{doi:https://doi.org/10.48550/arXiv.2211.02491}}.

\bibitem{Munker_2019}
M.~Munker, et~al., {Simulations of CMOS pixel sensors with a small collection electrode, improved for a faster charge collection and increased radiation tolerance}, JINST 14~(05) (2019) C05013.
\newblock \href {https://doi.org/10.1088/1748-0221/14/05/C05013} {\path{doi:10.1088/1748-0221/14/05/C05013}}.

\bibitem{Dyndal_2020}
M.~Dyndal, et~al., {Mini-MALTA: radiation hard pixel designs for small-electrode monolithic CMOS sensors for the High Luminosity LHC}, JINST 15~(02) (2020) P02005.
\newblock \href {https://doi.org/10.1088/1748-0221/15/02/P02005} {\path{doi:10.1088/1748-0221/15/02/P02005}}.

\bibitem{TPSCo65_2023}
G.~{Aglieri Rinella}, et~al., {Optimization of a 65 nm CMOS imaging process for monolithic CMOS sensors for high energy physics}, PoS Pixel2022 (2023) 083.
\newblock \href {https://doi.org/https://doi.org/10.22323/1.420.0083} {\path{doi:https://doi.org/10.22323/1.420.0083}}.

\bibitem{Sarritzu_2023}
V.~Sarritzu, et~al., {A readout system for monolithic pixel sensor prototypes towards the upgrade of the ALICE Inner Tracking System}, JINST 18~(01) (2023) C01047.
\newblock \href {https://doi.org/10.1088/1748-0221/18/01/C01047} {\path{doi:10.1088/1748-0221/18/01/C01047}}.

\bibitem{2023_david}
D.~Schledewitz, \href{http://www.physi.uni-heidelberg.de/Publications/MasterThesis_DavidSchledewitz.pdf}{{Characterization of APTS, an analog MAPS test structure fabricated in 65 nm CMOS technology}}, Master Thesis University of Heidelbergh (2023).
\newline\urlprefix\url{http://www.physi.uni-heidelberg.de/Publications/MasterThesis_DavidSchledewitz.pdf}

\bibitem{SPS}
D.~Banerjee, et~al., \href{https://cds.cern.ch/record/2774716}{{The North Experimental Area at the Cern Super Proton Synchrotron}} (2021).
\newline\urlprefix\url{https://cds.cern.ch/record/2774716}

\bibitem{optimizer}
M.~Mager, \href{https://mmager.web.cern.ch/telescope/tracking.html}{{The Telescope Optimiser}}.
\newline\urlprefix\url{https://mmager.web.cern.ch/telescope/tracking.html}

\bibitem{corryvreckan}
D.~Dannheim, et~al., {Corryvreckan: a modular 4D track reconstruction and analysis software for test beam data}, JINST 16~(03) (2021) P03008.
\newblock \href {https://doi.org/https://doi.org/10.1088/1748-0221/16/03/P03008} {\path{doi:https://doi.org/10.1088/1748-0221/16/03/P03008}}.

\bibitem{blobel2006new}
V.~Blobel, A new fast track-fit algorithm based on broken lines, NIMA 566~(1) (2006) 14--17.
\newblock \href {https://doi.org/https://doi.org/10.1016/j.nima.2006.05.156} {\path{doi:https://doi.org/10.1016/j.nima.2006.05.156}}.

\bibitem{turchetta1993spatial}
R.~Turchetta, Spatial resolution of silicon microstrip detectors, NIMA 335~(1-2) (1993) 44--58.
\newblock \href {https://doi.org/https://doi.org/10.1016/0168-9002(93)90255-G} {\path{doi:https://doi.org/10.1016/0168-9002(93)90255-G}}.

\bibitem{bugiel2021high}
R.~Bugiel, et~al., High spatial resolution monolithic pixel detector in soi technology, NIMA 988 (2021) 164897.
\newblock \href {https://doi.org/https://doi.org/10.1016/j.nima.2020.164897} {\path{doi:https://doi.org/10.1016/j.nima.2020.164897}}.

\bibitem{Riegler_2017}
W.~Riegler, G.~{Aglieri Rinella}, {Time resolution of silicon pixel sensors}, JINST 12~(11) (2017) P11017.
\newblock \href {https://doi.org/10.1088/1748-0221/12/11/P11017} {\path{doi:10.1088/1748-0221/12/11/P11017}}.

\bibitem{caribou2020}
T.~Vanat, {Caribou \textendash{} A versatile data acquisition system}, PoS TWEPP2019 (2020) 100.
\newblock \href {https://doi.org/10.22323/1.370.0100} {\path{doi:10.22323/1.370.0100}}.

\bibitem{DESY_beam}
R.~Diener, et~al., {The DESY II test beam facility}, NIMA 922 (2019) 265--286.
\newblock \href {https://doi.org/https://doi.org/10.1016/j.nima.2018.11.133} {\path{doi:https://doi.org/10.1016/j.nima.2018.11.133}}.

\bibitem{eudet_2016}
H.~Jansen, et~al., {Performance of the EUDET-type beam telescopes}, EPJ Tech. Instrum. 3~(1) (2016) 7.
\newblock \href {https://doi.org/10.1140/epjti/s40485-016-0033-2} {\path{doi:10.1140/epjti/s40485-016-0033-2}}.

\bibitem{Simancas_2023}
A.~Simancas, et~al., {Developing a Monolithic Silicon Sensor in a 65 nm CMOS Imaging Technology for Future Lepton Collider Vertex Detector}, IEEE Nuclear Science Symposium and Medical Imaging Conference (NSS/MIC) (2022) 1--7.~\href {https://doi.org/https://doi.org/10.1109/NSS/MIC44845.2022.10398964} {\path{doi:https://doi.org/10.1109/NSS/MIC44845.2022.10398964}}.

\bibitem{aida2019}
P.~Baesso, et~al., {The AIDA-2020 TLU: a flexible trigger logic unit for test beam facilities}, JINST 14~(09) (2019) P09019.
\newblock \href {https://doi.org/10.1088/1748-0221/14/09/P09019} {\path{doi:10.1088/1748-0221/14/09/P09019}}.

\bibitem{AGLIERIRINELLA2023168589}
G.~{Aglieri Rinella}, et~al., {Digital pixel test structures implemented in a 65 nm CMOS process}, NIMA 1056 (2023) 168589.
\newblock \href {https://doi.org/https://doi.org/10.1016/j.nima.2023.168589} {\path{doi:https://doi.org/10.1016/j.nima.2023.168589}}.

\bibitem{scope}
\href{https://www.picotech.com/download/datasheets/picoscope-6000e-series-data-sheet.pdf}{Picotech, picoscope 6000e series}.
\newline\urlprefix\url{https://www.picotech.com/download/datasheets/picoscope-6000e-series-data-sheet.pdf}

\bibitem{fano}
B.~Lowe, {Measurements of Fano factors in silicon and germanium in the low-energy X-ray region}, NIMA 399~(2) (1997) 354--364.
\newblock \href {https://doi.org/https://doi.org/10.1016/S0168-9002(97)00965-0} {\path{doi:https://doi.org/10.1016/S0168-9002(97)00965-0}}.

\end{thebibliography}

\newpage

\appendix
\section{Effect of different I$_\text{reset}$ and readout sampling frequency}
\label{app:reset_readout}
To study the dependency of the spectra parameters on the readout system, measurements were performed on two different setups: one with the standard readout system with a 4~MHz bandwidth, already described in this paper, and one connected to a picoscope \cite{scope} with up to 500~MHz bandwidth and 5~GS/s sampling rate.
It is important to understand the impact on the performance of different $I_\text{reset}$ values, as in Figure \ref{fig:fe55_summary} spectra with different $I_\text{reset}$ are compared. 
Figure \ref{fig:ireset_comparison} shows a slight increase in capacitance (left) and a degradation of energy resolution (right) for increasing $I_\text{reset}$, taken with the standard readout.

\begin{figure}[!hbt]
    \centering
	   \includegraphics[width=0.8\textwidth]{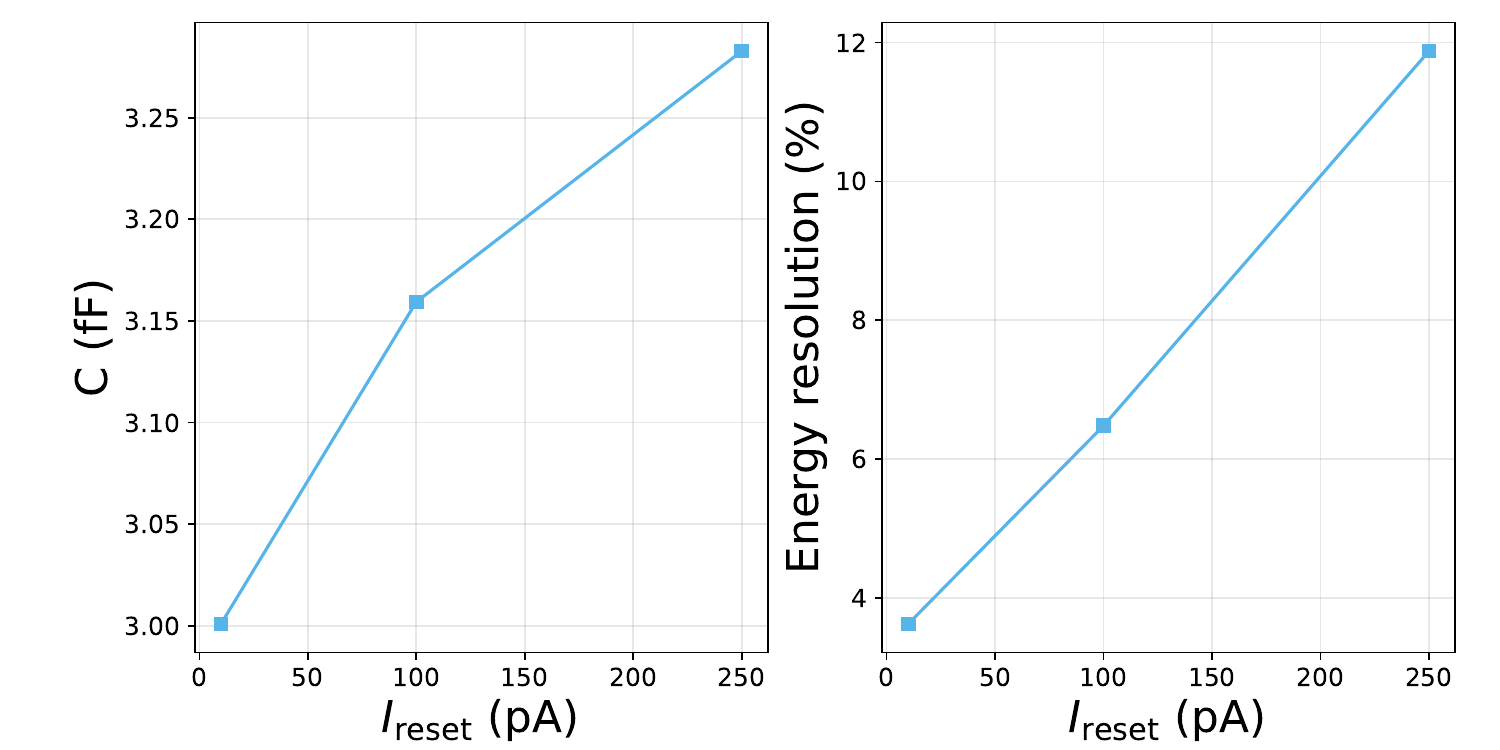}	
    \caption{Capacitance (left) and energy resolution (right) comparison between different $I_\text{reset}$. APTS with 15~\textmu m pitch, modified with gap, split 4, reference variant, $V_\text{sub}$ = -1.2~V, standard readout.}
    \label{fig:ireset_comparison}
\end{figure}


The picoscope measurements confirm the capacitance result and the degradation of the energy resolution with increasing $I_\text{reset}$.


Nevertheless, the degradation of the energy resolution in the picoscope setup is less important; an energy resolution of about $3.4\%$ and $3.9\%$ has been obtained for $I_\text{reset}$ = 10~pA ($\sim 5.3\%$ and $12.7\%$ for $I_\text{reset}$ = 250~pA), for the picoscope and standard readout, respectively.
In comparison, the  intrinsic energy resolution of the detector is $\sim 2\%$, obtained with a Fano factor F = 0.1161 \cite{fano}. 

Therefore, in this paper, spectra which strongly benefit from high energy resolution were obtained with a $I_\text{reset}$ of 10 pA (e.g. Fig. \ref{fig:spectrum_calib}). Other measurements were carried out at 100 pA except for radiation levels of $2\times 10^{15}$~1~MeV~n$_\text{eq}$~cm$^{-2}$ and higher that had to be carried out with a $I_\text{reset}$ of 250 pA to compensate the increased leakage currents.

\newpage

\section{Plots for supplementary material}
The plots reported here will NOT be part of the paper but will be put in supplementary material as a single pdf (\href{https://www.sciencedirect.com/journal/nuclear-instruments-and-methods-in-physics-research-section-a-accelerators-spectrometers-detectors-and-associated-equipment/publish/guide-for-authors}{NIMA link}). Each Figure will have no description if not the one reported in the caption.\\

Additional plots from measurements using a $^{55}$Fe source (Section \ref{sec:app_fe}) and a beam test setup (Section \ref{sec:app_bt}). Refer to the paper for details.

\subsection{\texorpdfstring{$^{55}$Fe measurements}{55Fe measurements}}
\label{sec:app_fe}
\begin{figure}[H]    
\centering
    \subfigure[]{{\includegraphics[width=0.8\textwidth]{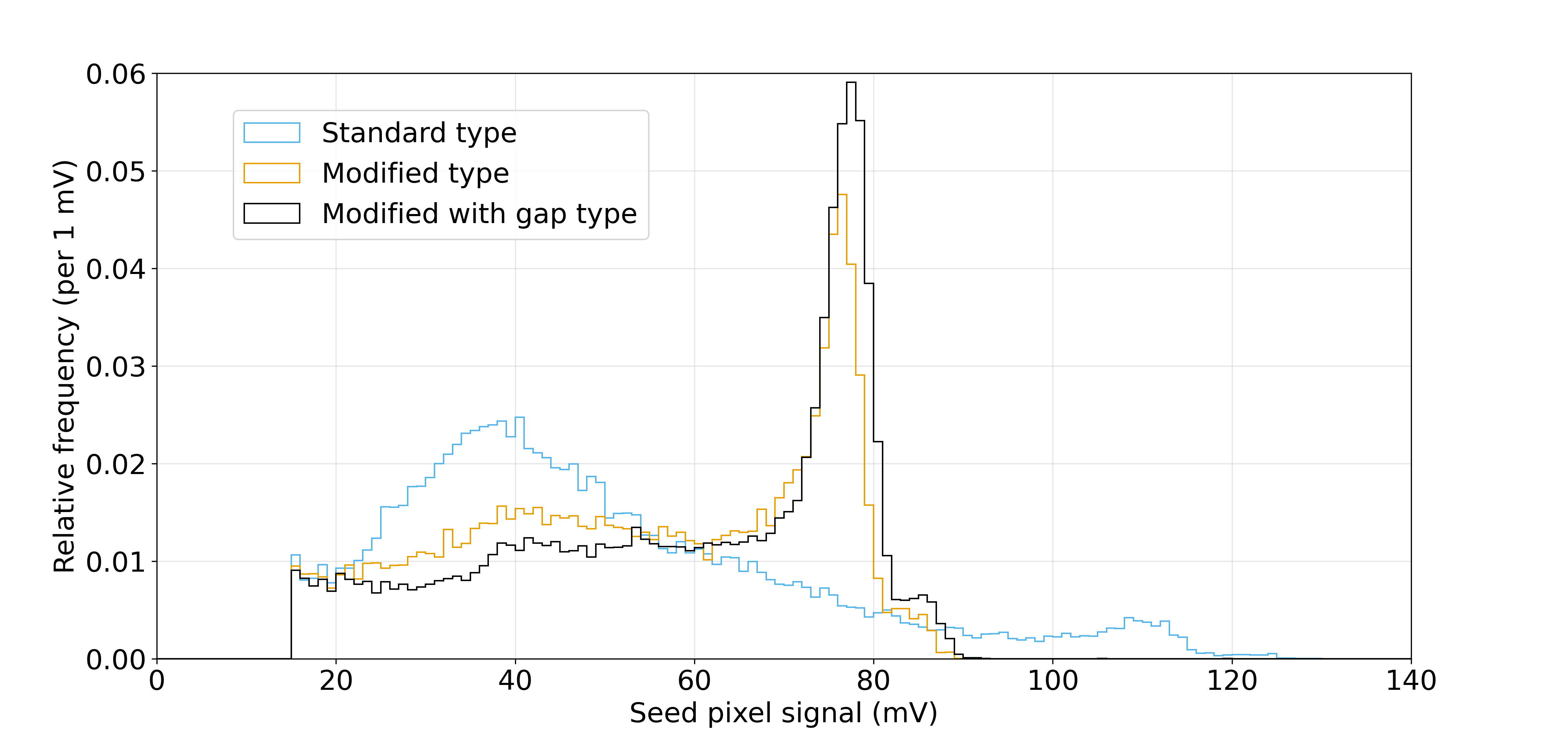} }}%
    \\
    \subfigure[]{{\includegraphics[width=0.8\textwidth]{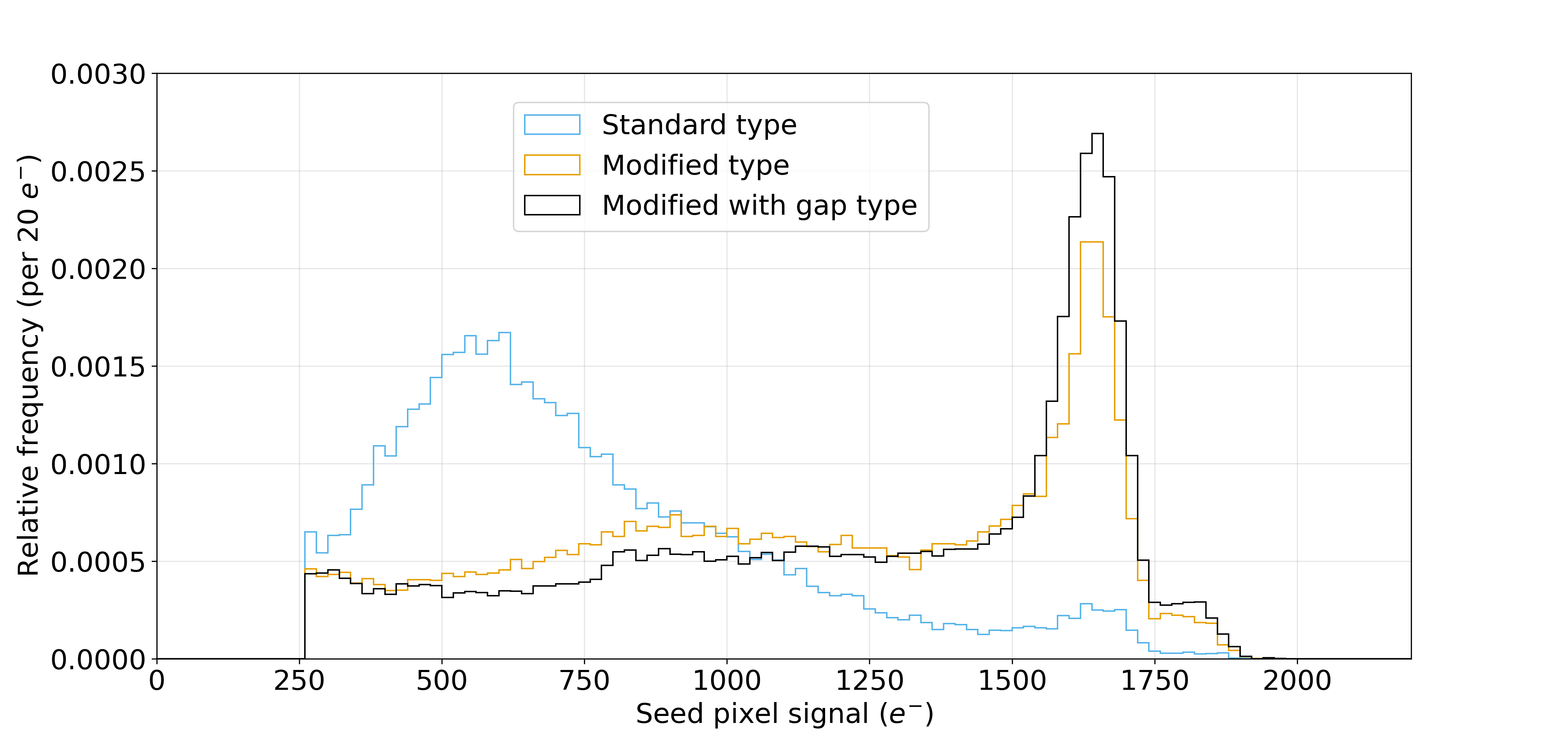} }}%
    \\
    \caption{$^{55}$Fe seed signal distribution comparison between different designs, in mV (a) and in electrons (b).  APTS with 15~\textmu m pitch, split 4, reference variant, $V_\text{sub}$ = -1.2~V.}%
\end{figure}

\begin{figure}[H]    
\centering
    \subfigure[]{{\includegraphics[width=0.8\textwidth]{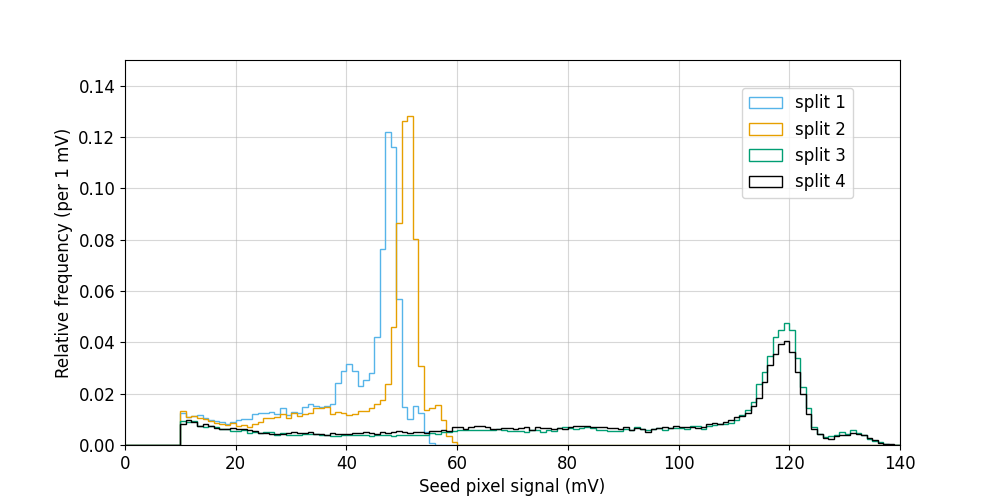} }}%
    \\
    \subfigure[]{{\includegraphics[width=0.8\textwidth]{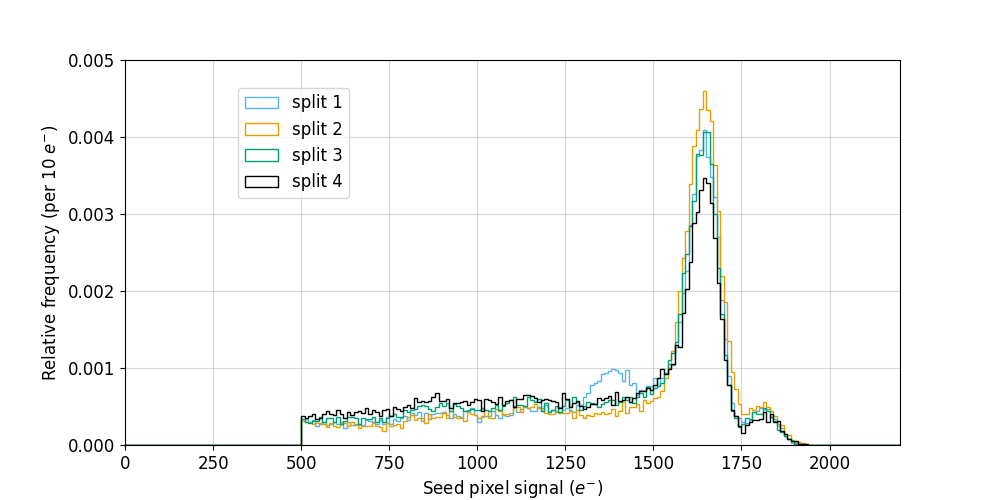} }}%
    \\
    \caption{$^{55}$Fe seed signal distribution comparison between different splits, in mV (a) and in electrons (b). APTS with 15~\textmu m pitch, modified with gap, reference variant, $V_\text{sub}$ = -4.8~V.}%
\end{figure}

\begin{figure}[H]    
\centering
    \subfigure[]{{\includegraphics[width=0.8\textwidth]{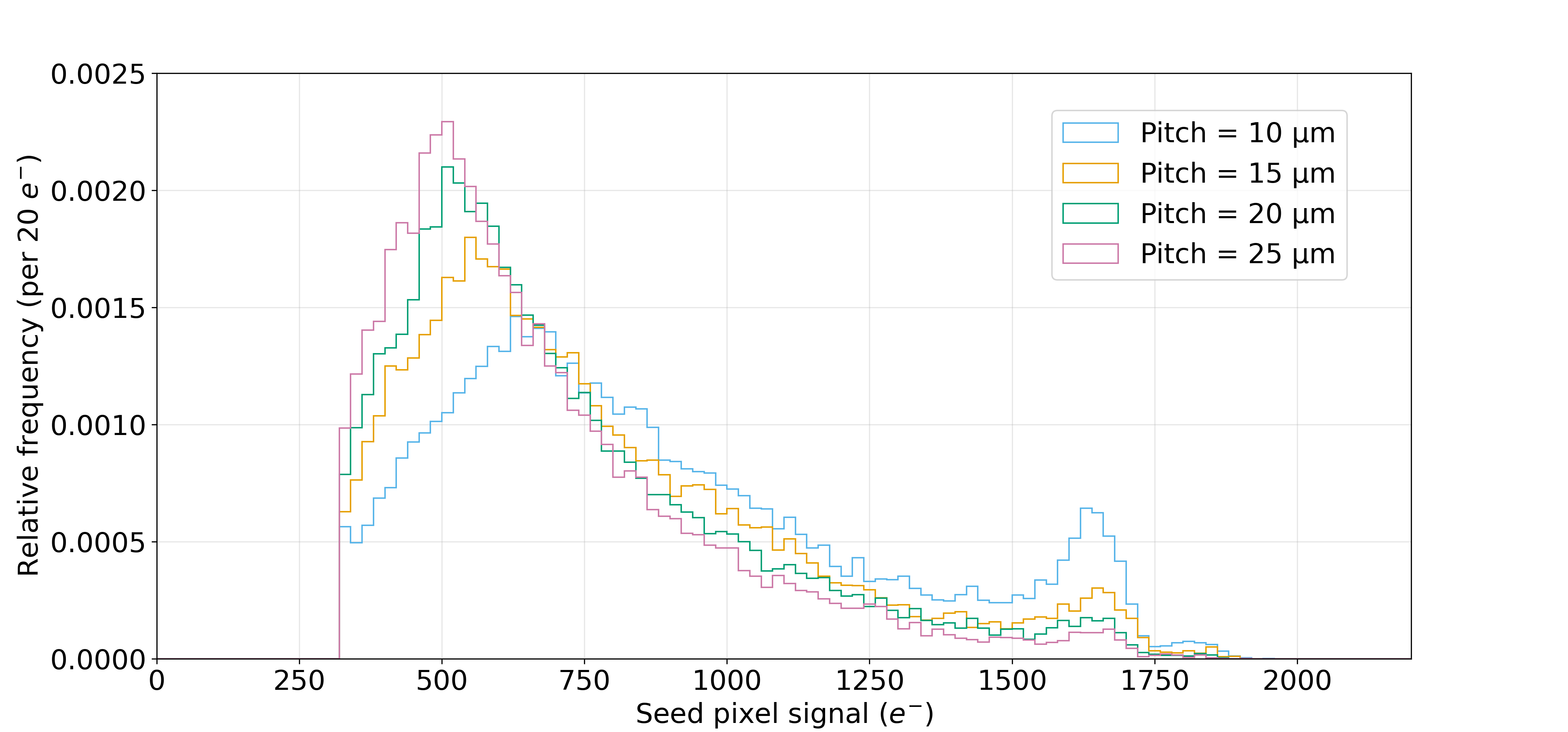} }}%
    \\
    \subfigure[]{{\includegraphics[width=0.8\textwidth]{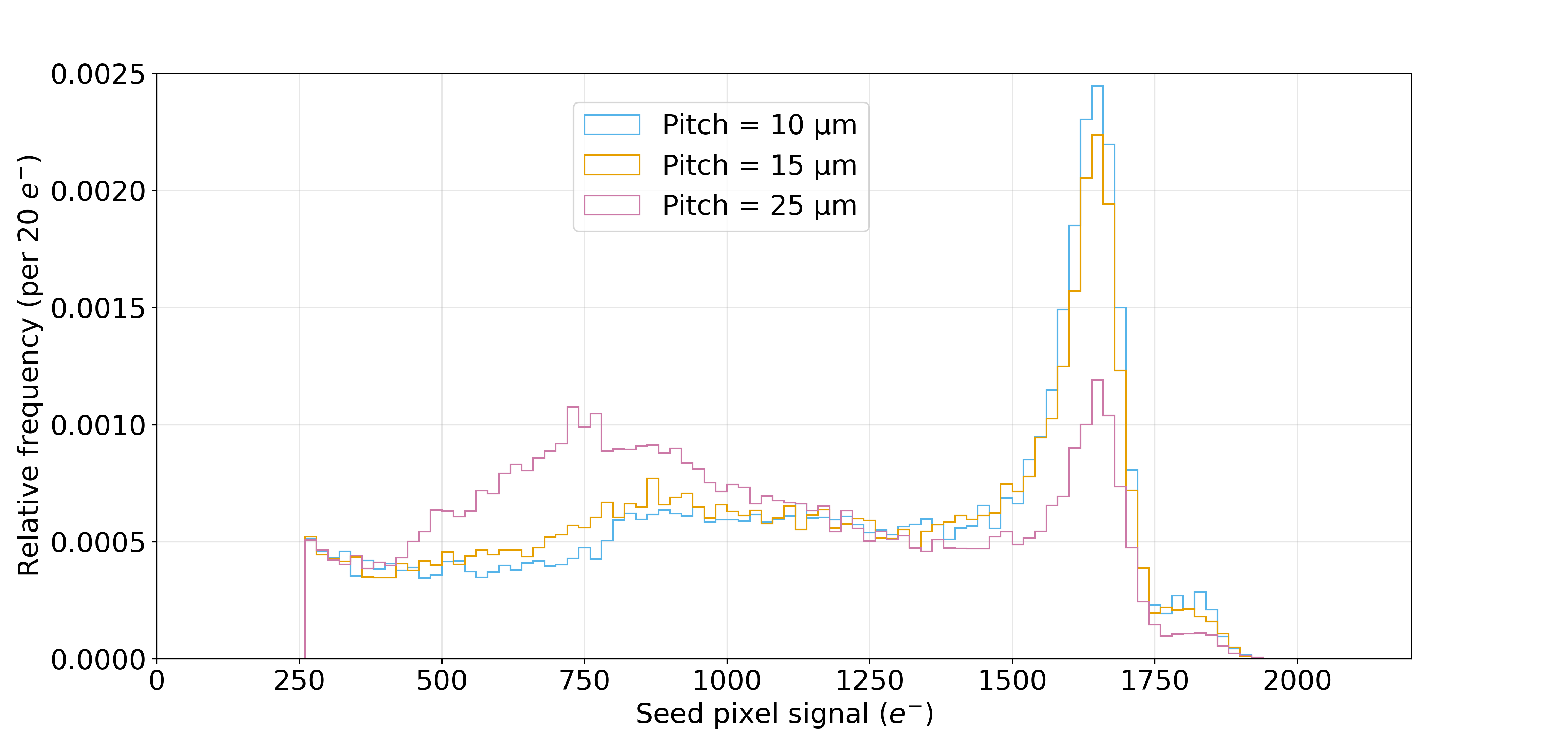} }}%
    \\
     \subfigure[]{{\includegraphics[width=0.8\textwidth]{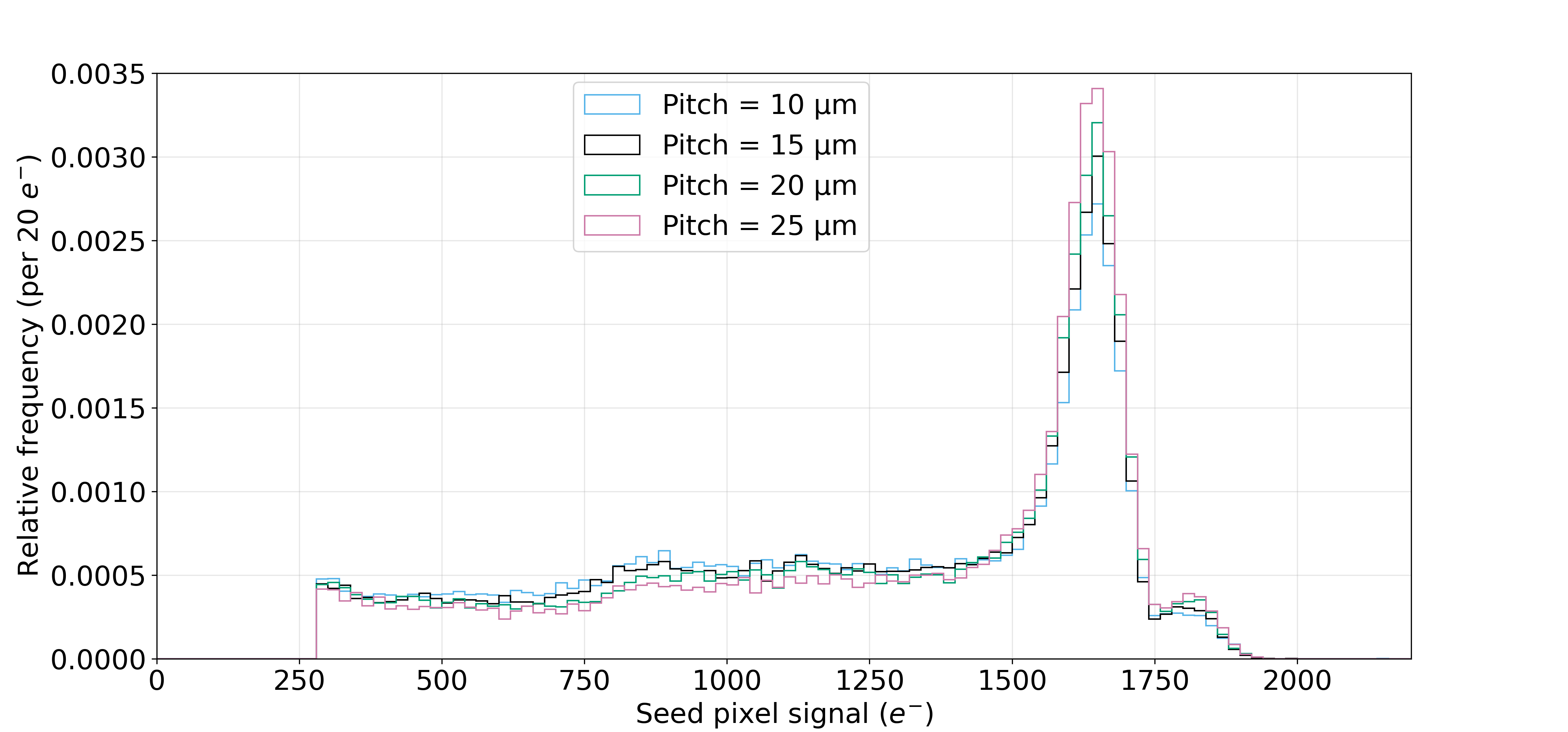} }}%
    \caption{$^{55}$Fe seed signal distribution in electrons compared between different pixel pitches for the standard (a), modified (b) and modified with gap (c) designs. APTS with split 4, reference variant, $V_\text{sub}$ = -1.2~V.}%
\end{figure}

\begin{figure}[H]    
\centering
    \subfigure[]{{\includegraphics[width=0.8\textwidth]{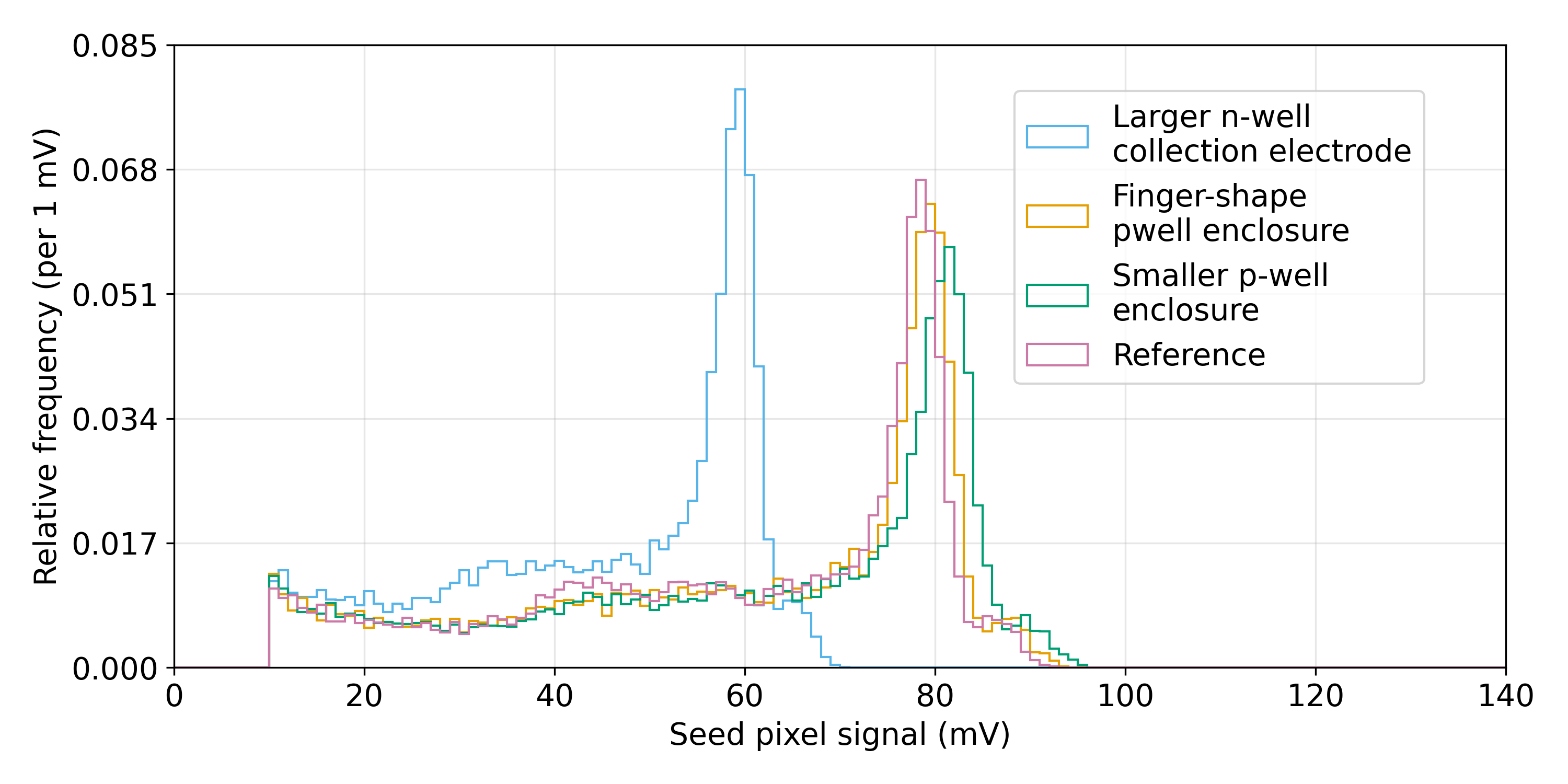} }}%
    \\
    \subfigure[]{{\includegraphics[width=0.8\textwidth]{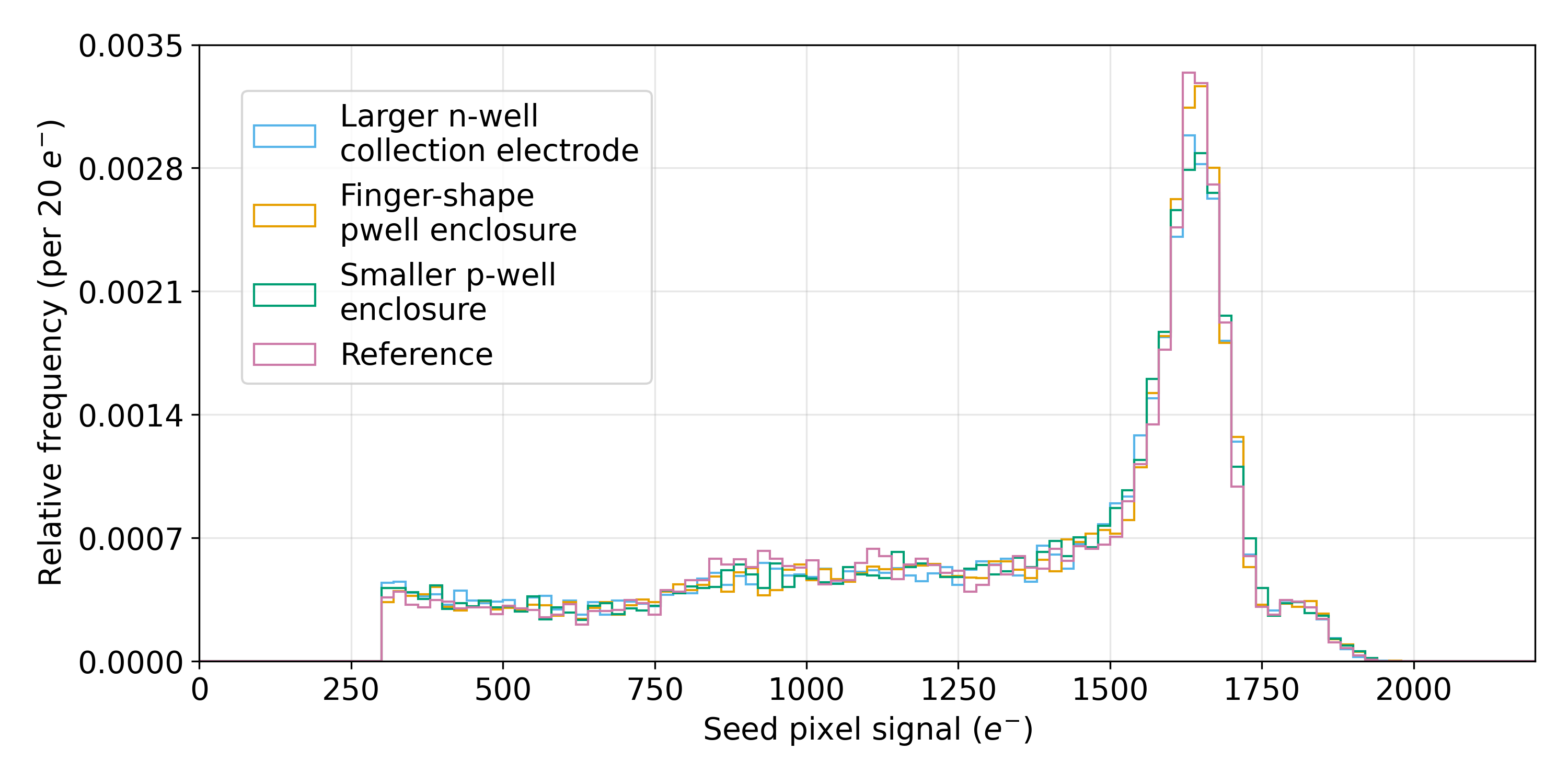} }}%
    \\
    \caption{$^{55}$Fe seed signal distribution comparison between different sensor variants: the reference one, the larger n-well collection electrode, the smaller p-well enclosure and the finger-shaped p-well enclosure, in mV (a) and in electrons (b). APTS with 20~\textmu m pitch, modified with gap, split 4, $V_\text{sub}$ = -1.2~V.}%
\end{figure}

\begin{figure}[H]    
\centering
    \subfigure[]{{\includegraphics[width=0.8\textwidth]{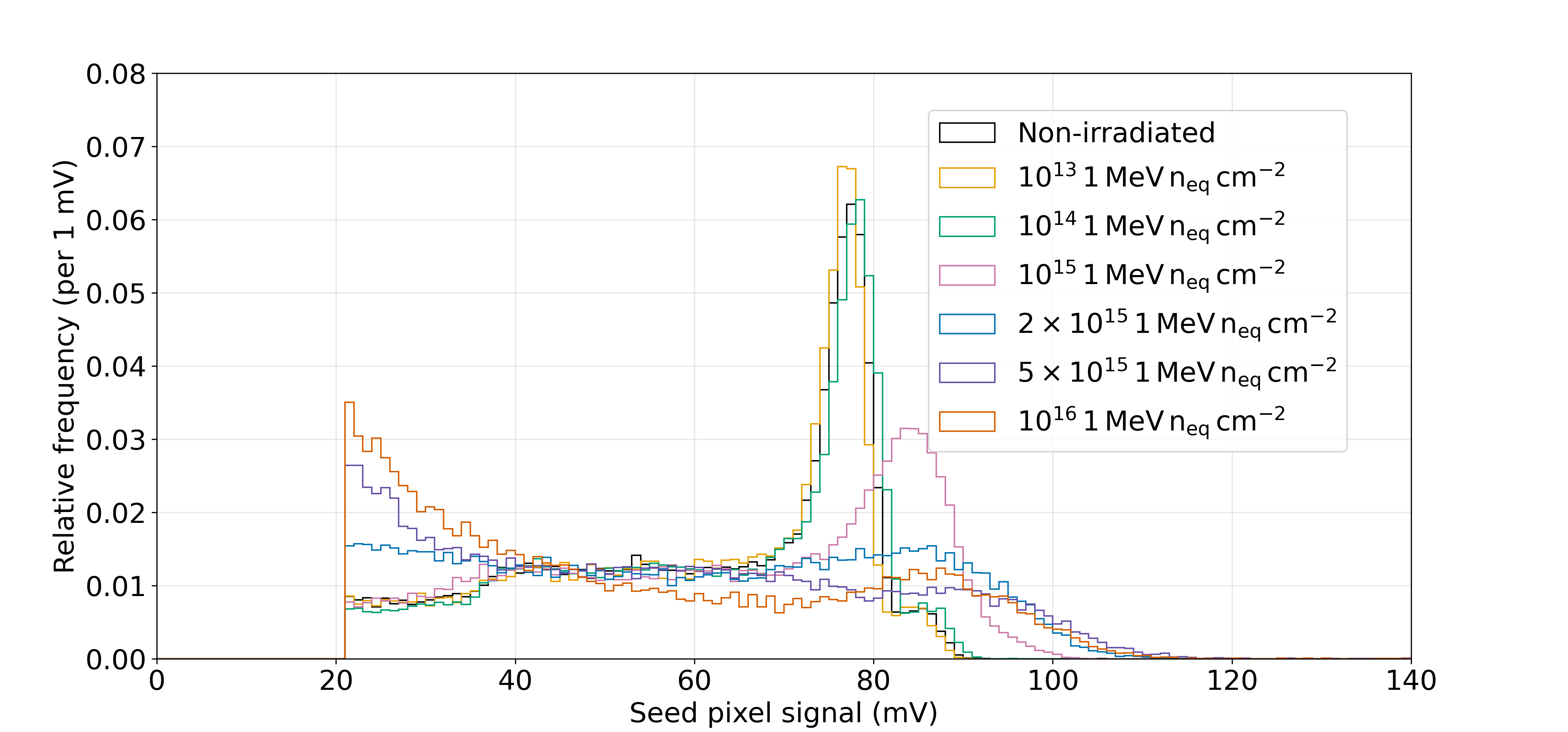} }}%
    \\
    \subfigure[]{{\includegraphics[width=0.8\textwidth]{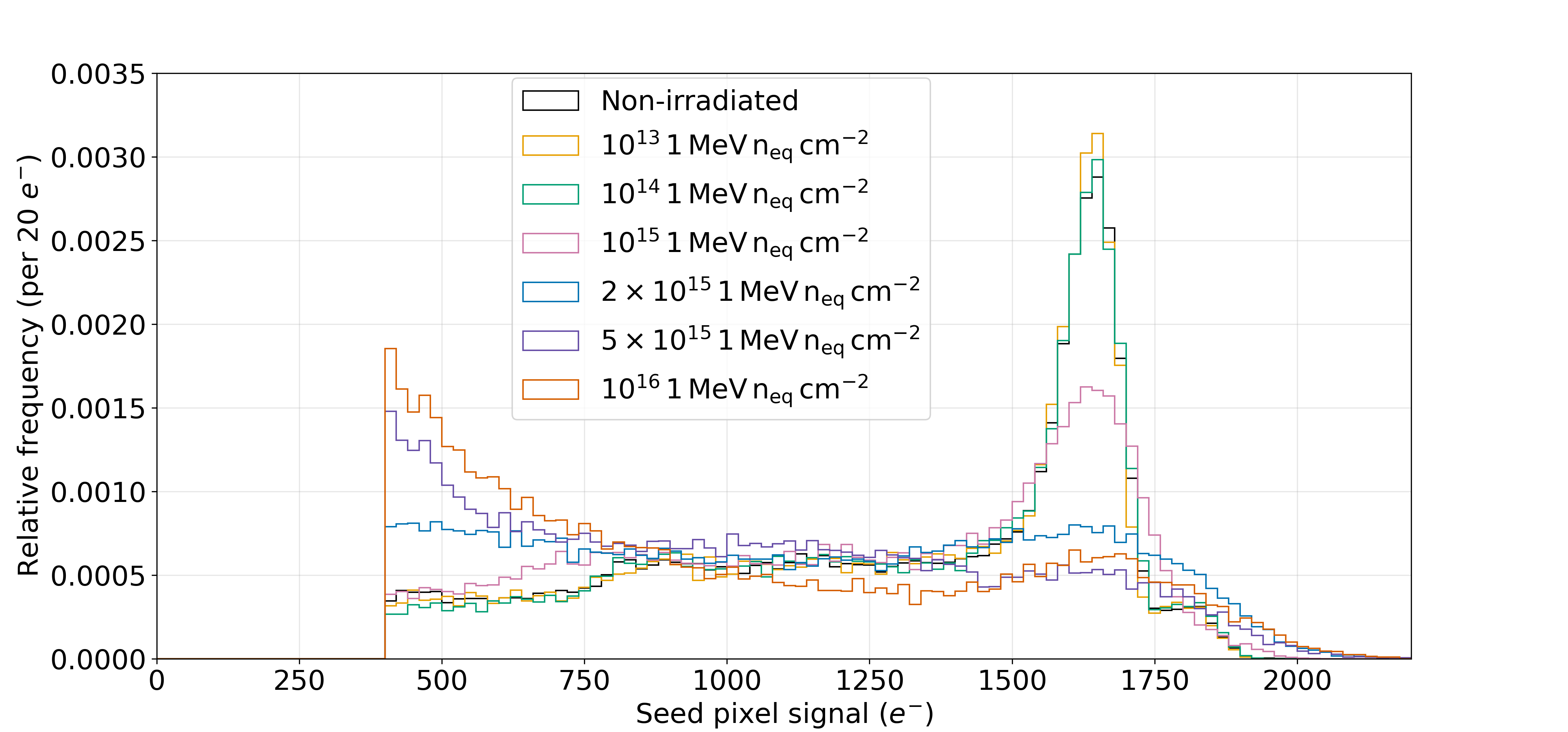} }}%
    \\
    \caption{$^{55}$Fe seed signal distribution comparison between different levels of NIEL irradiation, in mV (a) and in electrons (b). For irradiation levels of 2 $\times$ $10^{15}$~1~MeV~n$_\text{eq}$~cm$^{-2}$ or higher,  the measurements have been taken with $I_\text{reset}$ = 250 pA. Chiller temperature was 15~$^\circ$C. APTS with 15~\textmu m pitch, modified with gap, split 4, reference variant, $V_\text{sub}$ = -1.2~V.}%
\end{figure}

\begin{figure}[H]    
\centering
    \subfigure[]{{\includegraphics[width=0.8\textwidth]{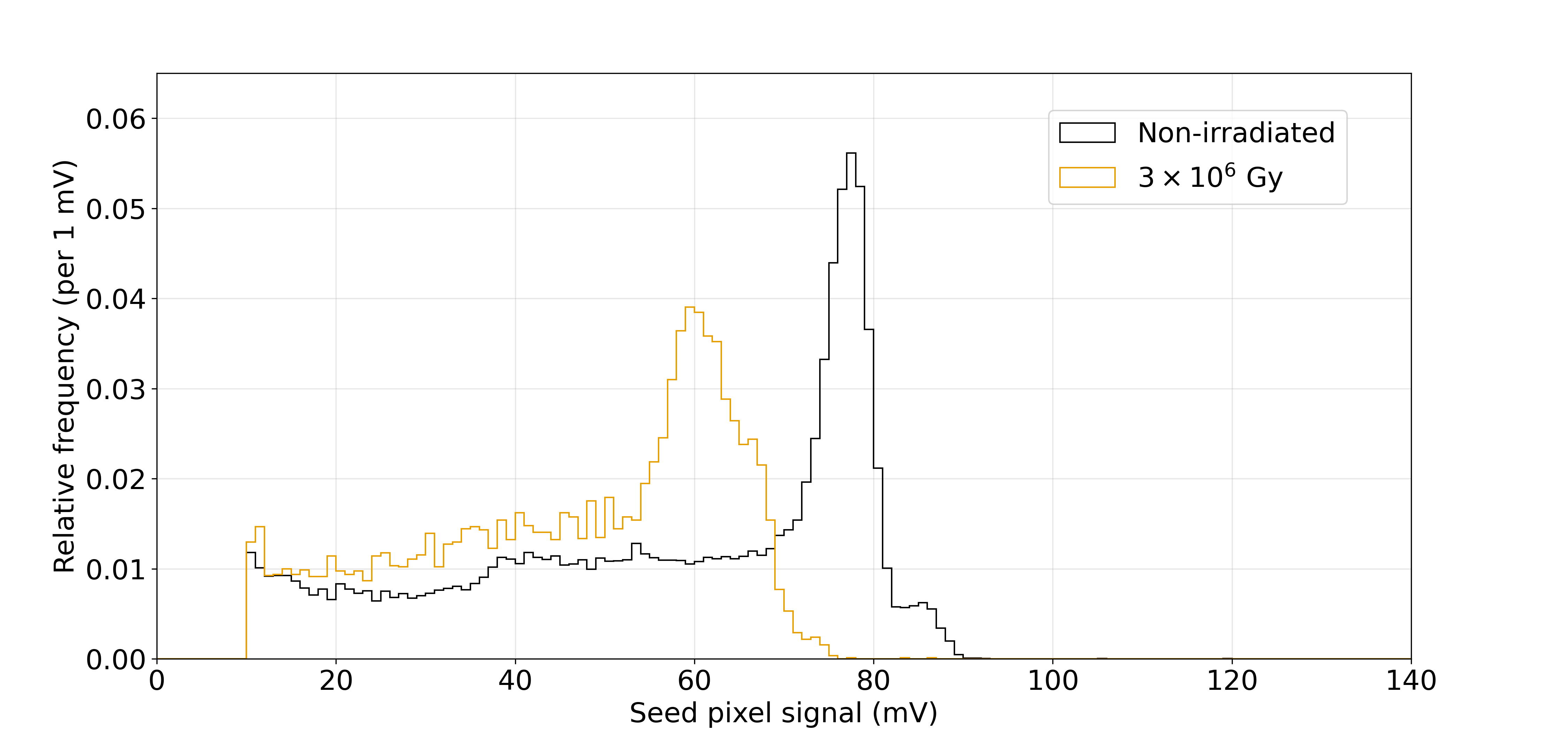} }}%
    \\
    \subfigure[]{{\includegraphics[width=0.8\textwidth]{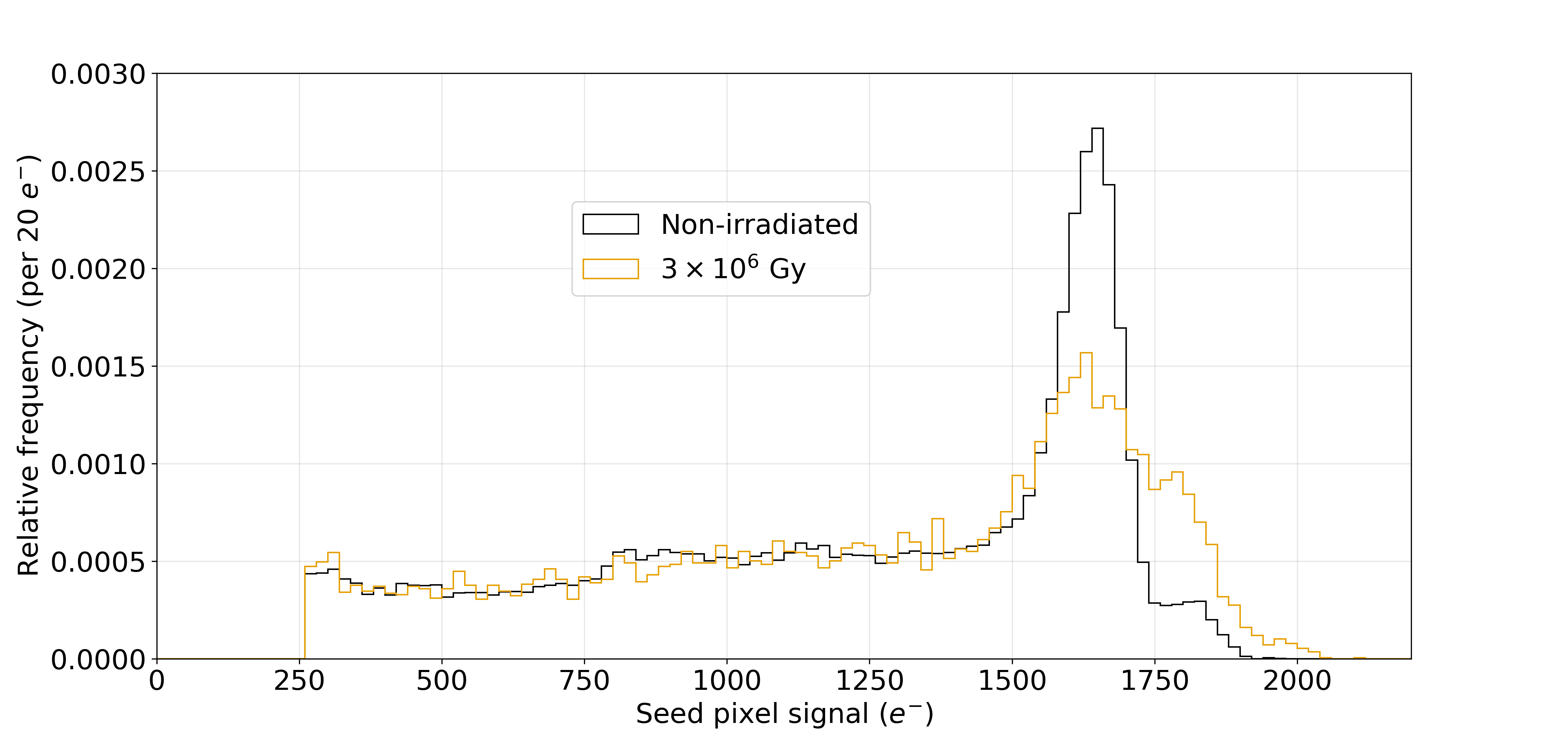} }}%
    \\
    \caption{$^{55}$Fe seed signal distribution comparison between different levels of TID irradiation, in mV (a) and in electrons (b). Chiller temperature was 15~$^\circ$C. APTS with 15~\textmu m pitch, modified with gap, split 4, reference variant, $V_\text{sub}$ = -1.2~V.}%
\end{figure}

\begin{figure}[H]    
\centering
    \subfigure[]{{\includegraphics[width=0.8\textwidth]{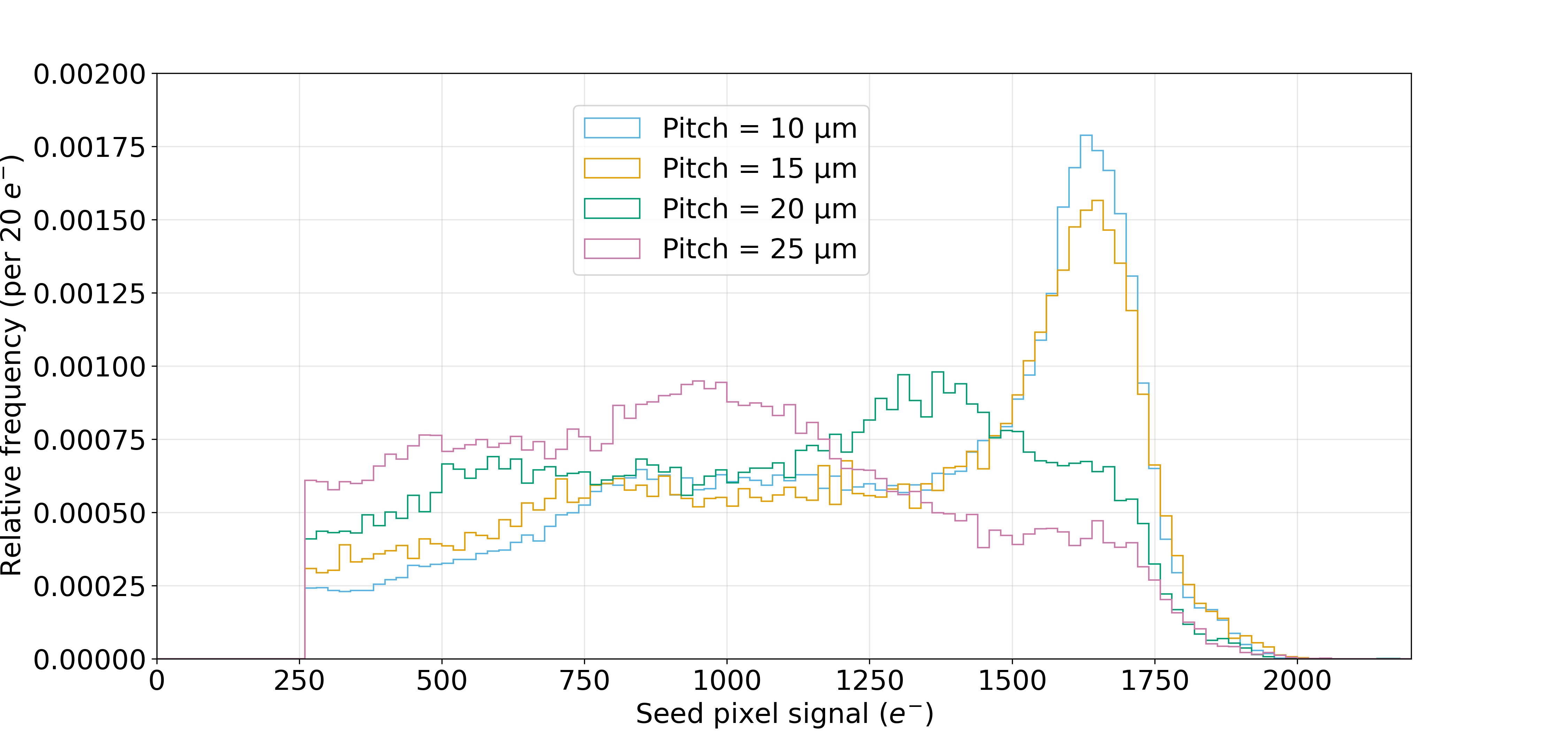} }}%
    \caption{$^{55}$Fe seed signal distribution comparison between different pixel pitches of a chip irradiated $10^{15}$~1~MeV~n$_\text{eq}$~cm$^{-2}$ in electrons. Chiller temperature was 15~$^\circ$C. APTS with modified with gap, split 4, reference variant, $V_\text{sub}$ = -1.2~V.}%
\end{figure}

\begin{figure}[H]    
\centering
	   \includegraphics[width=0.8\textwidth]{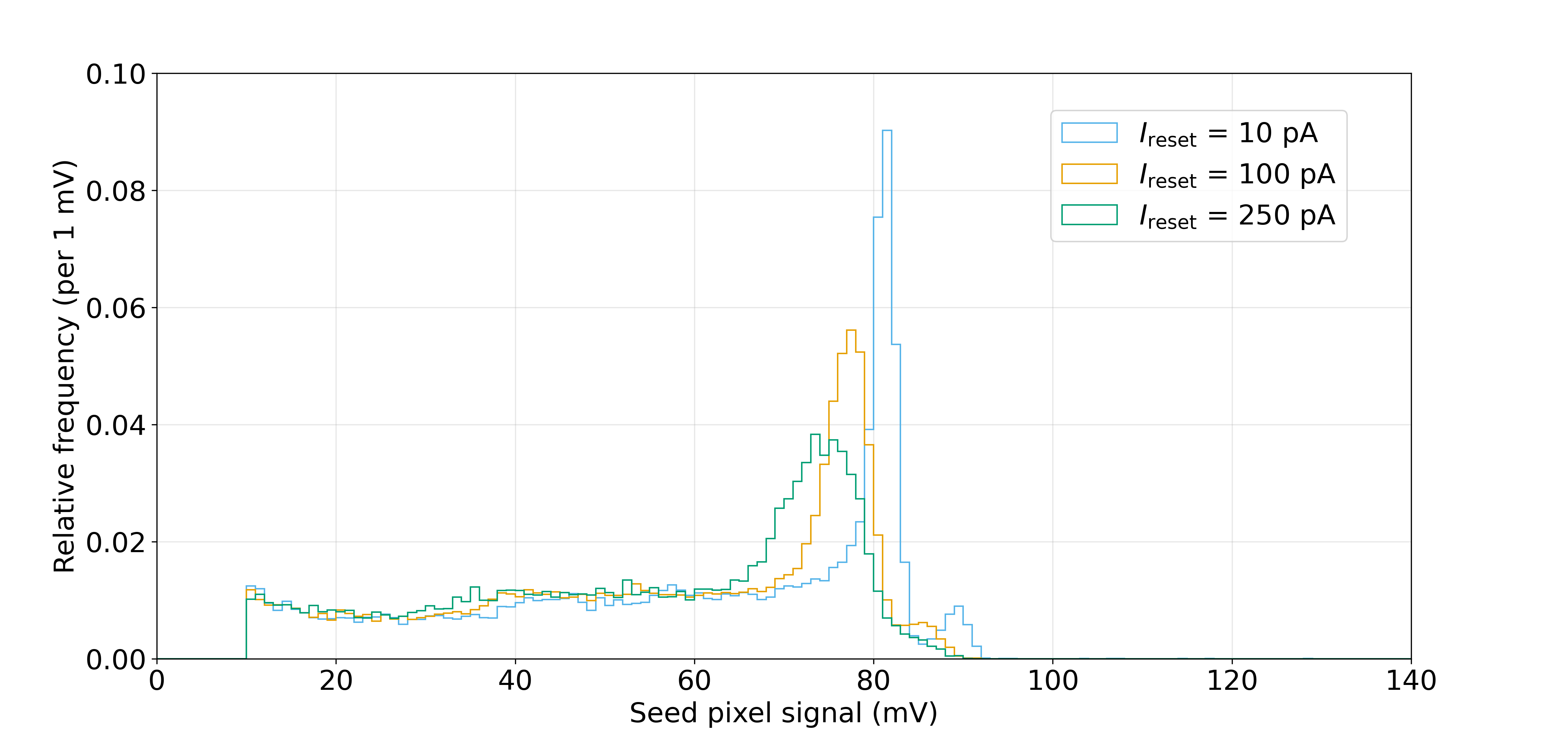}	
    \caption{$^{55}$Fe seed signal distribution comparison between different $I_\text{reset}$. APTS with 15~\textmu m pitch, modified with gap, split 4, reference variant, $V_\text{sub}$ = -1.2~V.}
\end{figure}

\clearpage
\subsection{Beam test results}
\label{sec:app_bt}

\begin{figure}[!hbt]
    \centering
	   \includegraphics[width=0.8\textwidth]{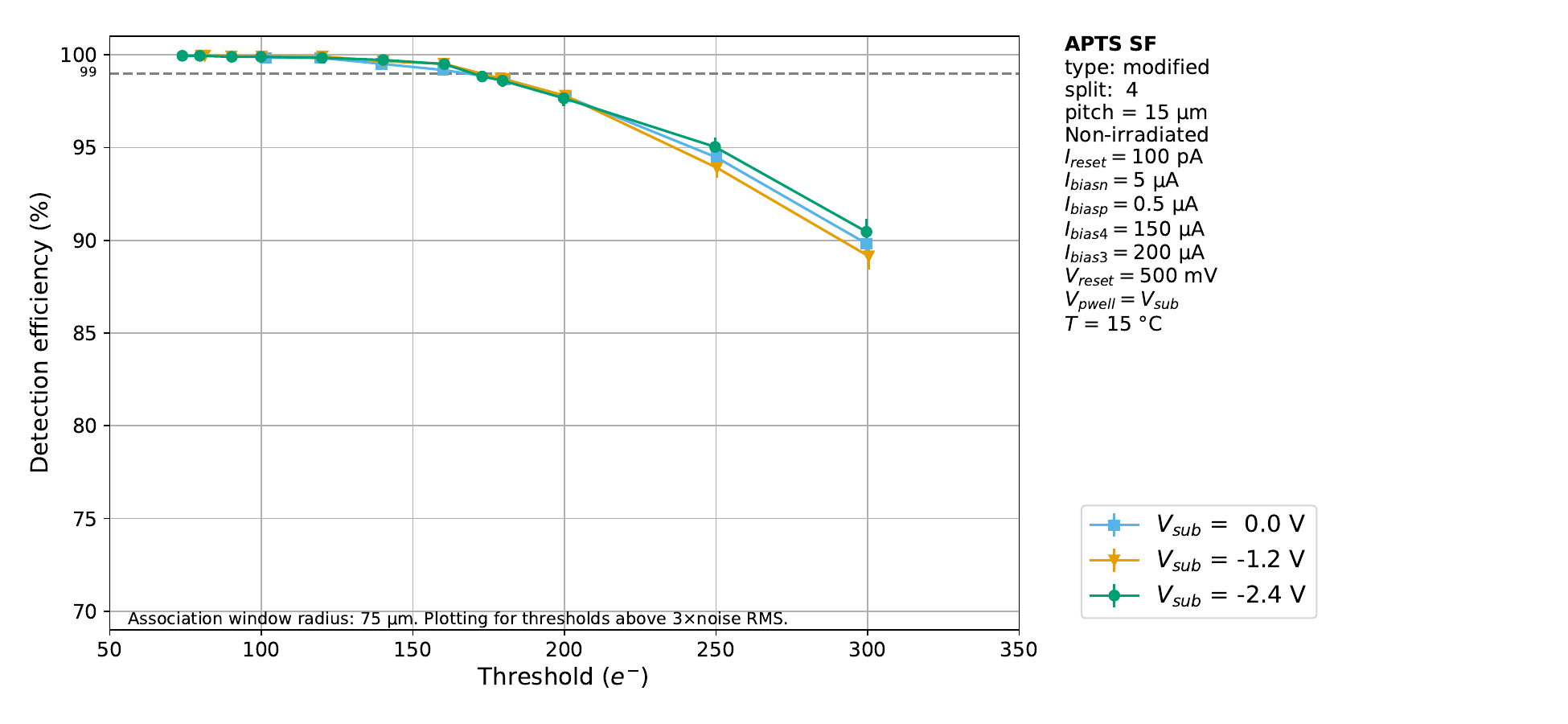 }	
    \caption{Efficiency comparison between different reverse substrate voltages as a function of the applied seed threshold. APTS with 15~\textmu m pitch, modified, split 4, reference variant.}
\end{figure}
\begin{figure}[!hbt]
    \centering
	   \includegraphics[width=0.8\textwidth]{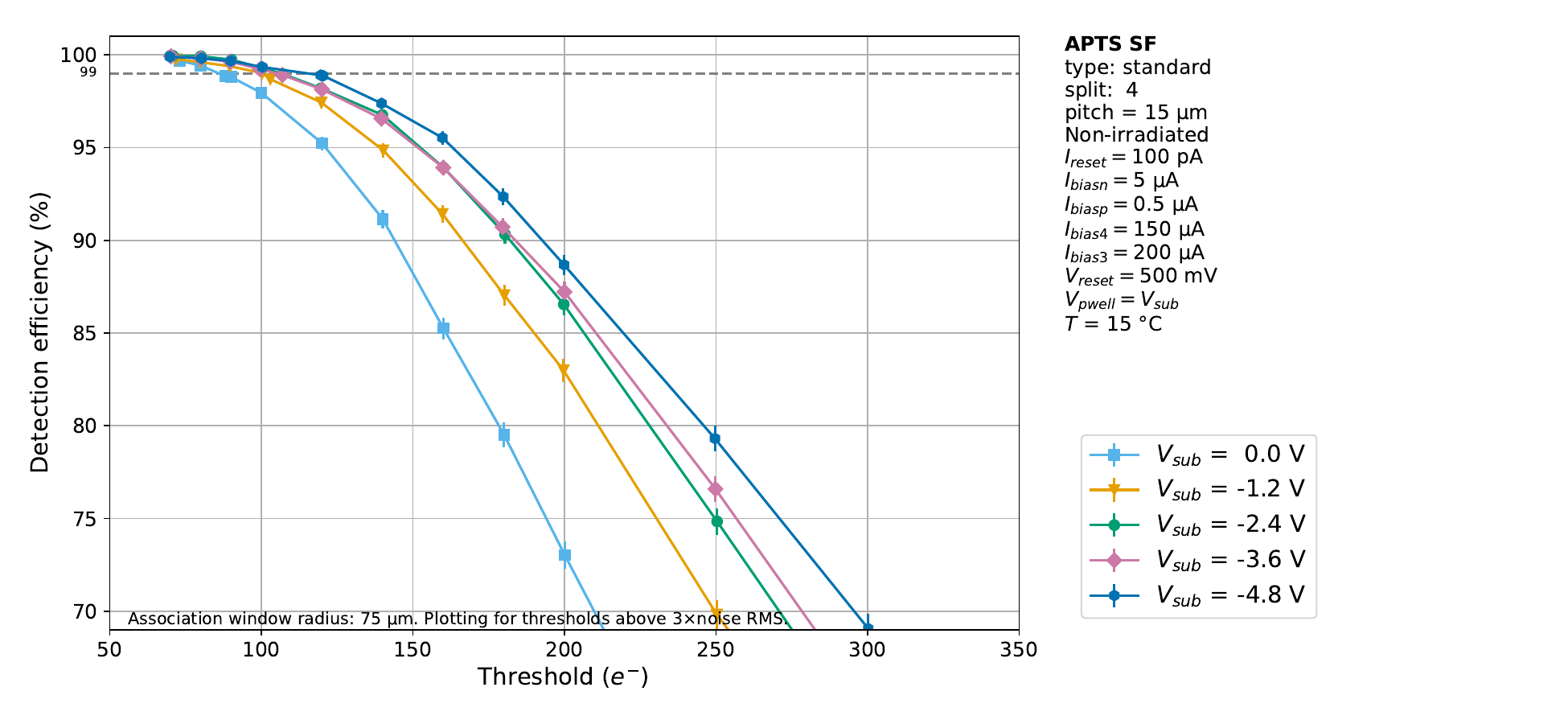 }	
    \caption{Efficiency comparison between different reverse substrate voltages as a function of the applied seed threshold. APTS with 15~\textmu m pitch, standard, split 4, reference variant.}
\end{figure}
\begin{figure}[!hbt]
    \centering
	   \includegraphics[width=0.8\textwidth]{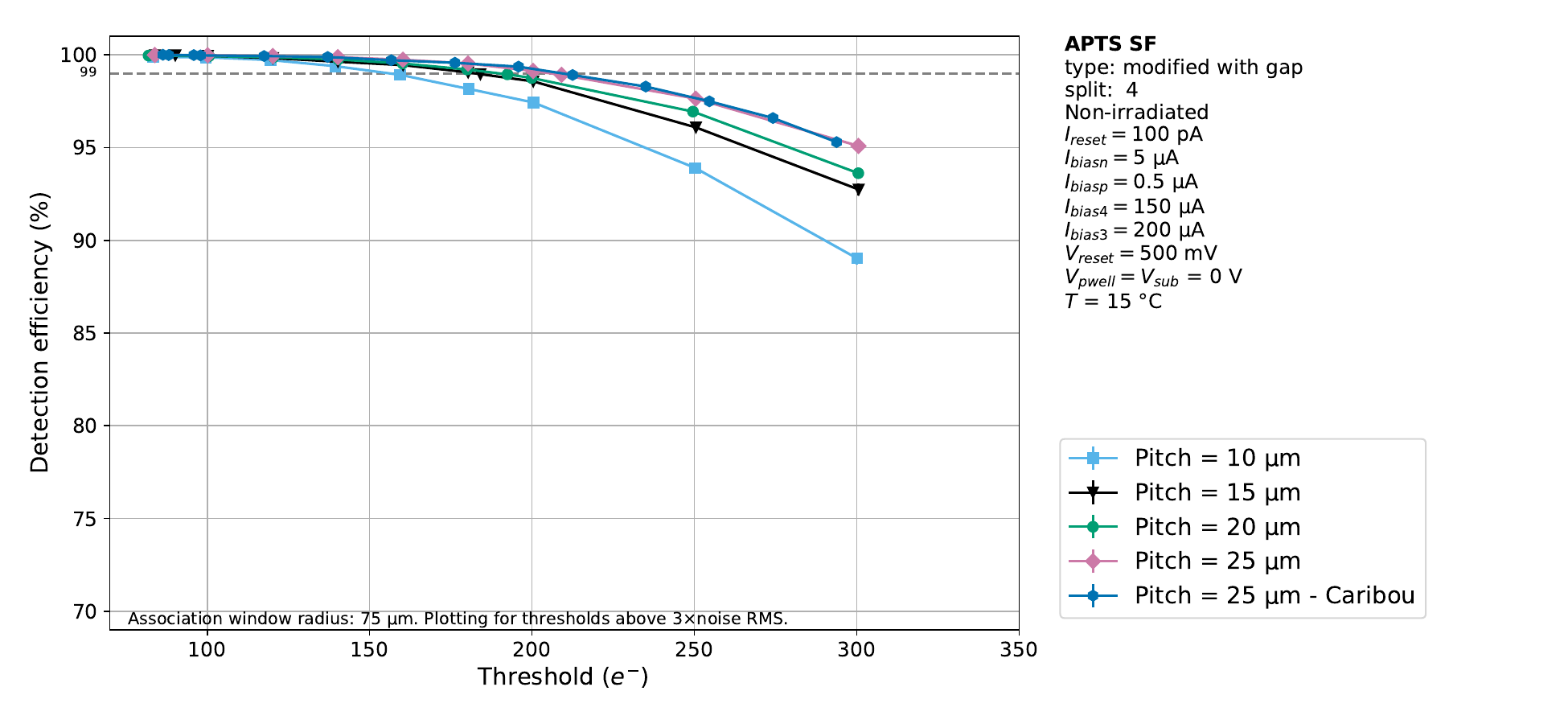 }	
    \caption{Efficiency comparison between different pitches as a function of the applied seed threshold. APTS with modified with gap, split 4, reference variant, $V_\text{sub}$ = -1.2~V.}
\end{figure}
\begin{figure}[!hbt]
    \centering
	   \includegraphics[width=0.8\textwidth]{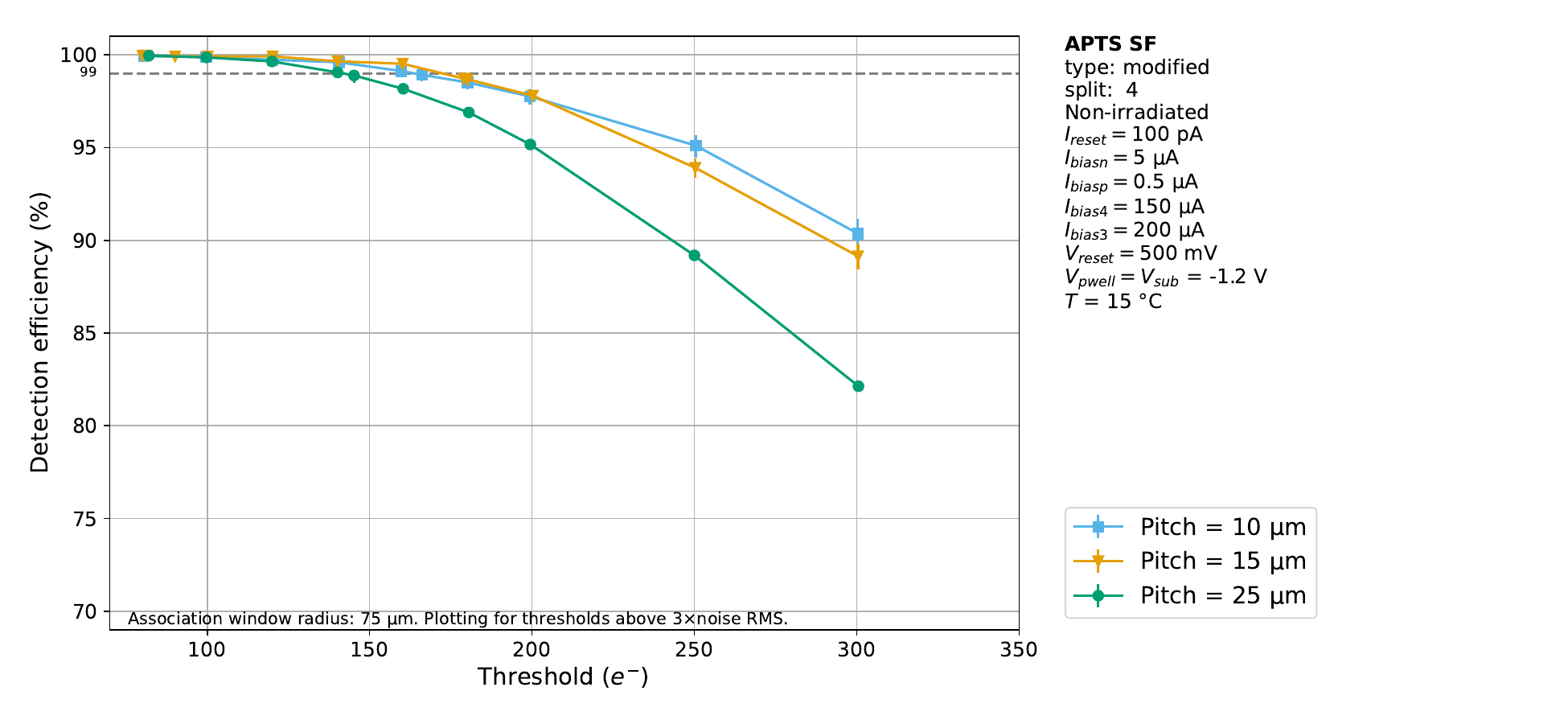 }	
    \caption{Efficiency comparison between different pitches as a function of the applied seed threshold. APTS with modified, split 4, reference variant, $V_\text{sub}$ = -1.2~V.}
\end{figure}
\begin{figure}[!hbt]
    \centering
	   \includegraphics[width=0.8\textwidth]{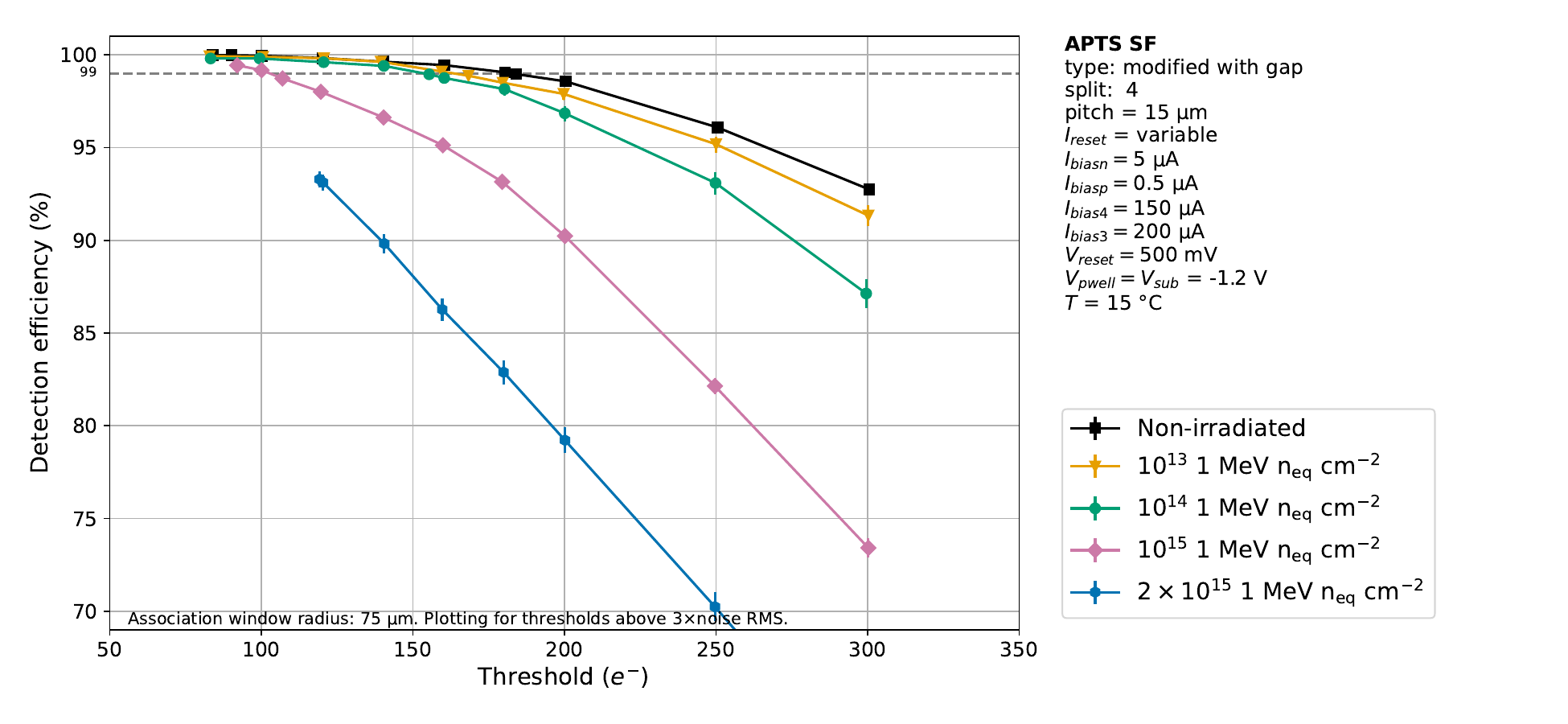}	
    \caption{Efficiency comparison between different NIEL irradiation levels as a function of the applied seed threshold. Chiller temperature was 15~$^\circ$C. APTS with 15~\textmu m pitch, modified with gap, split 4, reference variant, $V_\text{sub}$ = -1.2~V.}
\end{figure}
\begin{figure}[!hbt]
    \centering
	   \includegraphics[width=0.8\textwidth]{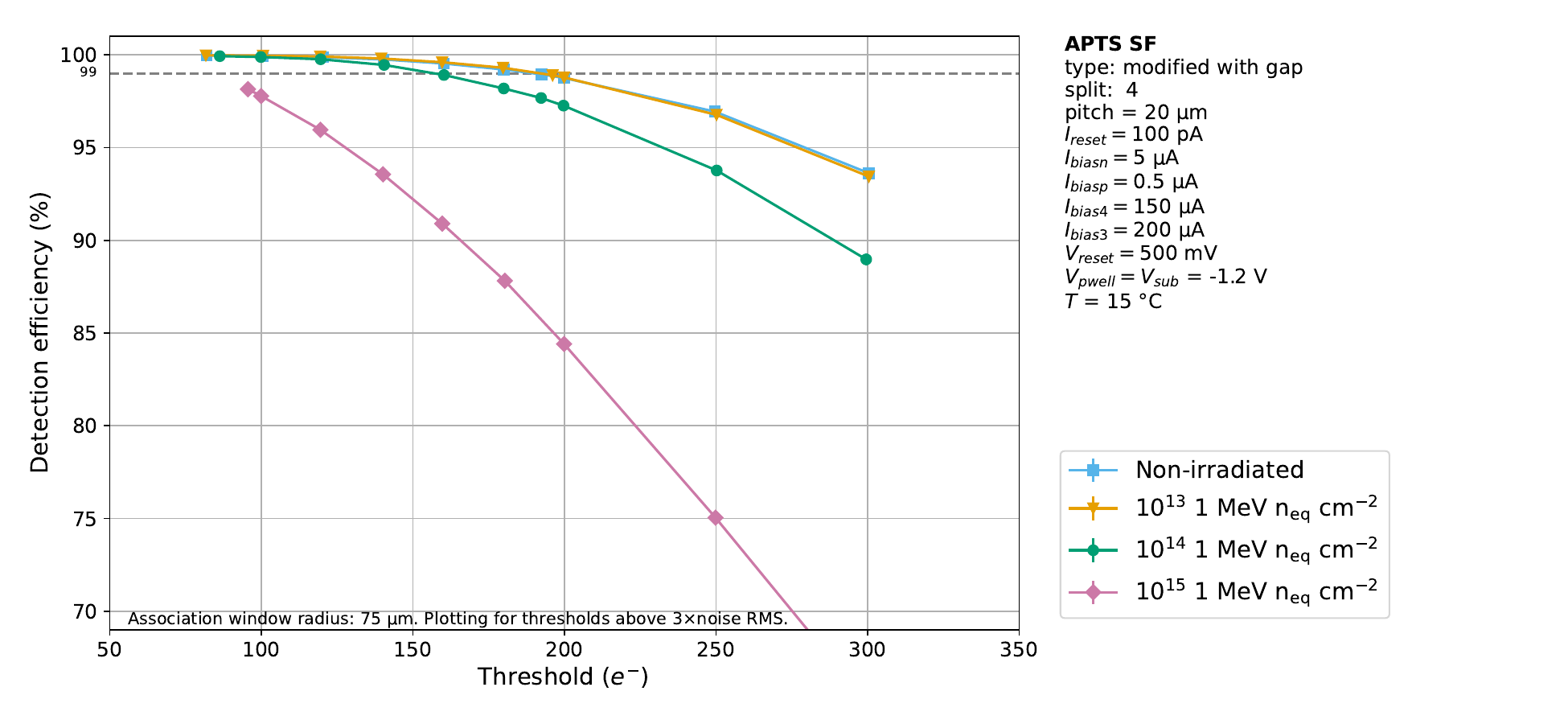}	
    \caption{Efficiency comparison between different NIEL irradiation levels as a function of the applied seed threshold. Chiller temperature was 15~$^\circ$C. APTS with 20~\textmu m pitch, modified with gap, split 4, reference variant, $V_\text{sub}$ = -1.2~V.}
\end{figure}
\begin{figure}[!hbt]
    \centering
	   \includegraphics[width=0.8\textwidth]{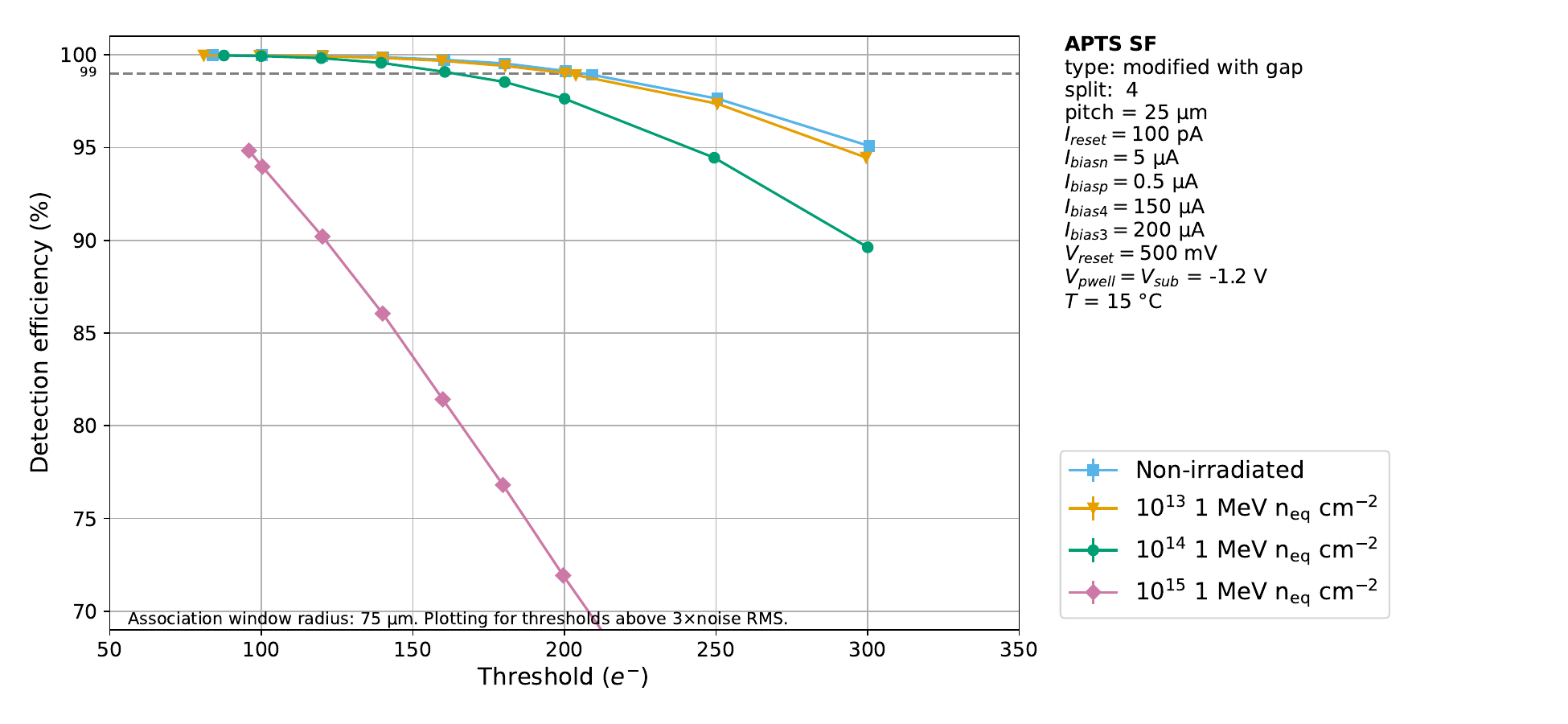}	
    \caption{Efficiency comparison between different NIEL irradiation levels as a function of the applied seed threshold. Chiller temperature was 15~$^\circ$C. APTS with 25~\textmu m pitch, modified with gap, split 4, reference variant, $V_\text{sub}$ = -1.2~V.}
\end{figure}

\begin{figure}[!hbt]
    \centering
	   \includegraphics[width=0.8\textwidth]{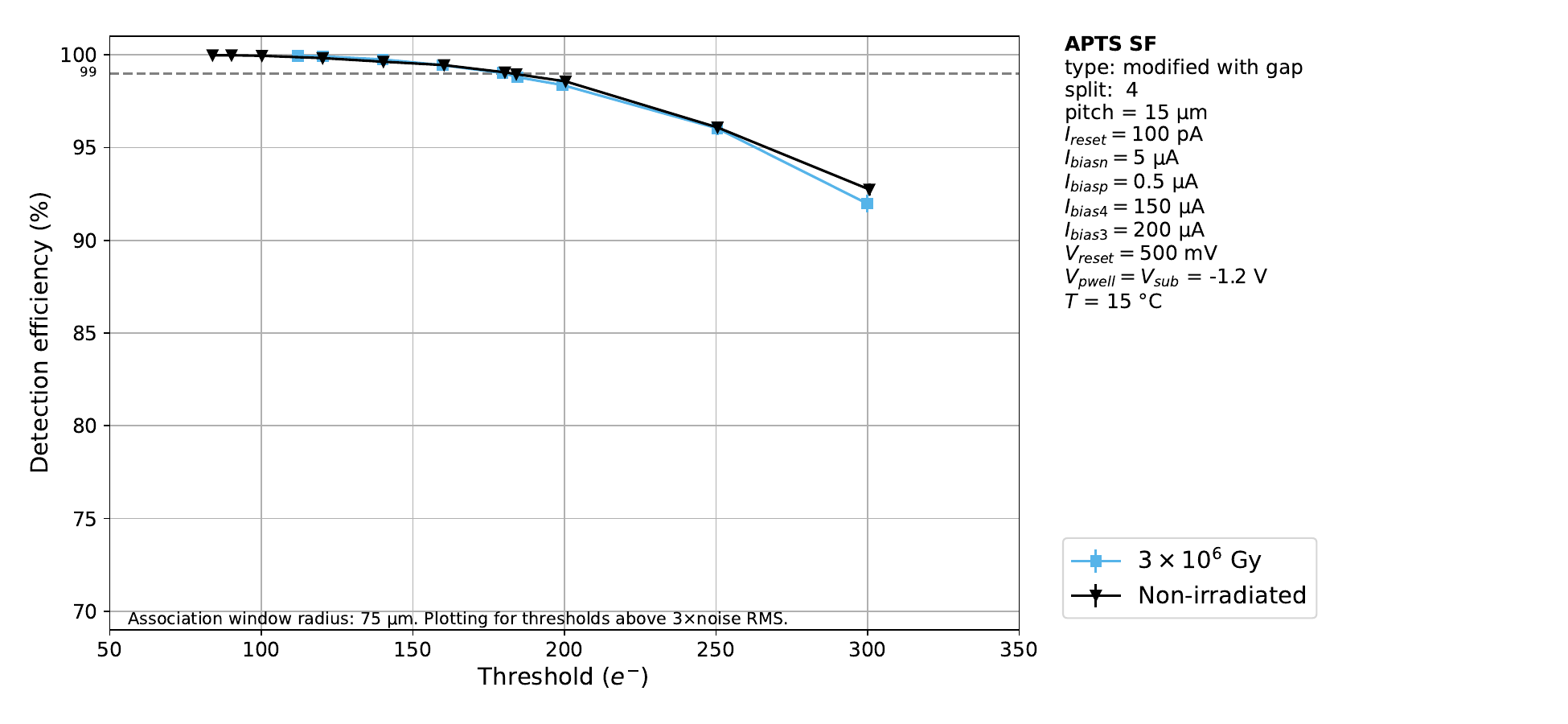}	
    \caption{Efficiency comparison with TID irradiatiad APTS as a function of the applied seed threshold. Chiller temperature was 15~$^\circ$C. APTS with 15~\textmu m pitch, modified with gap, split 4, reference variant, $V_\text{sub}$ = -1.2~V.}
\end{figure}

\begin{figure}[!hbt]
    \centering
	   \includegraphics[width=0.8\textwidth]{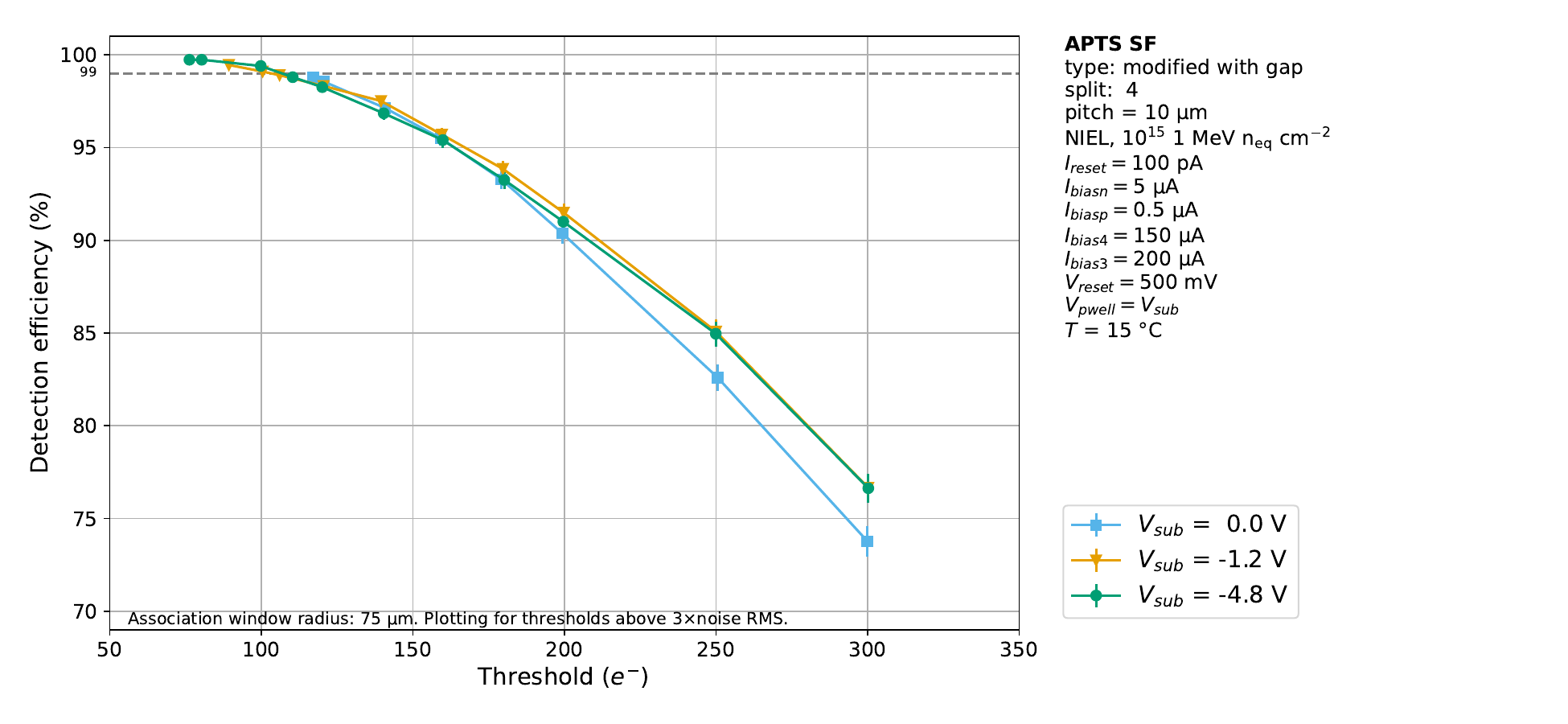}	
    \caption{Efficiency comparison between different substrate reverse biases as a function of the applied seed threshold for a NIEL irradiation level of $10^{15}$~1~MeV~n$_\text{eq}$~cm$^{-2}$. Chiller temperature was 15~$^\circ$C. APTS with 10~\textmu m pitch, modified with gap, split 4, reference variant, $V_\text{sub}$ = -1.2~V.}
\end{figure}

\begin{figure}[!hbt]
    \centering
	   \includegraphics[width=0.8\textwidth]{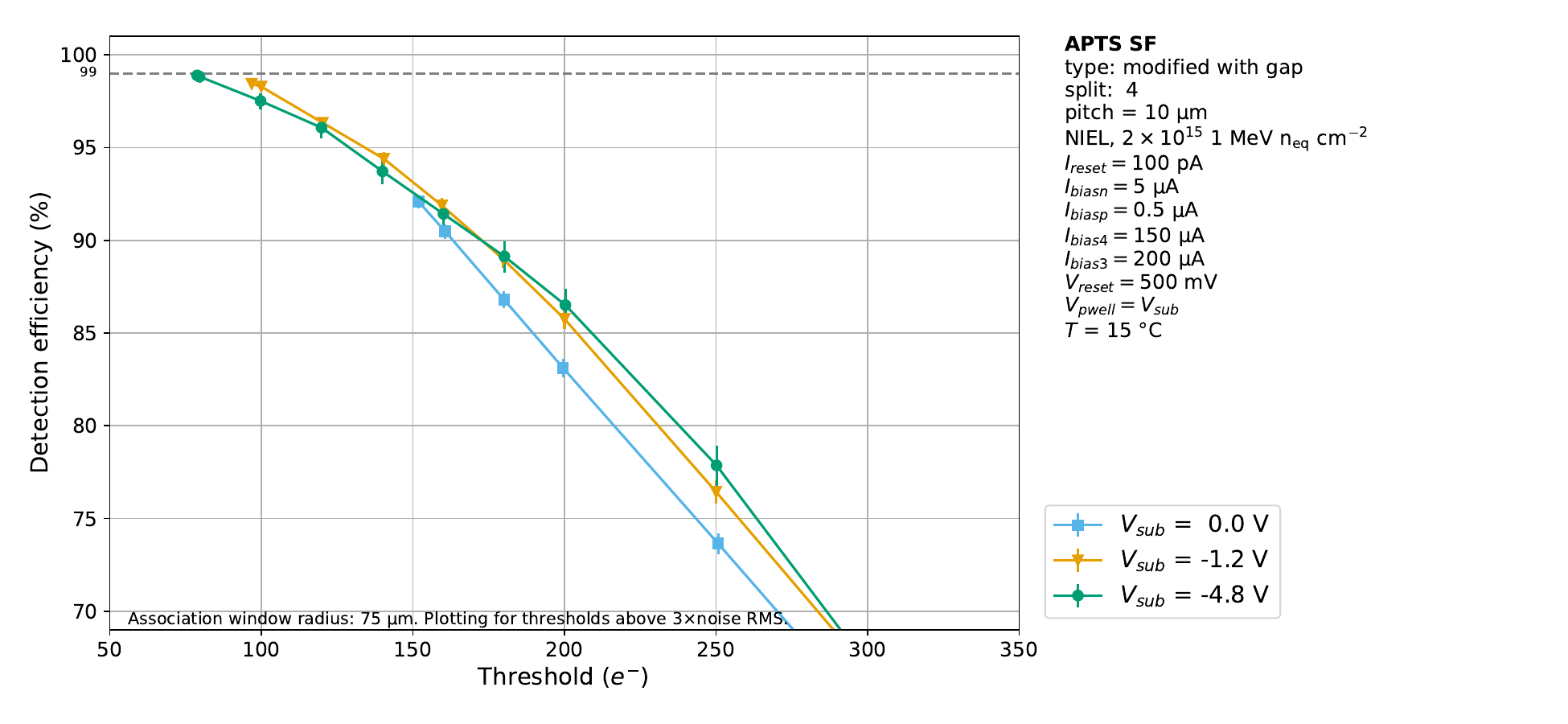}	
    \caption{Efficiency comparison between different substrate reverse biases as a function of the applied seed threshold for a NIEL irradiation level of 2 $\times$ $10^{15}$~1~MeV~n$_\text{eq}$~cm$^{-2}$. Chiller temperature was 15~$^\circ$C. APTS with 10~\textmu m pitch, modified with gap, split 4, reference variant, $V_\text{sub}$ = -1.2~V.}
\end{figure}

\begin{figure}[!hbt]
    \centering
	   \includegraphics[width=0.8\textwidth]{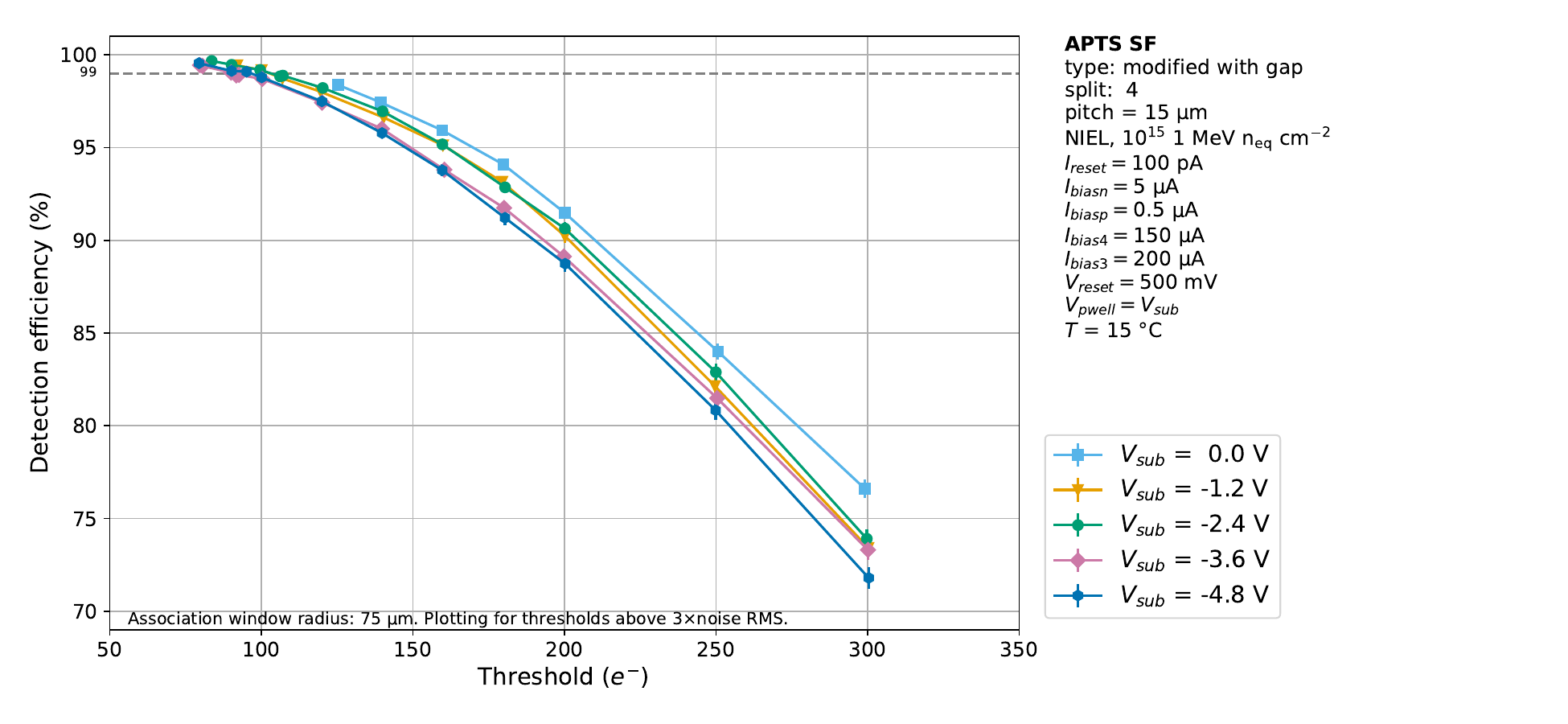}	
    \caption{Efficiency comparison between different substrate reverse biases as a function of the applied seed threshold for a NIEL irradiation level of $10^{15}$~1~MeV~n$_\text{eq}$~cm$^{-2}$. Chiller temperature was 15~$^\circ$C. APTS with 15~\textmu m pitch, modified with gap, split 4, reference variant, $V_\text{sub}$ = -1.2~V.}
\end{figure}

\begin{figure}[!hbt]
    \centering
	   \includegraphics[width=0.8\textwidth]{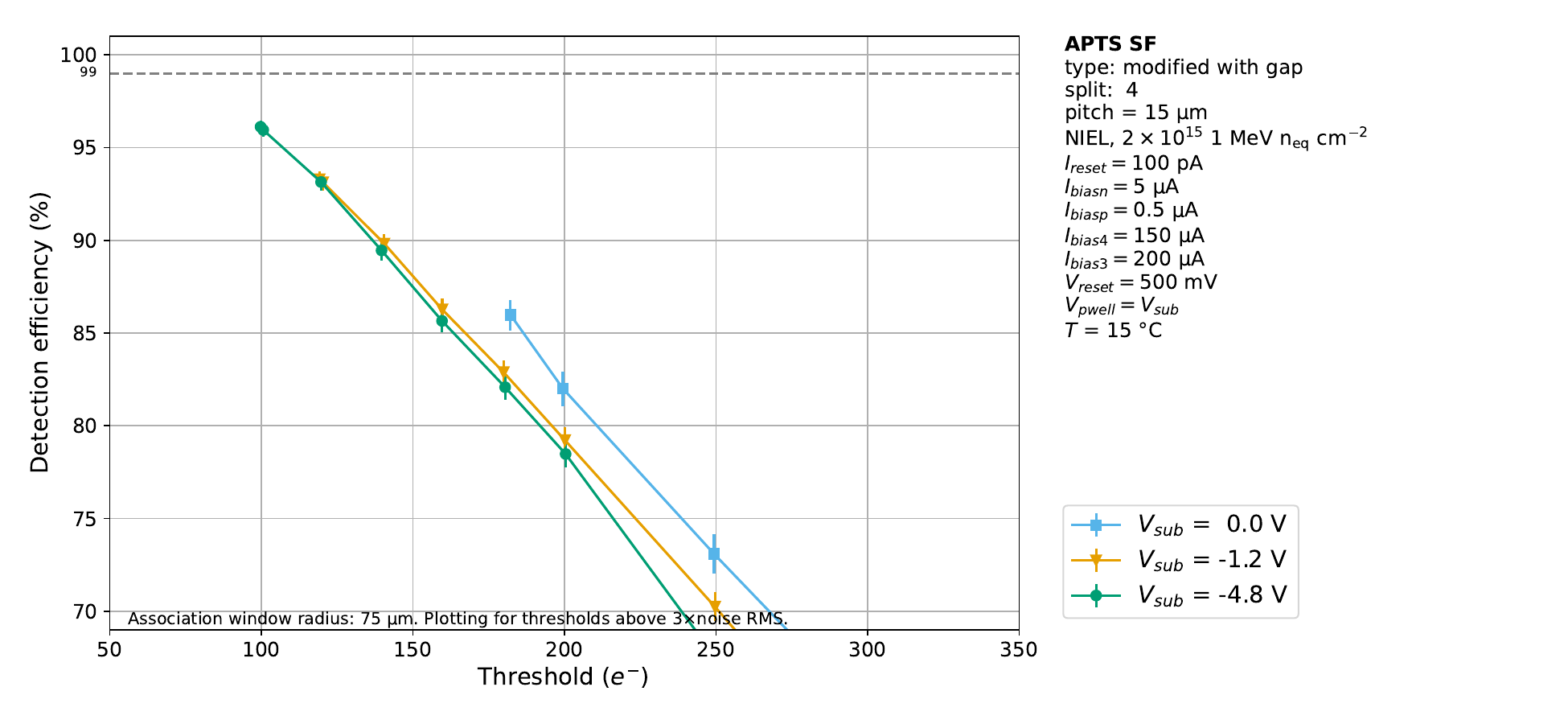}	
    \caption{Efficiency comparison between different substrate reverse biases as a function of the applied seed threshold for a NIEL irradiation level of 2 $\times$ $10^{15}$~1~MeV~n$_\text{eq}$~cm$^{-2}$. Chiller temperature was 15~$^\circ$C. APTS with 15~\textmu m pitch, modified with gap, split 4, reference variant, $V_\text{sub}$ = -1.2~V.}
\end{figure}

\begin{figure}[!hbt]
    \centering
	   \includegraphics[width=0.8\textwidth]{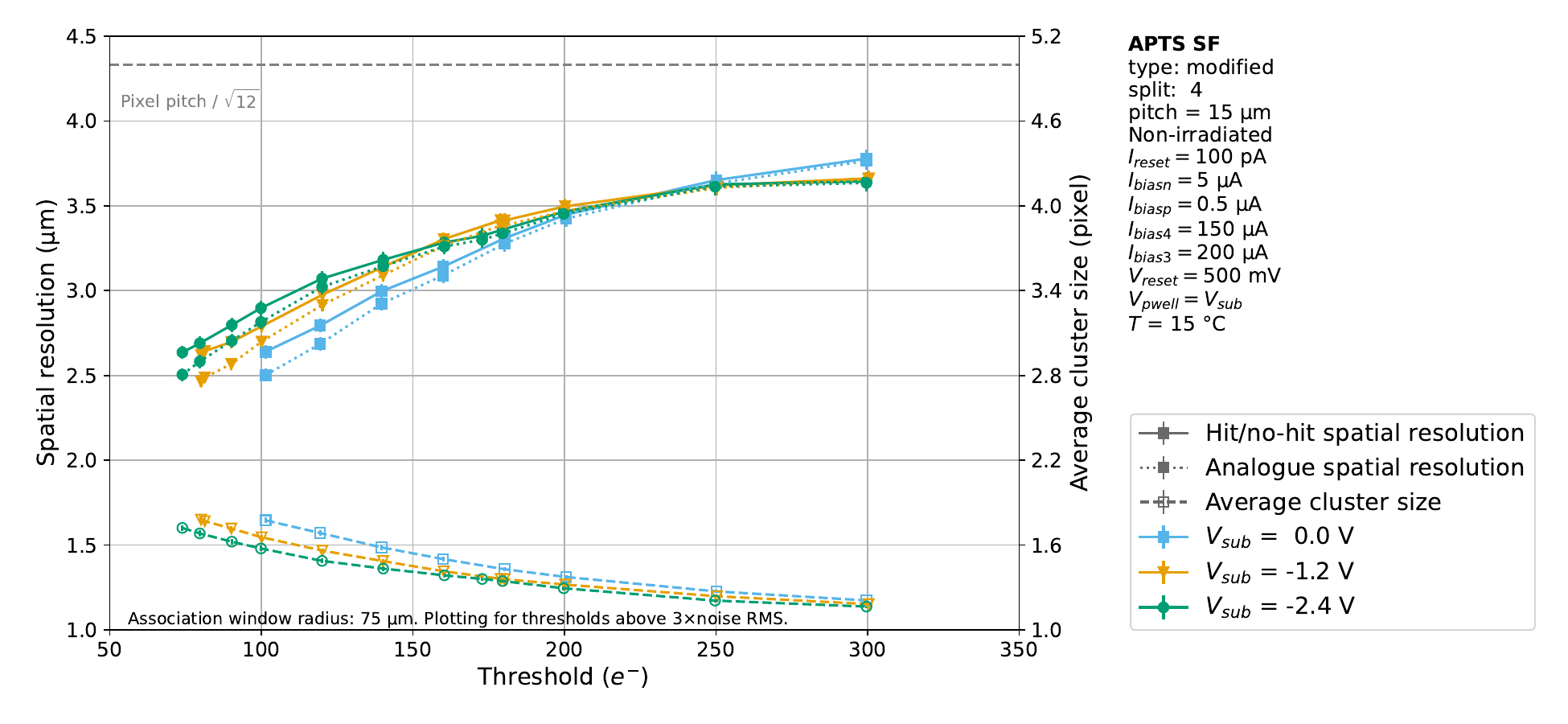 }	
    \caption{Resolution comparison between different reverse substrate voltages as a function of the applied seed threshold. APTS with 15~\textmu m pitch, modified, split 4, reference variant.}
\end{figure}
\begin{figure}[!hbt]
    \centering
	   \includegraphics[width=0.8\textwidth]{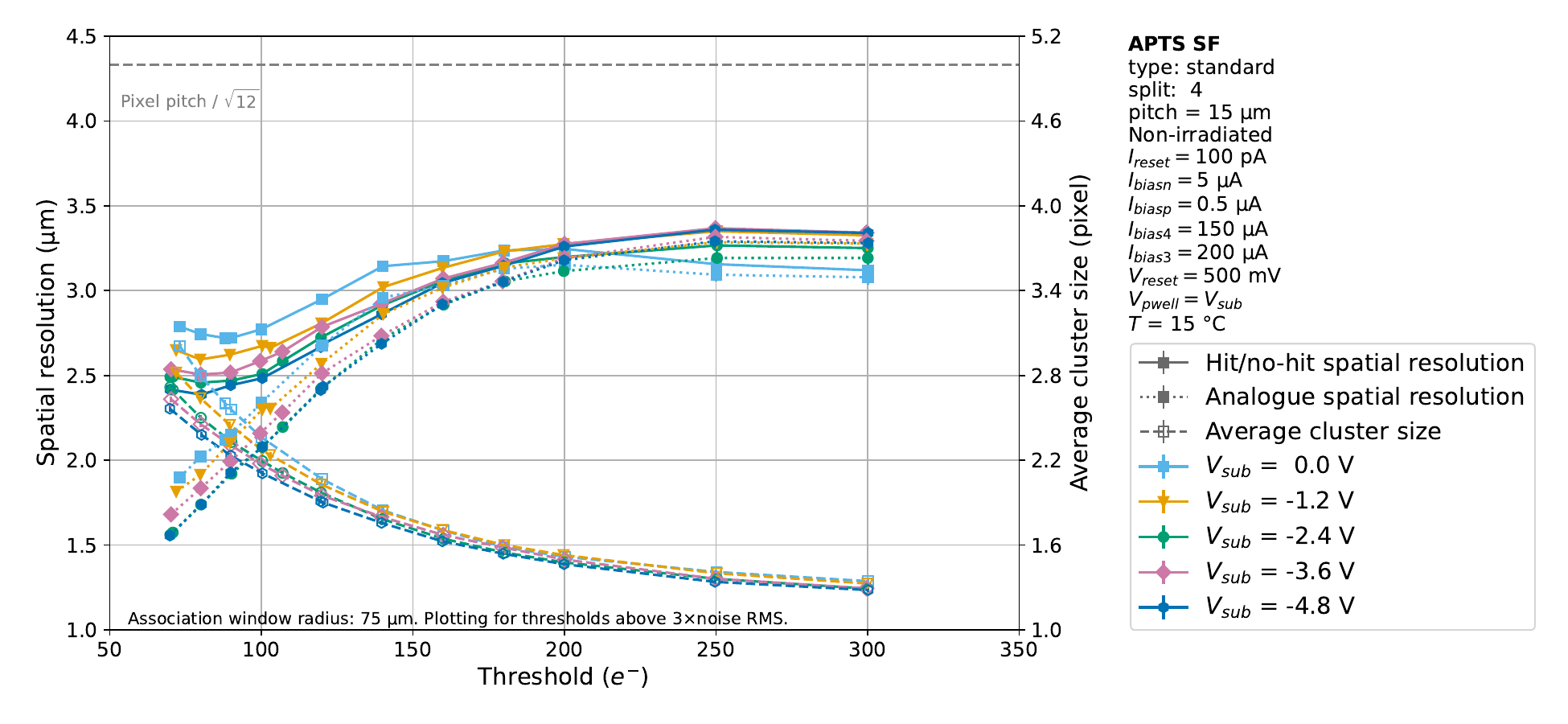 }	
    \caption{Resolution comparison between different reverse substrate voltages as a function of the applied seed threshold. APTS with 15~\textmu m pitch, standard, split 4, reference variant.}
\end{figure}
\begin{figure}[!hbt]
    \centering
	   \includegraphics[width=0.8\textwidth]{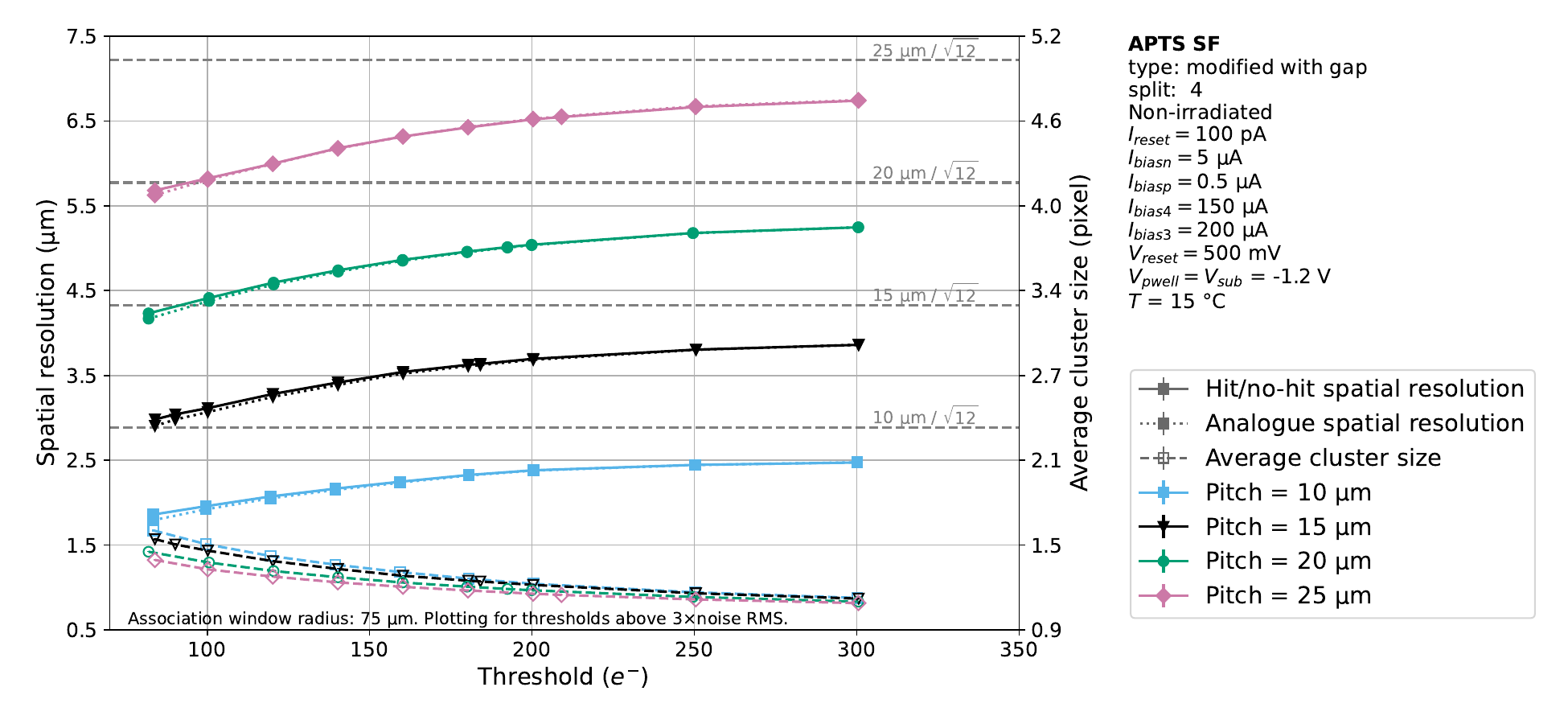 }	
    \caption{Resolution comparison between different pitches as a function of the applied seed threshold. APTS with modified with gap, split 4, reference variant, $V_\text{sub}$ = -1.2~V.}
\end{figure}
\begin{figure}[!hbt]
    \centering
	   \includegraphics[width=0.8\textwidth]{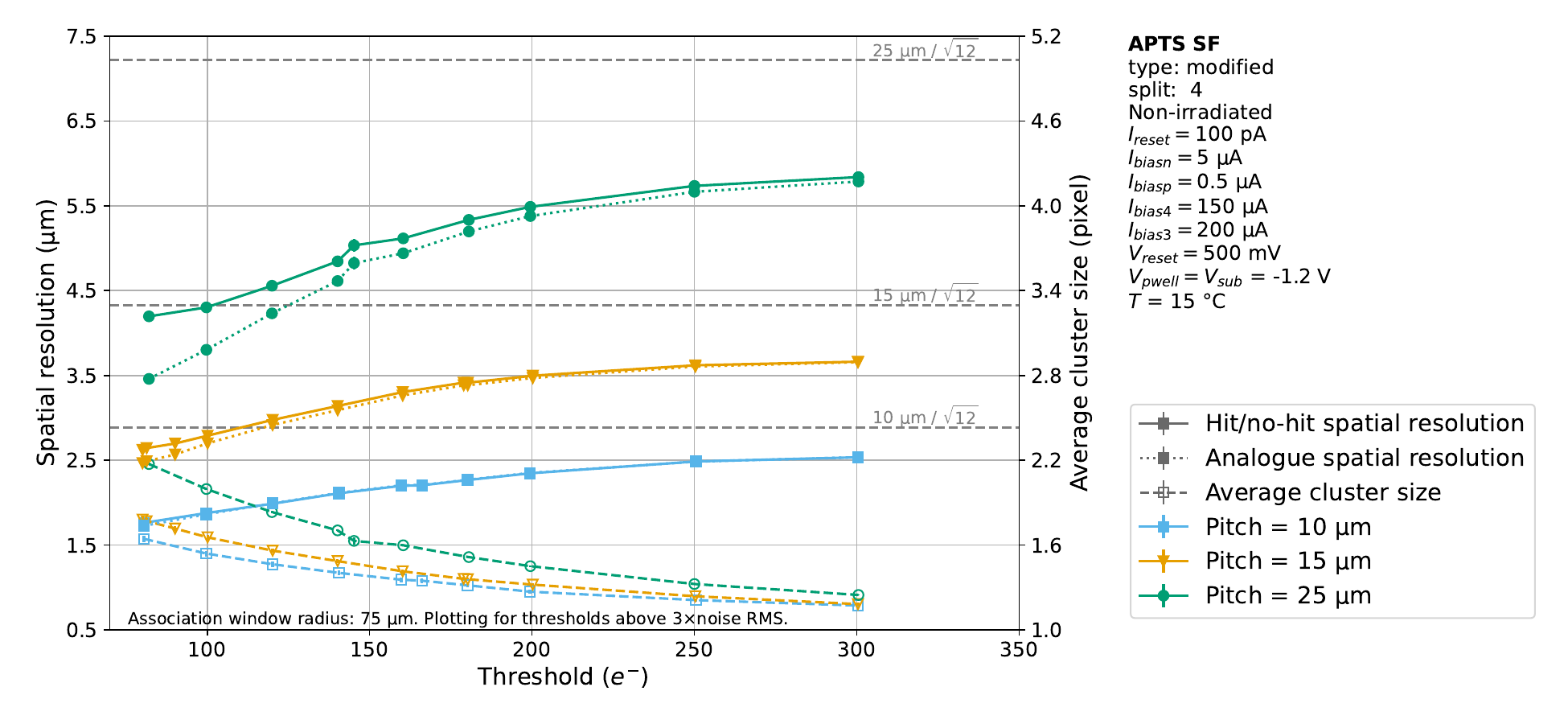 }	
    \caption{Resolution comparison between different pitches as a function of the applied seed threshold. APTS with modified, split 4, reference variant, $V_\text{sub}$ = -1.2~V.}
\end{figure}
\begin{figure}[!hbt]
    \centering
	   \includegraphics[width=0.8\textwidth]{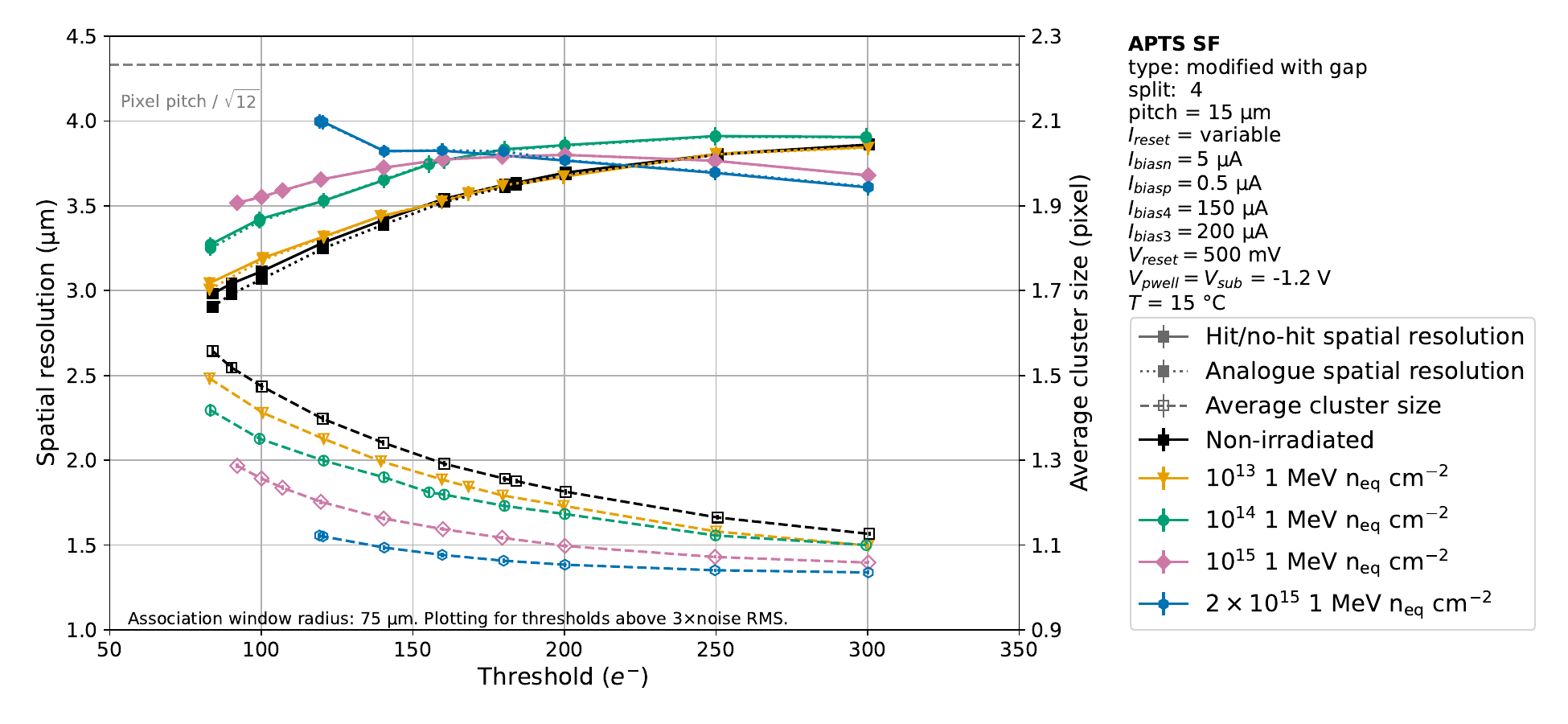}	
    \caption{Resolution comparison between different NIEL irradiation levels as a function of the applied seed threshold. Chiller temperature was 15~$^\circ$C. APTS with 15~\textmu m pitch, modified with gap, split 4, reference variant, $V_\text{sub}$ = -1.2~V.}
\end{figure}
\begin{figure}[!hbt]
    \centering
	   \includegraphics[width=0.8\textwidth]{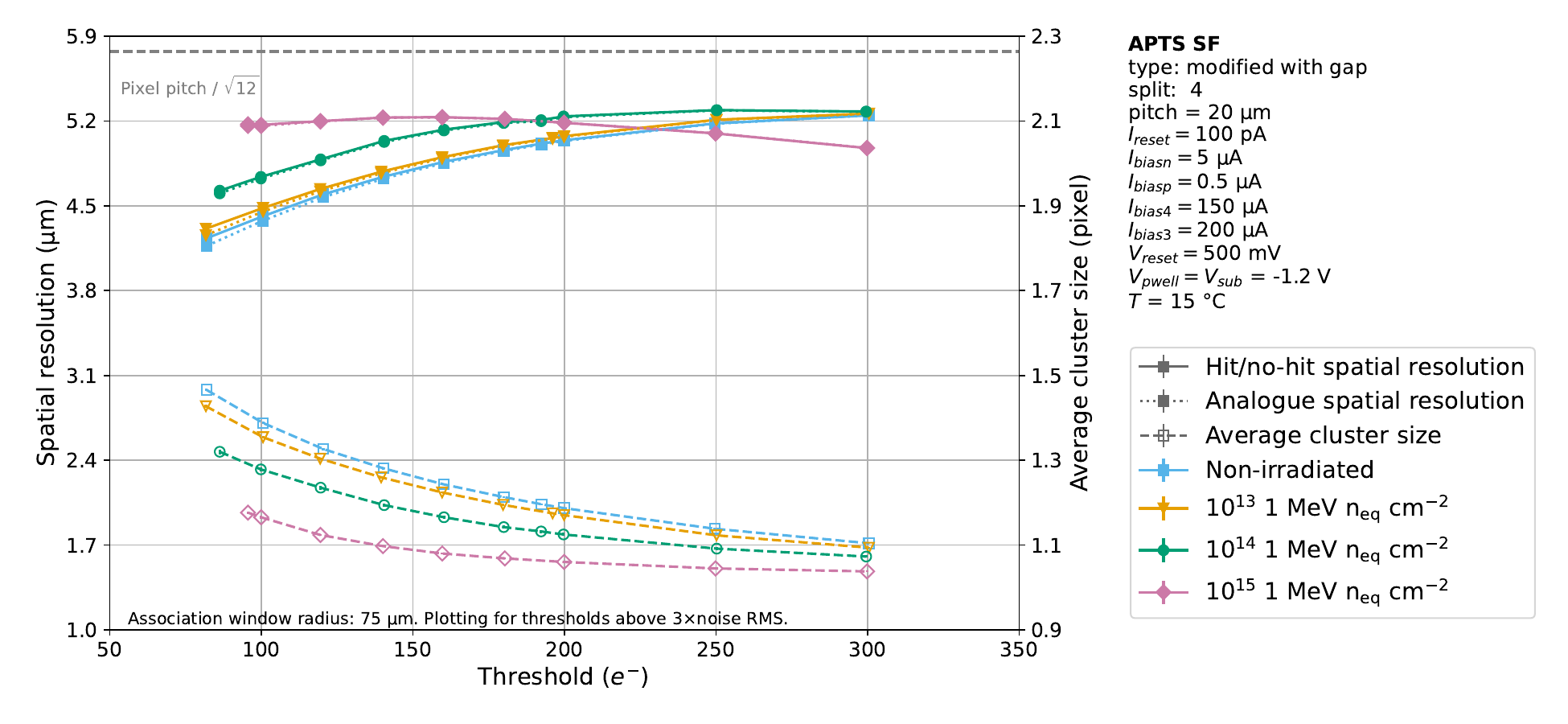}	
    \caption{Resolution comparison between different NIEL irradiation levels as a function of the applied seed threshold. Chiller temperature was 15~$^\circ$C. APTS with 20~\textmu m pitch, modified with gap, split 4, reference variant, $V_\text{sub}$ = -1.2~V.}
\end{figure}
\begin{figure}[!hbt]
    \centering
	   \includegraphics[width=0.8\textwidth]{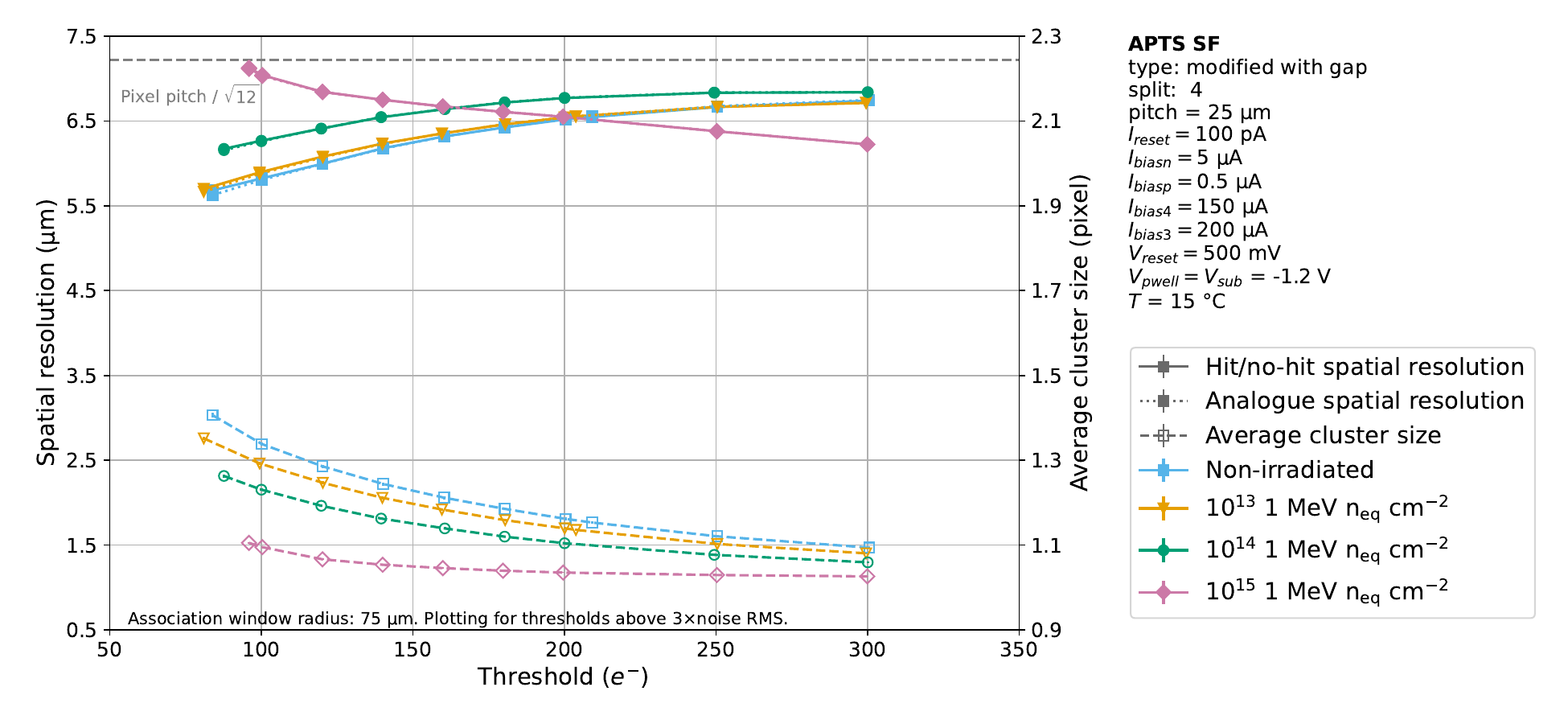}	
    \caption{Resolution comparison between different NIEL irradiation levels as a function of the applied seed threshold. Chiller temperature was 15~$^\circ$C. APTS with 25~\textmu m pitch, modified with gap, split 4, reference variant, $V_\text{sub}$ = -1.2~V.}
\end{figure}

\begin{figure}[!hbt]
    \centering
	   \includegraphics[width=0.8\textwidth]{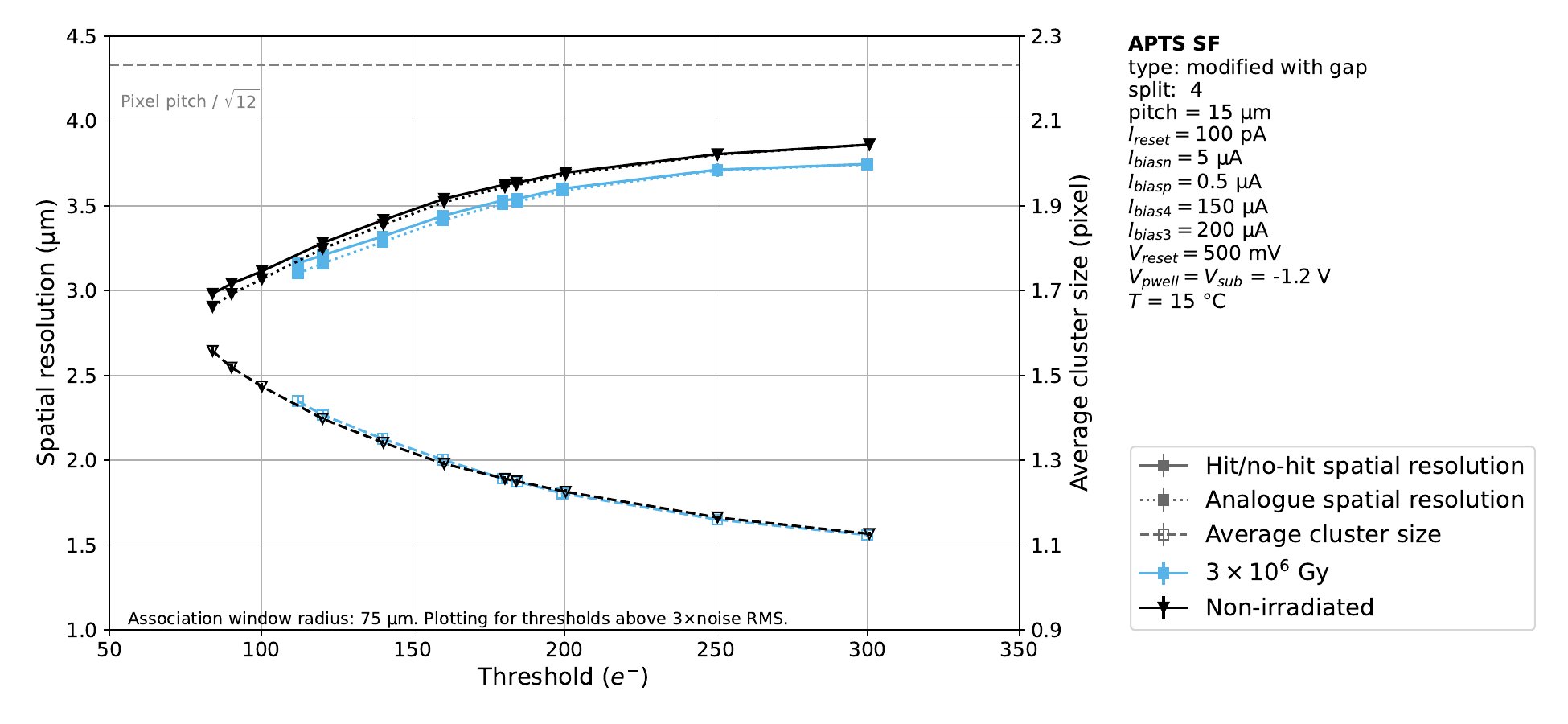}	
    \caption{Resolution comparison with TID irradiatiad APTS as a function of the applied seed threshold. Chiller temperature was 15~$^\circ$C.  APTS with 15~\textmu m pitch, modified with gap, split 4, reference variant, $V_\text{sub}$ = -1.2~V.}
\end{figure}

\begin{figure}[!hbt]
    \centering
	   \includegraphics[width=0.8\textwidth]{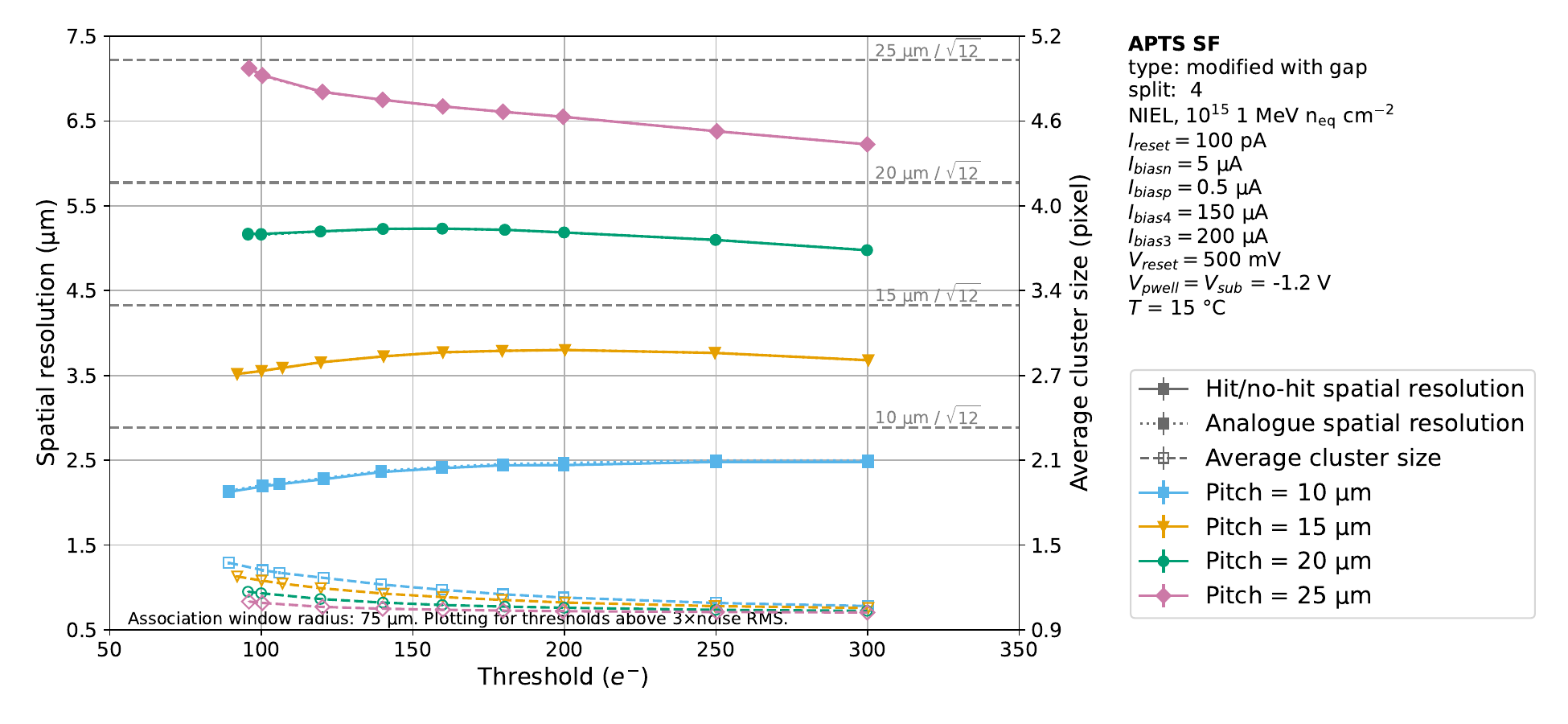}	
    \caption{Resolution comparison between different pitches as a function of the applied seed threshold for a NIEL irradiation level of $10^{15}$~1~MeV~n$_\text{eq}$~cm$^{-2}$. Chiller temperature was 15~$^\circ$C. APTS with modified with gap, split 4, reference variant, $V_\text{sub}$ = -1.2~V.}
\end{figure}








\end{document}